\documentclass[11pt,letterpaper]{article}
\usepackage{amsmath,amsthm,nicefrac}
\usepackage{amsfonts, amstext}
\usepackage{bm}

\usepackage{subcaption}
\usepackage{hhline}
\usepackage{soul}
\usepackage[table,dvipsnames]{xcolor}

\usepackage{typearea}
\typearea{14}
\usepackage[font={small,it}]{caption}

\usepackage{setspace}
\usepackage[compact]{titlesec}

\usepackage[shortlabels]{enumitem}
\usepackage[breaklinks]{hyperref}
\hypersetup{colorlinks=true,
            citebordercolor={.6 .6 .6},linkbordercolor={.6 .6 .6},
citecolor=blue,urlcolor=blue,linkcolor=blue,pagecolor=black,breaklinks}

\usepackage[nameinlink]{cleveref}
\Crefname{algocf}{Algorithm}{Algorithms}
\crefname{algocfline}{line}{lines}
\Crefname{invariant}{Invariant}{Invariants}
\Crefname{claim}{Claim}{Claims}
\Crefname{corollary}{Corollary}{Corollaries}
\Crefname{subclaim}{Subclaim}{Subclaims}

\usepackage{epsfig}
\usepackage{amsthm,amssymb}

\usepackage{mathrsfs}
\usepackage{xspace}
\usepackage{soul}
\usepackage{latexsym}
\usepackage{bbm}
\usepackage{dsfont}

\usepackage{accents}

\usepackage{framed}
\renewenvironment{leftbar}[1][\hsize]
{
    
    \MakeFramed{\hsize#1\advance\hsize-\width\FrameRestore}
}
{\endMakeFramed}

\definecolor{DarkGray}{rgb}{0.66, 0.66, 0.66}
\definecolor{DarkPowderBlue}{rgb}{0.0, 0.2, 0.6}
\definecolor{fluorescentyellow}{rgb}{0.8, 1.0, 0.0}
\definecolor{cerulean}{rgb}{0.0, 0.48, 0.65}
\definecolor{bleudefrance}{rgb}{0.19, 0.55, 0.91}

\usepackage[ruled,vlined,algonl]{algorithm2e}
\SetEndCharOfAlgoLine{}
\SetKwComment{Comment}{\footnotesize$\triangleright$\ }{}

\SetCommentSty{mycommfont}

\usepackage{thmtools,thm-restate}

\allowdisplaybreaks

\newtheorem{theorem}{Theorem}[section]
\newtheorem{lemma}[theorem]{Lemma}
\newtheorem{claim}[theorem]{Claim}
\newtheorem{fact}[theorem]{Fact}
\newtheorem{corollary}[theorem]{Corollary}

\newtheorem{proposition}[theorem]{Proposition}
\newtheorem{observation}[theorem]{Observation}

\theoremstyle{definition}
\newtheorem{defn}[theorem]{Definition}

\theoremstyle{remark}

\usepackage{tikz}
\usepackage{pgfplots}
\pgfplotsset{compat=newest}

\usetikzlibrary{decorations.pathreplacing}
\usetikzlibrary{calc,tikzmark}
\usetikzlibrary{positioning,shapes,arrows}
\usetikzlibrary{matrix}
\usetikzlibrary{patterns,patterns.meta}
\usetikzlibrary{decorations.pathreplacing,calligraphy,backgrounds}
\usetikzlibrary{snakes}

\usepackage{comment}

\usepackage[
 backend=biber,
 style=alphabetic,
 citetracker,
 hyperref=auto,
 maxcitenames=5,
 sortcites,
 sorting=nyt,
 maxbibnames=12,
 date=year,
 isbn=false,
 url=false,
 doi=false,
 eprint=false,
]{biblatex}

\newbibmacro{string+doiurlisbn}[1]{
  \iffieldundef{doi}{
    \iffieldundef{url}{
      #1
    }{
      \href{\thefield{url}}{#1}
    }
  }{
    \href{http://dx.doi.org/\thefield{doi}}{#1}
  }
}

\DeclareFieldFormat
[article,inbook,incollection,inproceedings,patent,thesis,unpublished]
  {title}{\usebibmacro{string+doiurlisbn}{#1}}

\usepackage{titlesec}
\usepackage{lipsum}
\usepackage{float}
\titlespacing{\paragraph}{
  0pt}{
  0.3\baselineskip}{
  1em}

\usepackage[normalem]{ulem}

\newcommand{\eps}{\varepsilon}
\newcommand{\sse}{\subseteq}

\newcommand{\factor}{\smash{\ceil{\nicefrac{1}{\varepsilon}}}}
\newcommand{\factorWithoutCeil}{\smash{\nicefrac{1}{\varepsilon}}}
\newcommand{\bfactor}{\smash{\bceil{\textstyle \nicefrac{1}{\varepsilon}}}}

\newcommand{\myAlfa}{(1-\eps)}
\newcommand{\oneminusA}{\eps}
\newcommand{\ainv}{(1-\eps)^{-1}}
\newcommand{\threshold}{\frac{\varepsilon}{1-\varepsilon}\xspace}

\newcommand{\ALG}{{\ensuremath{\mathrm{SLF}}}\xspace}
\newcommand{\SLF}{\ALG}
\newcommand{\SETF}{{\ensuremath{\mathrm{SETF}}}\xspace}
\newcommand{\SETFI}{{\ensuremath{\mathrm{SETFI}}}\xspace}

\newcommand{\ALGc}{\SETF}
\newcommand{\ALGi}{\SETFI}

\newcommand{\alg}{\ensuremath{\mathrm{SLF}}\xspace}
\newcommand{\opt}{\ensuremath{\mathrm{OPT}}\xspace}
\newcommand{\OPT}{\opt}

\newcommand{\argmax}{\operatorname{argmax}}

\newcommand{\minsuffix}{\operatorname{min-suffix}}

\newcommand{\vol}{\operatorname{vol}}

\newcommand{\greedy}{\textsc{GreedyMatching}}

\newcommand{\bceil}[1]{\big\lceil #1 \big\rceil}
\newcommand{\ceil}[1]{\left\lceil #1 \right\rceil}

\renewcommand{\emptyset}{\varnothing}

\newcommand{\floor}[1]{\lfloor#1\rfloor}

\newcommand{\cA}{{\mathcal {A}}}
\newcommand{\cE}{{\mathcal {E}}}

\newcommand{\cJ}{{\mathcal {J}}}
\newcommand{\cO}{{\mathcal {O}}}

\newcommand{\calJ}{\mathcal{J}}
\newcommand{\calA}{\mathcal{A}}

\newcommand{\calI}{\mathcal{I}}

\newcommand{\calB}{\mathcal{B}}

\newcommand{\ind}{\ensuremath{\mathds{1}}}
\newcommand{\EX}{\mathbb{E}}
\newcommand{\VAR}{\mathbb{V}}
\newcommand{\pr}{\mathrm{Pr}}

\newcommand{\early}{early-arriving\xspace}

\newcommand{\leader}{L}

\newcommand{\nf}{\nicefrac}

\usepackage{xcolor}
\definecolor{job1}{HTML}{6D9DC5}
\definecolor{job2}{HTML}{E0CC74}
\definecolor{job3}{HTML}{649884}
\definecolor{job4}{HTML}{C9809C}
\definecolor{job5}{HTML}{F39B6D}
\definecolor{job6}{HTML}{547781}
\definecolor{job7}{HTML}{1c8591}

\tikzset{
vertex/.style={circle, draw, fill=black, black, inner sep=0pt, minimum width=7pt},
asg/.style={line width=1pt},
}

\newcommand\machine{}
\def\machine[#1](#2,#3)(#4,#5){
  \draw[black,#1,line width=2pt] (#4,#3) -- (#2,#3) -- (#2,#5) -- (#4,#5);
}

\newcommand\interval{}
\def\interval[#1](#2,#3){
  \draw[line width=1pt,dotted,black,#1] (#2,0) -- (#2,#3);
}

\newcommand\rectjob{}
\def\rectjob[#1](#2,#3)(#4,#5){
  \draw[black,fill=#1,line width=0.7pt] (#2,#3) rectangle ++(#4,#5);
}

\newcommand\rectjobT{}
\def\rectjobT[#1](#2,#3)(#4,#5)(#6){
  \draw[black,fill=#1,line width=0.7pt] (#2,#3) rectangle ++(#4,#5);
  \node[#1,black] at (#2+#4/2.0,#3+#5/2.0) {{#6}};
}

\newcommand\matchingjob{}
\def\matchingjob(#1,#2)(#3,#4){
  \draw[black,fill=none,line width=1.2pt] (#1,#2) rectangle ++(#3,#4);
}

\newcommand\partialjob{}
\def\partialjob(#1,#2)(#3,#4){
  \draw[black,dashed,fill=none,line width=1.2pt] (#1,#2) rectangle ++(#3,#4);
}

\newcommand\rectjobnob{}
\def\rectjobnob[#1](#2,#3)(#4,#5){
  \draw[white,fill=#1,line width=0.0pt] (#2,#3) rectangle ++(#4,#5);
}

\newcommand{\convexpath}[2]{
  [
  create hullcoords/.code={
    \global\edef\namelist{#1}
    \foreach [count=\counter] \nodename in \namelist {
      \global\edef\numberofnodes{\counter}
      \coordinate (hullcoord\counter) at (\nodename);
    }
    \coordinate (hullcoord0) at (hullcoord\numberofnodes);
    \pgfmathtruncatemacro\lastnumber{\numberofnodes+1}
    \coordinate (hullcoord\lastnumber) at (hullcoord1);
  },
  create hullcoords
  ]
  ($(hullcoord1)!#2!-90:(hullcoord0)$)
  \foreach [
  evaluate=\currentnode as \previousnode using \currentnode-1,
  evaluate=\currentnode as \nextnode using \currentnode+1
  ] \currentnode in {1,...,\numberofnodes} {
    let \p1 = ($(hullcoord\currentnode) - (hullcoord\previousnode)$),
    \n1 = {atan2(\y1,\x1) + 90},
    \p2 = ($(hullcoord\nextnode) - (hullcoord\currentnode)$),
    \n2 = {atan2(\y2,\x2) + 90},
    \n{delta} = {Mod(\n2-\n1,360) - 360}
    in
    {arc [start angle=\n1, delta angle=\n{delta}, radius=#2]}
    -- ($(hullcoord\nextnode)!#2!-90:(hullcoord\currentnode)$)
  }
}

\title{A Little Clairvoyance Is All You Need}

\author{
   Anupam Gupta\thanks{
       New York University, \texttt{anupam.g@nyu.edu}. 
   } \and Haim Kaplan\thanks{
       Tel Aviv University,
       \texttt{haimk@post.tau.ac.il}
   }  \and Alexander Lindermayr\thanks{
       University of Bremen,
       \texttt{linderal@uni-bremen.de}
   }  \and Jens Schlöter\thanks{
       Centrum Wiskunde \& Informatica (CWI),
       \texttt{jens.Schloter@cwi.nl}
   } \and
        Sorrachai Yingchareonthawornchai\thanks{
       Institute for Theoretical Studies, ETH Zürich,
       \texttt{sorrachai.yingchareonthawornchai@eth-its.ethz.ch}
   }}

\date{}
\begin{document}

\maketitle
\begin{abstract}
  We revisit the classical problem of minimizing the total \emph{flow
    time} of jobs on a single machine in the online setting where jobs
  arrive over time. It has long been known that the Shortest Remaining
  Processing Time (SRPT) algorithm is optimal (i.e., $1$-competitive)
  when the job sizes are known up-front~[Schrage, 1968]. But in the
  non-clairvoyant setting where job sizes are revealed only when the
  job finishes, no algorithm can be constant-competitive~[Motwani,
  Phillips, and Torng, 1994].

  \medskip  We consider the $\eps$-\emph{clairvoyant} setting, where
  $\eps \in [0,1]$, and each job's processing time becomes known once
  its remaining processing time equals an $\eps$ fraction of its
  processing time. This captures settings where the system user uses
  the initial $(1-\eps)$ fraction of a job's processing time to learn
  its true length, which it can then reveal to the algorithm. The
  model was proposed by Yingchareonthawornchai and Torng (2017), and
  it smoothly interpolates between the clairvoyant setting (when
  $\eps = 1$) and the non-clairvoyant setting (when $\eps = 0$).  In
  a concrete sense, we are asking: \emph{how much knowledge is
    required to circumvent the hardness of this problem?}

  \medskip We show that \emph{a little knowledge is enough}, and that
  a constant competitive algorithm exists for every constant $\eps >
  0$. More precisely, for all $\eps \in (0,1)$, we present a
  deterministic $\factor$-competitive algorithm, which is optimal for
  deterministic algorithms. We also present a matching lower bound (up
  to a constant factor) for randomized algorithms.

  \medskip Our algorithm to achieve this bound is remarkably simple and
  applies the ``optimism in the face of
  uncertainty'' principle.
  For each job, we form an optimistic estimate of its length,
  based on the information revealed thus far and run SRPT on
  these optimistic estimates.
The proof relies on maintaining a matching between the jobs in \OPT's queue and the algorithm's queue,  with small prefix expansion. We achieve this
by carefully choosing a set of jobs \emph{to
  arrive earlier than their release times} without changing the
  algorithm, and possibly helping the adversary.
 These early arrivals allow us to maintain structural properties inductively, giving us the tight guarantee.

\end{abstract}

\newpage
\setcounter{tocdepth}{2}
\tableofcontents
\clearpage

\pagenumbering{arabic}
\setcounter{page}{1}
\newpage

\newcommand{\ph}{\hat{p}}

\section{Introduction}

We consider the classical problem of minimizing the total \emph{flow
  time} (or \emph{response time}) of jobs on a single machine in the
online setting where jobs arrive over time and preemption is allowed. The flow time is the total
time the job spends in the system
and we consider
the objective of minimizing the total (or equivalently, the average)
flow time of all the jobs.
The \emph{Shortest
  Remaining Processing Time} (SRPT) algorithm has long been known to solve
the problem optimally. SRPT is a preemptive algorithm that, at any
point in time, processes the available job with the shortest remaining
processing time. If the total processing required by the job $j$ is
denoted by $p_j$, and a job $j$ has received $e_j(t)$ amount of
processing by time $t$ (i.e., its \emph{elapsed time} is $e_j(t)$),
then
the remaining processing is
$r_j(t) := p_j - e_j(t)$. The SRPT algorithm processes a job
minimizing $r_j(t)$. Note
that
the algorithm
requires us to know the processing requirement $p_j$ of each job $j$. But
what if this processing requirement is not known (e.g., in schedulers
for operating systems)?

A first attempt was to
consider the \emph{non-clairvoyant} setting in which
the length $p_j$ of the job is only learned when the job completes,
i.e., once it has received $p_j$ amount of processing.
This very demanding setting was  considered by Motwani, Phillips, and
Torng~\cite{MotwaniPT94}, who showed that every deterministic
algorithm for total flow time has competitive ratio $\Omega(n^{1/3})$ in the non-clairvoyant setting (where $n$ is the number of jobs), and every randomized algorithm is
$\Omega(\log n)$-competitive.

How can we sidestep these strong lower bounds in the purely
non-clairvoyant setting?
What if we have \emph{partial information} about $p_j$? The
\emph{semi-clairvoyant} model of Bender, Muthukrishnan, and
Rajaraman~\cite{BenderMR04} considers the setting where the ``class'', defined as
$\floor{\log_\mu p_j}$, of  job $j$ is revealed when $j$ is released (for
some parameter~$\mu$). More recently, the \emph{predicted processing
  time} model of Azar, Leonardi, and Touitou~\cite{AzarLT21,AzarLT22} considers the
case where the algorithm is given a prediction~$\hat{p}_j$ for the
true processing time $p_j$, such that $\hat{p}_j/p_j \in [\nf1\mu,\mu]$
for some (known or unknown) parameter $\mu$. These predictions can be
naturally motivated: e.g., a web server can approximate the download
time, based on document size, traffic bandwidth, recent download
times, etc. These works and others give algorithms whose performance
smoothly improves as the ``distortion'' $\mu$ gets smaller,
showing that access to rough estimates of the job size \emph{upon job arrival}
suffices to give
improved guarantees.

Our work asks: \emph{what algorithms can we use
  if we have no \underline{a priori} estimates of the job size, but we
  are able to learn the job sizes as we process the jobs?} This
question, which is natural in the age of machine learning, was made concrete by
Yingchareonthawornchai and Torng~\cite{YingchareonthawornchaiT17} in
their model:\footnote{They call the model \emph{delayed-clairvoyant},
  but we suggest using $\eps$-clairvoyant.}
\begin{quote}
  \textbf{$\eps$-Clairvoyant Model}. Fix some constant
  $\eps \in [0,1]$. When a job $j$ arrives, we know nothing about its
  processing time $p_j$. However, we start to learn the job size as it is being
  processed: the value $p_j$ is revealed when an $\eps$ fraction of
  the job $j$ remains to be processed.
\end{quote}
This model gives a different interpolation between clairvoyant and
non-clairvoyant models, so that $0$-clairvoyant is the non-clairvoyant
model of \cite{MotwaniPT94}, and $1$-clairvoyant is the classical
(``fully'' clairvoyant) model.
This abstract model captures practical
scheduling situations. In the context of job
profiling~\cite{hilman2018task,xie2021two,weerasiri2017taxonomy} and
scheduling with predictions, we can imagine that our prediction
mechanism gets better estimates of a job's runtime as we process a
larger fraction of it until we have a perfect prediction at some
point.  Finally, as a toy example relevant to researchers, if the
scheduler is a (busy) researcher and the jobs are research projects,
then the processing time of every job is initially unknown to the
researcher. Still, after spending time on a job, the researcher has a
much better understanding of the remaining time, whether it will take
a while or is about to be
completed.

Observe that any algorithm now has to alternate between
\emph{exploring} (to learn the processing times of jobs) and
\emph{exploiting} (finishing jobs whose processing times are now known
to be small). How should we schedule jobs to minimize the total flow
time in the $\eps$-clairvoyant model? \cite{YingchareonthawornchaiT17}
proposed an algorithm based on interleaving Round-Robin and SRPT that
achieves constant-competitiveness for values of $\eps$ close to $1$
(when the job lengths are revealed early), but their approaches fail
when $\eps \leq \nf12$. They ask the question:
  Can we achieve $f(\eps)$-competitive algorithms in the
  $\eps$-clairvoyant model, for all $\eps \in (0,1)$?

\subsection{Our Results}

Our main result answers this question in the affirmative, and shows
that \emph{even a little clairvoyance is enough!} In fact, we give
optimal results for this model. Let us call an algorithm
$\eps$-clairvoyant if it operates in the $\eps$-clairvoyant model.

\begin{theorem}[Main Theorem]
    \label{thm:main1}
    For any constant $\eps \in (0,1]$, there is a deterministic
    $\eps$-clairvoyant algorithm \ALG (Shortest Lower-bound First)
    that is $\factor$-competitive for the objective of minimizing the
    total flow time.
\end{theorem}

The \ALG algorithm is particularly simple and natural, and can be seen
as
following the paradigm of ``optimism in the face of uncertainty''.
For each job $j$, the algorithm maintains an optimistic prediction of
its true length---\emph{the minimum possible length consistent
  with the observed facts}.  Specifically, if the job length $p_j$ has
not yet been revealed, the prediction is
\[ \ph_j(t) = e_j(t)/(1-\eps). \] Recall that $e_j(t)$ is the elapsed
time for job $j$---the amount of processing $j$ has received by
time~$t$. Of course, once the length of job $j$ is revealed, we define
$\ph_j(t) = p_j$. Now we run SRPT with these predictions: \emph{we run
  a job $j$ that minimizes $\ph_j(t) - e_j(t)$; i.e., any job that
minimizes the lower bound on its remaining processing time.}

To understand this algorithm, consider the case when $\eps = 1$, i.e.\
the lengths of the jobs are known upon arrival. In this case, the
algorithm is identical to SRPT, which is optimal. On the other hand,
if $\eps \to 0$, the algorithm behaves like SETF (Shortest Elapsed
Time First) and executes the jobs with the smallest elapsed times,
leading to doing Round-Robin or ``water filling'' on the jobs that
have received the least processing so far.
Note that the SETF
algorithm does not have a bounded competitive ratio
(independent of $n$) since it is a non-clairvoyant algorithm.

Our next result considers a special case where all jobs arrive at time $0$;
for this setting, the SETF algorithm (which in this setting corresponds to Round-Robin) is
$2$-competitive~\cite{MotwaniPT94}. We
show how \ALG smoothly
interpolates between SRPT and SETF in this simultaneous arrival
setting.

\begin{theorem}[Simultaneous Arrivals]
    \label{thm:main2}
  For all $\eps \in [0,1]$, the algorithm \ALG is
  $(2-\eps)$-competitive for the objective of total flow time in the $\eps$-clairvoyant setting if all
  the jobs arrive at time $0$. Moreover, every randomized algorithm in this case has a competitive ratio of at least $2-\eps$.
\end{theorem}

\paragraph{Tightness of Our Results.}
Note that unlike the simultaneous arrivals case, the competitive ratio
of $\factor$ in \Cref{thm:main1} is 1 when $\eps = 1$, but becomes $2$
as soon as $\eps$ drops even slightly below $1$.  Since
$\nicefrac{1}{\eps} = (2 - \eps) + \sum_{i \geq 2} (1-\eps)^i$, it
would be natural to expect the ``right'' bound to be smooth, and that
there should be an algorithm that is $\nicefrac{1}{\eps}$-competitive,
without the ceiling. However, this is not the
case.
We show that the bound $\factor$ is
\emph{exactly} optimal for deterministic
algorithms.
\begin{theorem}[Lower Bounds]
    \label{thm:main3}
    For all $\eps \in (0,1]$, every deterministic $\eps$-clairvoyant
    algorithm has a competitive ratio of at least $\factor$ for the
    objective of total flow time.
    Moreover, every  randomized $\eps$-clairvoyant algorithm has a competitive ratio of $\Omega(\factorWithoutCeil)$.
\end{theorem}

The asymptotic lower bounds of $\Omega(\factorWithoutCeil)$ for algorithms can be obtained by extending standard constructions (e.g.,  \cite{MotwaniPT94,AzarLT22}), but these constructions do not give us the tight deterministic lower bound: this construction is more delicate and requires additional care.

\paragraph{Relating to Resource Augmentation.}

The speed augmentation approach of \textcite{KalyanasundaramP00}
compares the performance of the algorithm with a speed of $(1+\eps)$
to the optimal solution for unit speed.
In this setting, \cite{KalyanasundaramP00} show that the $\SETF$
algorithm is $(1+\nicefrac{1}{\varepsilon})$-competitive.
As a by-product of our analysis of the $\eps$-clairvoyant algorithm
$\ALG$, we obtain an alternate proof of $\SETF$'s competitiveness with similar bounds via a black-box reduction.

\begin{restatable}{lemma}{lemReduction}
  \label{lem:reduction}
  If there is some $\eps$-clairvoyant algorithm $\calA$ for total flow
  time that is $f(\eps)$-competitive for all $\eps \in [0,1]$, then
  there exists a non-clairvoyant algorithm $\calB$ that is
  $(1+\delta)$-speed $f(\frac{\delta}{1+\delta})$-competitive for all
  $\delta > 0$.  Moreover, if the algorithm $\calA$ is \ALG, then we
  can choose $\calB$ to be \SETF.
\end{restatable}

The proof of the first part of the statement is immediate: consider
the schedule produced by $\calA$. For each job, $\calB$ starts
processing it exactly like $\calA$. Giving $\calB$ a speed of
$(1+\delta) = 1/(1-\eps)$ means that each job completes exactly
when its size becomes known in $\calA$. And now the algorithm $\calB$
can idle during the remaining $\eps p_j$ time that $\calA$ processes
the job $j$. Observe that $\calB$ is non-clairvoyant, since it processes
each job while its size is still unknown, and then idles based on the
known size. Applying this proof with $\calA = \ALG$ gives us an algorithm $\calB$ that looks like \SETF, but leaves ``forced idle times'' in its schedule. To get the second part of the statement, we relate this
back to \SETF in \Cref{sec:reduction}.
This implies that $\SETF$ is $(1+\eps)$-speed
$(1+\factor)$-competitive.

This reduction suggests an intriguing connection between the
$\eps$-clairvoyant and the speed-augmentation models in settings beyond
flow time and/or a single machine. If a non-clairvoyant algorithm in the
speed setting is a form of SETF (e.g., \cite{ChekuriGKK04}) or
Round-Robin (e.g., \cite{ImKM18}), can we replace speed augmentation with
$\eps$-clairvoyance in these settings, thereby giving us results with an
apples-to-apples comparison
with the optimum? On the other hand, if we give an $\eps$-clairvoyant
algorithm, we may be able to get new non-clairvoyant
algorithms
in the speed-augmentation
setting.

\subsection{Our Proof Technique}

At a very high level, the bad examples for the non-clairvoyant model
are built around the adversary taking advantage of any algorithm
making mistakes and switching from the current job to the new job,
even if the old job is almost finished. The $\eps$-clairvoyant model
ensures that we know the size of the job when only a small amount of
it remains. But this does not suffice by itself to prove competitiveness---the
\ALG algorithm still switches to SETF to explore the size of new jobs as they arrive. Thus,
our analysis needs to be considerably more careful.

\paragraph{Local Competitiveness and Valid Assignments.}
As is typical for (deterministic) flow-time problems,
we prove \emph{local competitiveness}, i.e., that the number of active jobs
in the algorithm at any time $t$ is at most $\factor$ times larger
than the number of active jobs in $\OPT$.
We introduce the notion of a
\emph{valid assignment} to show local competitiveness. Namely, for any integer $k$, we consider the
$k$ jobs in $\OPT$ with the largest sum of remaining processing
times. We show that this sum is at most that of the $\bfactor k$ jobs
with the largest remaining processing time in $\ALG$. (This is similar
to the optimality proof for SRPT given by Schrage~\cite{Schrage68}, and used in several other contexts since.)

Specifically, a valid assignment is a fractional transportation between
the jobs in $\OPT$ and the jobs in $\ALG$, such that
when jobs are sorted by non-increasing order of
remaining processing times,
the neighborhood of
 any prefix of jobs in $\OPT$ is at most $\bfactor$ times as large. In this case, we say that the \emph{prefix expansion} of the assignment is
at most $\factor$. It is not difficult to show that if there is a
valid assignment, then the ``canonical'' greedy assignment that
matches the volume of jobs in \OPT to those in \ALG greedily, from
highest to lowest remaining processing times, is also valid
(\Cref{lem:valid imp canonical}). For this section, whenever we refer
to a valid assignment, it may be convenient to keep this canonical
assignment in mind.

\paragraph{Inductively Maintaining a Valid Assignment.}
A valid assignment can be easily maintained
when a new job arrives. This job is active in both \OPT and \ALG and therefore can be matched to itself, without increasing
the prefix expansion, since
$\factor \geq 1$.  We call a job \emph{known} if its processing time
has been revealed to the algorithm, and \emph{unknown} otherwise.
When the algorithm works on a known job, this job has
the smallest remaining processing time. Hence, both $\OPT$ (which is
just SRPT) and $\ALG$ are processing the
jobs in their queues\footnote{The queue of $\OPT$ at time $t$ is the set of active jobs in $\OPT$'s schedule at time $t$. We denote this set by $\OPT(t)$, and define $\ALG(t)$ analogously. See \Cref{sec:prelim}.}  with the smallest remaining processing time, so we can maintain the
validity of the assignment easily. (This is exactly like the proof of optimality of SRPT, and other proofs in the literature.) The main
challenge arises when the algorithm switches to the
$\SETF$ mode, and hence reduces the size of multiple jobs which may reside anywhere in its queue: this plays havoc with the prefix expansion
property. In particular, the key challenge is to find the ``correct''
inductive invariants to maintain.

\paragraph{The Fast-Forward Lemma.}
In our argument, we focus on a single target time $t$ for which we want to
show a valid assignment, and show how to maintain a sequence of valid
assignments for times $t_0 = 0 < t_1 < t_2 < \ldots < t_k = t$. Crucially, this sequence depends on the target time $t$, and
the argument becomes specific to time $t$, and close-by times may have very different assignments. The crux of the analysis of
$\ALG$ is the following idea (\Cref{lem:nontrivial update
  assignment}): Suppose we have a valid assignment for time
$s$. Suppose a set of jobs $J_{new}$ arrives right after $s$, let the
``leader'' $L(s)$ be the largest job in $J_{new}$. Now consider some
time $\ell > s$ such that
\begin{itemize} [nosep]
\item during $(s,\ell]$ the algorithm $\ALG$ does not work on the
  unknown jobs in its queue at time $s$,
\item $\ALG$ touches (i.e.~works on) the job $L(s)$ at time $\ell$, and $L(s)$ is
  still an unknown job, and
\item no other jobs arrive during $(s,\ell]$.
\end{itemize}
Then, we can ``fast-forward'' time from $s$ to obtain a valid
assignment for time $\ell$.

Of course, these three conditions are very restrictive and may not
help us advance very far in the timeline. Indeed, consider a
``magical'' time $s$, such that we do not work on the jobs unknown at
time $s$ during the entire period $(s,t]$: this ensures the first
condition. We would like to fast-forward to some later magical time
$\ell$, but the other conditions may be violated for this time.  The
chief culprit is the last condition: new jobs may arrive during
$(s,\ell]$. This is where the next idea comes to play: we introduce a
new notion of \emph{\early jobs}. We carefully reduce the arrival
times of some jobs, in a way that leaves the algorithm's state at
time $t$ unchanged. (This is where we use the fact that we fixed the target time $t$.) Of
course, reducing the arrival times of jobs can only help the
adversary. Now, if many jobs arrive during $(s,\ell]$, we show that
(under some technical conditions) it is ``safe'' to pretend that those
jobs arrive at time $s$ instead. This may change the leader for time
$s$: so we repeat this process, making more jobs \early, until we can
apply the Fast-Forward Lemma to get a valid assignment for some
magical time $\ell > s$, thereby making progress. This argument
appears in \Cref{sec:proof-valid-theorem}, with a more detailed
intuition given in \Cref{sec:proof-idea}.

\paragraph{A Proof Idea of the Fast-Forward Lemma.}

Let us view the valid assignment $\sigma$ at time $s$ as some fractional matching $H_\sigma$
between the jobs of $\OPT$ and jobs of \ALG with some added degree and
expansion properties. Since all the jobs arriving in $(s,\ell]$ must
arrive at time $s$ by the assumption of the lemma, we can add a
perfect matching between these new jobs to get a valid fractional
matching $H'$ immediately after these arrivals. Henceforth, both
$\OPT$ and \ALG work on their jobs during $(s,\ell]$, and our goal is
to transform $H'$ into a valid assignment at time~$\ell$.

The proof uses the following simple but crucial property of \ALG (which is formally proven in \Cref{appendix:omitted proofs}).
\begin{observation}[New jobs join SETF or complete] \label{lem:new job
    joins RR} Suppose job $i$ is in \ALG's queue and unknown at both
  times $t$ and $t' > t$, and suppose job $j$ arrives at time
  $t$. Suppose \ALG works on $i$ at time $t'$, then either $j$ has the
  same elapsed time as $i$, or else $j$ has been completed.
\end{observation}

The main idea is that we can explicitly track the work done by both OPT and \ALG during $(s,\ell]$:
\begin{enumerate}[nosep]
    \item OPT works on the smallest jobs in its queue.
    \item Suppose \ALG has done $\gamma$ work on the leader by time $\ell$. By \Cref{lem:new job joins RR},  all the new jobs that arrived at time $s$ either have finished by time $\ell$, or have exactly $\gamma$ work done on them. Moreover, a suffix of the smallest known jobs in \ALG at time $s$ have been finished by time $\ell$; in fact, these are exactly the known jobs with remaining sizes that were at most $\threshold\cdot \gamma$
    at time $s$. (Recall the precondition in the statement of the Fast-Forward Lemma that
    no job in $U(s)$ is touched during $(s, \ell]$ where $U(s)$ is the set of unknown jobs at time $s$.)
\end{enumerate}
Given this explicit description of the state at time $\ell$ and its
relationship to that at time $s$, we show how to alter the fractional
matching $H'$ into one at time $\ell$, and hence get the desired valid
assignment. Next,  we discuss how to perform this alteration in a special case in \Cref{sec:toy}.

\subsection{An Illustrative Special Case of the Fast-Forward Lemma}
\label{sec:toy}

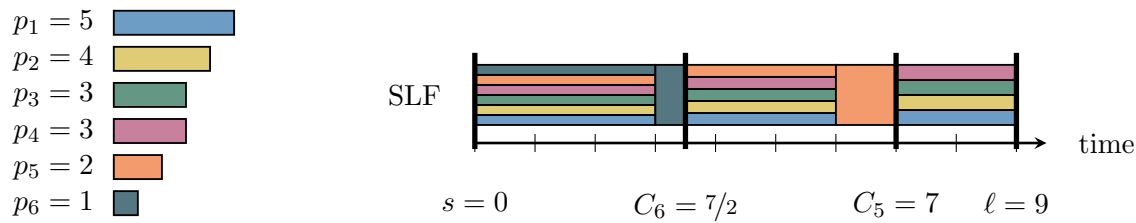
\begin{figure}
    \centering
    \begin{tikzpicture}[scale=0.8]
    \node at (-1,0.5) {$\ALG$};

  \rectjob[job1](0,0)(3,1/6);
  \rectjob[job2](0,1/6)(3,1/6);
  \rectjob[job3](0,2/6)(3,1/6);
  \rectjob[job4](0,3/6)(3,1/6);
  \rectjob[job5](0,4/6)(3,1/6);
  \rectjob[job6](0,5/6)(3,1/6);

  \rectjob[job6](3,0)(1/2,1);

  \rectjob[job1](3+1/2,0)(5/2,1/5);
  \rectjob[job2](3+1/2,1/5)(5/2,1/5);
  \rectjob[job3](3+1/2,2/5)(5/2,1/5);
  \rectjob[job4](3+1/2,3/5)(5/2,1/5);
  \rectjob[job5](3+1/2,4/5)(5/2,1/5);

  \rectjob[job5](6,0)(1,1);

  \rectjob[job1](7,0)(2,1/4);
  \rectjob[job2](7,1/4)(2,1/4);
  \rectjob[job3](7,2/4)(2,1/4);
  \rectjob[job4](7,3/4)(2,1/4);

    \node at (10.5,-0.3) {time};
  \draw[black,line width=1pt,->,>=stealth] (0,-0.3) -- (9.5,-0.3);
      \foreach \x in {0,1,...,9}{
          \draw[black] (\x,-0.3-0.15) -- (\x,-0.3+0.15);
              }

  \draw[line width=2pt] (0,-0.4) -- node[below=30pt] {$s=0$} (0,1.2);
  \draw[line width=2pt] (9,-0.4) -- node[below=30pt] {$\ell=9$} (9,1.2);
  \draw[line width=2pt] (7/2,-0.4) -- node[below=30pt] {$C_6=\nicefrac{7}{2}$} (7/2,1.2);
  \draw[line width=2pt] (7,-0.4) -- node[below=30pt] {$C_5=7$} (7,1.2);

  \begin{scope}[shift={(-6,-1.5)},scale=0.4]

    \node at (-2.5,8) {$p_1=5$};
    \node at (-2.5,6.5) {$p_2=4$};
    \node at (-2.5,5) {$p_3=3$};
    \node at (-2.5,3.5) {$p_4=3$};
    \node at (-2.5,2) {$p_5=2$};
    \node at (-2.5,0.5) {$p_6=1$};

    \rectjob[job1](0,7.5)(5,1);
    \rectjob[job2](0,6)(4,1);
    \rectjob[job3](0,4.5)(3,1);
    \rectjob[job4](0,3)(3,1);
    \rectjob[job5](0,1.5)(2,1);
    \rectjob[job6](0,0)(1,1);
  \end{scope}

\end{tikzpicture}
    \caption{On the left side there is an example instance with $\eps = \nicefrac{1}{2}$, which we consider throughout this section.
    On the right side, the schedule produced by $\ALG$ until time $\ell=9$ is depicted.
    In this example, we have $\gamma=\nicefrac{3}{2}$.
    }
    \label{fig:toy-example}
\end{figure}

Consider the special case where  both
\ALG and OPT have no existing jobs at time $s=0$, and a set
$J_{new} = \{1,\ldots,k\}$ of $k$ jobs arrives at this
time. Let $p_i$ be the processing time of job $i$, and suppose
$p_1 > p_2 > \ldots > p_k$, so that job $1$ is the leader at time
$s = 0$. \Cref{fig:toy-example}
shows this instance and the schedule of $\ALG$.

\begin{figure}
    \centering
    \begin{tikzpicture}[yscale=0.4,xscale=0.7]

  \node at (-1.5,4.5) {$\opt(0)$};
  \pgfmathsetmacro\sum{0}
  \pgfmathsetmacro\index{1}
  \pgfplotsforeachungrouped \x in {5,4,3,3,2,1} {
      \rectjobnob[job\index](\sum,4)(\x,1);
      \matchingjob(\sum,4)(\x,1);
      \node[vertex] (o\index) at (\sum + 0.5 * \x ,3.5) {};
      \pgfmathsetmacro\sum{\sum + \x};
      \pgfmathsetmacro\index{int(\index + 1)};
  }

    \node at (-1.5,0.5) {$\alg(0)$};

    \pgfmathsetmacro\sum{0}
    \pgfmathsetmacro\index{1}
    \pgfplotsforeachungrouped \x in {5,4,3,3,2,1} {
      \rectjobnob[job\index](\sum,0)(\x,1);
      \matchingjob(\sum,0)(\x,1);
        \node[vertex] (a\index) at (\sum + 0.5 * \x ,1.5) {};
        \pgfmathsetmacro\sum{\sum + \x};
        \pgfmathsetmacro\index{int(\index + 1)};
    }

    \draw[line width=1.1pt] (o1) -- node[midway,right] {$5$} (a1);
    \draw[line width=1.1pt] (o2) -- node[midway,right] {$4$} (a2);
    \draw[line width=1.1pt] (o3) -- node[midway,right] {$3$} (a3);
    \draw[line width=1.1pt] (o4) -- node[midway,right] {$3$} (a4);
    \draw[line width=1.1pt] (o5) -- node[midway,right] {$2$} (a5);
    \draw[line width=1.1pt] (o6) -- node[midway,right] {$1$} (a6);

\draw [decorate,decoration={brace,amplitude=10pt}] (0,5) -- (5,5) node[midway,above=10pt] {$p_1$};
\draw [decorate,decoration={brace,amplitude=10pt}] (5,5) -- (9,5) node[midway,above=10pt] {$p_2$};

\end{tikzpicture}
    \caption{The initial perfect matching $M$. The widths of the rectangles correspond to the initial processing times.  Job $1$ to $6$ are ordered from left to right in both queues.}
    \label{fig:toy-initial-matching}
\end{figure}
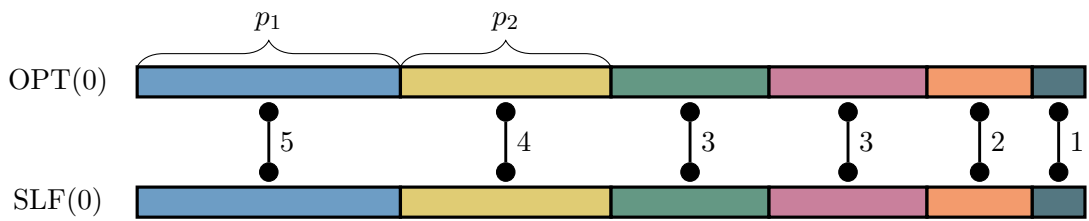

Upon arrival of jobs in $J_{new}$, we create a perfect matching $M$
where each edge $(i,i)$ has weight $p_i$ for all $1 \leq i \leq k$. The
first coordinate refers to the job in \ALG, whereas the second
coordinate refers to the job in OPT.
\Cref{fig:toy-initial-matching} illustrates this matching for our instance.
This is a valid assignment at
time $0$, and we now want to transform it into a valid assignment at
time $\ell > 0$. Suppose at time $\ell$, \ALG has worked on the leader for a
total amount equal to
$\gamma$.

\begin{leftbar}
\textbf{Simplifying Assumption.} \OPT has no partial jobs at time $\ell$: all jobs in its queue are  completed or untouched.
\end{leftbar}

We now describe the state of $\alg(\ell)$ and $\OPT(\ell)$.
Let
$e(i)$ and $e^*(i)$ be the elapsed time of job $i$ at time $\ell$ in
$\alg$'s and $\OPT$'s queue, respectively. For each $1\leq i \leq k$, we
define the following.
\begin{enumerate} [nosep]
    \item $\Delta(i) := \min\{e(i),e^*(i)\}$, i.e., the common work done by \ALG and $\OPT$ on job $i$ at time $\ell$.
    \item $\tau(i) := \max\{e(i)-e^*(i),0\}$, i.e., the amount of work   \ALG exceeds $\OPT$ on job $i$ at time $\ell$.
    \item $\tau^*(i) := \max\{e^*(i)-e(i),0\}$, i.e., the amount of work   $\OPT$ exceeds \ALG on job $i$ at time $\ell$.
\end{enumerate}

Observe that   $\sum_{i} \big(\Delta(i) + \tau (i)\big) = \sum_{i} \big(\Delta(i) + \tau^*(i)\big)$. In particular, $\sum_{i} \tau(i) = \sum_i \tau^*(i)$.

\begin{leftbar}
  \textbf{Step 1.} We update the matching $M$ by reducing the weight
  of every edge $(j,j)$ by $\Delta(j)$.
\end{leftbar}

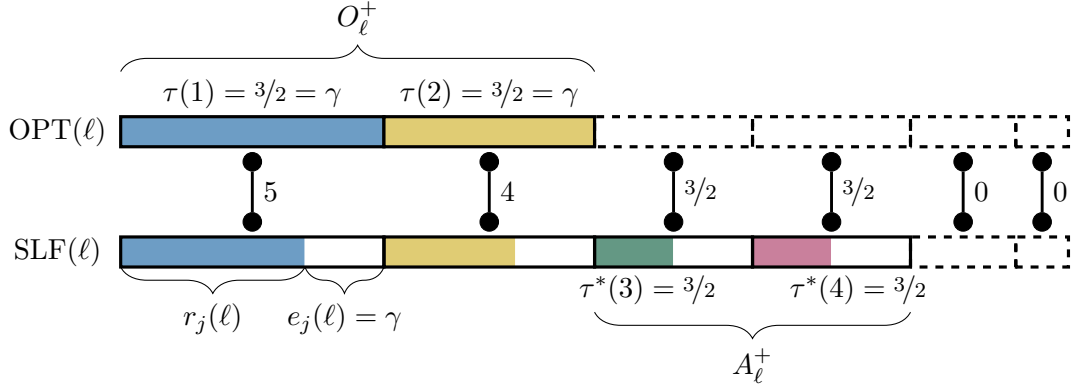
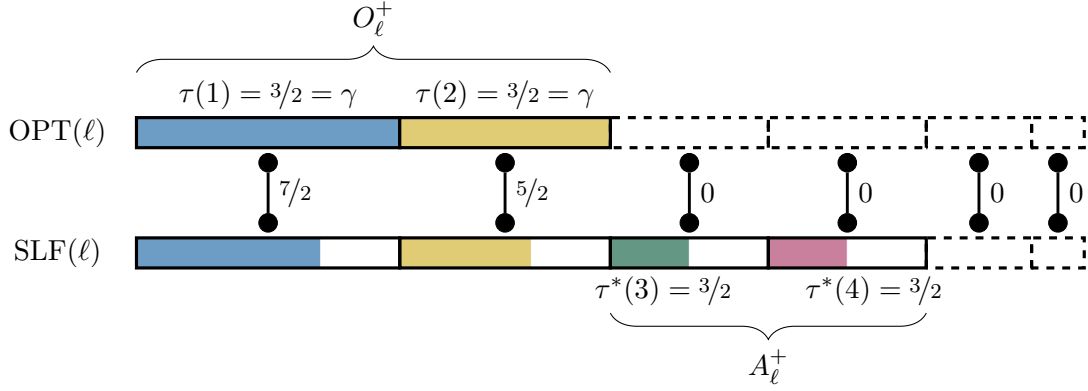
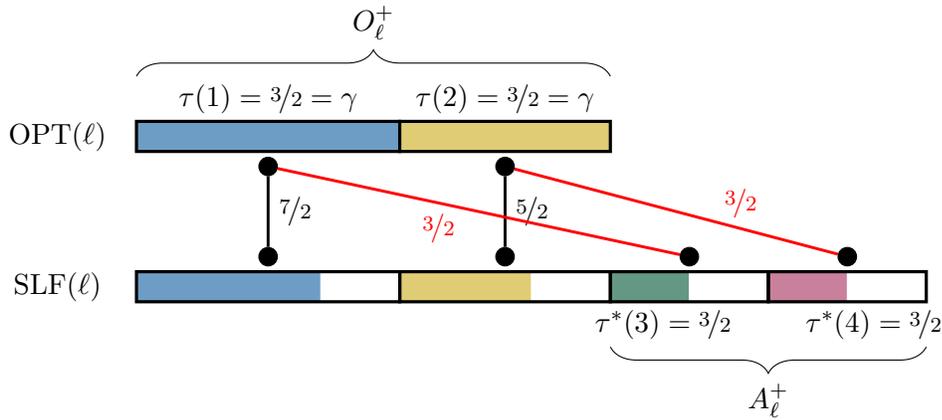
\begin{figure}
    \begin{subfigure}[c]{\textwidth}
    \begin{tikzpicture}[yscale=0.4,xscale=0.7]

  \node at (-1.2,4.5) {$\opt(\ell)$};
  \pgfmathsetmacro\sum{0}
  \pgfmathsetmacro\index{1}
  \pgfplotsforeachungrouped \x in {5,4} {
      \rectjobnob[job\index](\sum,4)(\x,1);
      \matchingjob(\sum,4)(\x,1);
      \node[vertex] (o\index) at (\sum + 0.5 * \x ,3.5) {};
      \pgfmathsetmacro\sum{\sum + \x};
      \pgfmathsetmacro\index{int(\index + 1)};
  }
  \pgfmathsetmacro\sum{9}
  \pgfmathsetmacro\index{3}
  \pgfplotsforeachungrouped \x in {3,3,2,1} {
      \partialjob(\sum,4)(\x,1);
      \node[vertex] (o\index) at (\sum + 0.5 * \x ,3.5) {};
      \pgfmathsetmacro\sum{\sum + \x};
      \pgfmathsetmacro\index{int(\index + 1)};
  }

    \rectjobnob[job1](0,0)(5-3/2,1);
    \rectjobnob[job2](5,0)(4-3/2,1);
    \rectjobnob[job3](9,0)(3-3/2,1);
    \rectjobnob[job4](12,0)(3-3/2,1);

    \node at (-1.2,0.5) {$\alg(\ell)$};

    \pgfmathsetmacro\sum{0}
    \pgfmathsetmacro\index{1}
    \pgfplotsforeachungrouped \x in {5,4,3,3} {
      \matchingjob(\sum,0)(\x,1);
        \node[vertex] (a\index) at (\sum + 0.5 * \x ,1.5) {};
        \pgfmathsetmacro\sum{\sum + \x};
        \pgfmathsetmacro\index{int(\index + 1)};
    }
    \pgfmathsetmacro\sum{15}
    \pgfmathsetmacro\index{5}
    \pgfplotsforeachungrouped \x in {2,1} {
      \partialjob(\sum,0)(\x,1);
      \node[vertex] (a\index) at (\sum + 0.5 * \x ,1.5) {};
      \pgfmathsetmacro\sum{\sum + \x};
      \pgfmathsetmacro\index{int(\index + 1)};
    }

    \draw[line width=1.1pt] (o1) -- node[midway,right] {$5$} (a1);
    \draw[line width=1.1pt] (o2) -- node[midway,right] {$4$} (a2);
    \draw[line width=1.1pt] (o3) -- node[midway,right] {{$\nicefrac{3}{2}$}} (a3);
    \draw[line width=1.1pt] (o4) -- node[midway,right] {{$\nicefrac{3}{2}$}} (a4);
    \draw[line width=1.1pt] (o5) -- node[midway,right] {{$0$}} (a5);
    \draw[line width=1.1pt] (o6) -- node[midway,right] {{$0$}} (a6);

    \draw [decorate,decoration={brace,amplitude=10pt}] (0,6.5) -- (9,6.5) node[midway,above=10pt] {$O_\ell^+$};
    \draw [decorate,decoration={brace,mirror,amplitude=10pt}] (9,-1.5) -- (15,-1.5) node[midway,below=10pt] {$A_\ell^+$};

    \draw [decorate,decoration={brace,mirror,amplitude=10pt}] (0,0) -- (3.5,0) node[midway,below=10pt] {$r_j(\ell)$};
    \draw [decorate,decoration={brace,mirror,amplitude=10pt}] (3.5,0) -- (5,0) node[midway,below=10pt] {$e_j(\ell) = \gamma$};

    \node at (2.5,5.75) {$\tau(1) = \nicefrac{3}{2} = \gamma$};
    \node at (7,5.75) {$\tau(2) = \nicefrac{3}{2} = \gamma$};

    \node at (10,-0.75) {$\tau^*(3) = \nicefrac{3}{2}$};
    \node at (14,-0.75) {$\tau^*(4) = \nicefrac{3}{2}$};

\end{tikzpicture}
    \caption{The matching $M$ after performing Step~1. The width of the rectangles correspond to the initial processing times, the width of the filled area of each rectangle corresponds to the remaining processing time at time $\ell$ in the corresponding schedule. Dashed rectangles indicate jobs that have been completed by time $\ell$. We have (from left to right) $\Delta(1) = \Delta(2) = 0$, $\Delta(3) = \Delta(4) = \nicefrac{3}{2}$, $\Delta(5) = 2 = p_5$, and $\Delta(6) = 1 = p_6$.}
    \label{fig:toy-step-1}
    \end{subfigure}
    \par\bigskip
    \begin{subfigure}[c]{\textwidth}
    \begin{tikzpicture}[yscale=0.4,xscale=0.7]

  \node at (-1.5,4.5) {$\opt(\ell)$};
  \pgfmathsetmacro\sum{0}
  \pgfmathsetmacro\index{1}
  \pgfplotsforeachungrouped \x in {5,4} {
      \rectjobnob[job\index](\sum,4)(\x,1);
      \matchingjob(\sum,4)(\x,1);
      \node[vertex] (o\index) at (\sum + 0.5 * \x ,3.5) {};
      \pgfmathsetmacro\sum{\sum + \x};
      \pgfmathsetmacro\index{int(\index + 1)};
  }
  \pgfmathsetmacro\sum{9}
  \pgfmathsetmacro\index{3}
  \pgfplotsforeachungrouped \x in {3,3,2,1} {
      \partialjob(\sum,4)(\x,1);
      \node[vertex] (o\index) at (\sum + 0.5 * \x ,3.5) {};
      \pgfmathsetmacro\sum{\sum + \x};
      \pgfmathsetmacro\index{int(\index + 1)};
  }

    \rectjobnob[job1](0,0)(5-3/2,1);
    \rectjobnob[job2](5,0)(4-3/2,1);
    \rectjobnob[job3](9,0)(3-3/2,1);
    \rectjobnob[job4](12,0)(3-3/2,1);

    \node at (-1.5,0.5) {$\alg(\ell)$};

    \pgfmathsetmacro\sum{0}
    \pgfmathsetmacro\index{1}
    \pgfplotsforeachungrouped \x in {5,4,3,3} {
      \matchingjob(\sum,0)(\x,1);
        \node[vertex] (a\index) at (\sum + 0.5 * \x ,1.5) {};
        \pgfmathsetmacro\sum{\sum + \x};
        \pgfmathsetmacro\index{int(\index + 1)};
    }
    \pgfmathsetmacro\sum{15}
    \pgfmathsetmacro\index{5}
    \pgfplotsforeachungrouped \x in {2,1} {
      \partialjob(\sum,0)(\x,1);
      \node[vertex] (a\index) at (\sum + 0.5 * \x ,1.5) {};
      \pgfmathsetmacro\sum{\sum + \x};
      \pgfmathsetmacro\index{int(\index + 1)};
    }

    \draw[line width=1.1pt] (o1) -- node[midway,right] {$\nicefrac{7}{2}$} (a1);
    \draw[line width=1.1pt] (o2) -- node[midway,right] {$\nicefrac{5}{2}$} (a2);
    \draw[line width=1.1pt] (o3) -- node[midway,right] {$0$} (a3);
    \draw[line width=1.1pt] (o4) -- node[midway,right] {$0$} (a4);
    \draw[line width=1.1pt] (o5) -- node[midway,right] {{$0$}} (a5);
    \draw[line width=1.1pt] (o6) -- node[midway,right] {{$0$}} (a6);

    \draw [decorate,decoration={brace,amplitude=10pt}] (0,6.5) -- (9,6.5) node[midway,above=10pt] {$O_\ell^+$};
    \draw [decorate,decoration={brace,mirror,amplitude=10pt}] (9,-1.5) -- (15,-1.5) node[midway,below=10pt] {$A_\ell^+$};

    \node at (2.5,5.75) {$\tau(1) = \nicefrac{3}{2} = \gamma$};
    \node at (7,5.75) {$\tau(2) = \nicefrac{3}{2} = \gamma$};

    \node at (10,-0.75) {$\tau^*(3) = \nicefrac{3}{2}$};
    \node at (14,-0.75) {$\tau^*(4) = \nicefrac{3}{2}$};

\end{tikzpicture}
    \caption{The matching $M$ after performing Step~2.}
    \label{fig:toy-step-2}
    \end{subfigure}
    \par\bigskip
    \begin{subfigure}[c]{\textwidth}
    \begin{tikzpicture}[yscale=0.4,xscale=0.7]

  \node at (-1.5,5.5) {$\opt(\ell)$};
  \pgfmathsetmacro\sum{0}
  \pgfmathsetmacro\index{1}
  \pgfplotsforeachungrouped \x in {5,4} {
      \rectjobnob[job\index](\sum,5)(\x,1);
      \matchingjob(\sum,5)(\x,1);
      \node[vertex] (o\index) at (\sum + 0.5 * \x ,4.5) {};
      \pgfmathsetmacro\sum{\sum + \x};
      \pgfmathsetmacro\index{int(\index + 1)};
  }

    \rectjobnob[job1](0,0)(5-3/2,1);
    \rectjobnob[job2](5,0)(4-3/2,1);
    \rectjobnob[job3](9,0)(3-3/2,1);
    \rectjobnob[job4](12,0)(3-3/2,1);

    \node at (-1.5,0.5) {$\alg(\ell)$};

    \pgfmathsetmacro\sum{0}
    \pgfmathsetmacro\index{1}
    \pgfplotsforeachungrouped \x in {5,4,3,3} {
      \matchingjob(\sum,0)(\x,1);
        \node[vertex] (a\index) at (\sum + 0.5 * \x ,1.5) {};
        \pgfmathsetmacro\sum{\sum + \x};
        \pgfmathsetmacro\index{int(\index + 1)};
    }

    \draw[line width=1.1pt] (o1) -- node[midway,right] {$\nicefrac{7}{2}$} (a1);
    \draw[line width=1.1pt] (o2) -- node[midway,right] {$\nicefrac{5}{2}$} (a2);

    \draw[line width=1.1pt,red] (o1) -- node[pos=0.4,below] {$\nicefrac{3}{2}$} (a3);
    \draw[line width=1.1pt,red] (o2) -- node[pos=0.7,above] {$\nicefrac{3}{2}$} (a4);

    \draw [decorate,decoration={brace,amplitude=10pt}] (0,7.5) -- (9,7.5) node[midway,above=10pt] {$O_\ell^+$};
    \draw [decorate,decoration={brace,mirror,amplitude=10pt}] (9,-1.5) -- (15,-1.5) node[midway,below=10pt] {$A_\ell^+$};

    \node at (2.5,6.75) {$\tau(1) = \nicefrac{3}{2} = \gamma$};
    \node at (7,6.75) {$\tau(2) = \nicefrac{3}{2} = \gamma$};

    \node at (10,-0.75) {$\tau^*(3) = \nicefrac{3}{2}$};
    \node at (14,-0.75) {$\tau^*(4) = \nicefrac{3}{2}$};

\end{tikzpicture}
    \caption{The matching $M$ (black) and the new greedy assignment (red) after performing Step~3. Note that $\tau^*(3) = \tau^*(4) = \nicefrac{3}{2} \geq \threshold \cdot \gamma = \gamma = \nicefrac{3}{2}$ and that the prefix expansion of the overall assignment is equal to $\nicefrac{1}{\eps}=2$.}
    \label{fig:toy-step-3}
    \end{subfigure}
    \caption{The update process of Steps 1 to 3.}
    \label{fig:toy-steps-1-3}
\end{figure}

After this step, all jobs that are both finished by $\OPT$ and \ALG by
time $\ell$ disappear from the matching as the corresponding edge weights are reduced to zero, and we focus on the symmetric
difference of work done by \ALG and $\OPT$ at time $\ell$. Let
$O_{\ell} $ be the set of jobs that remain in $\OPT$ at time $\ell$
and similarly, $A_{\ell}$ be the set of jobs that remain in \ALG at
time $\ell$. Furthermore, we define $O_{\ell}^+$ as the set of jobs
$j$ in $O_{\ell}$ such that $\tau(j) > 0$ and $A_{\ell}^+$ as the set
of jobs $j$ in $A_{\ell}$ such that $\tau^*(j) > 0$. By definition,
$O_{\ell}^+ \cap A_{\ell}^+ = \emptyset$; by our simplifying assumption $O_\ell =
O_\ell^+$. See \Cref{fig:toy-step-1} for
an illustration.

\begin{leftbar}
  \textbf{Step 2.}  We update the matching $M$ as follows:
  \begin{itemize} [nosep]
  \item For every job $j \in O_{\ell}^+$, we reduce the weight of edge
    $(j,j)$ by $\tau(j)$.
  \item For every job $j \in A_{\ell}^+$, we reduce the weight of edge
    $(j,j)$ by $\tau^*(j)$.
  \end{itemize}
\end{leftbar}

We view the resulting matching as a weighted bipartite graph $H = (V,V^*,E,w)$ where $V$ is the set of jobs in \ALG and $V^*$ is the set of jobs in $\OPT$ at time $\ell$.

After these two steps,
\begin{itemize} [nosep]
\item For every job $j \in O_{\ell}^+$, we have
  $\vol_H(j) := \sum_{j'} w(j,j') = r_{j}(\ell)$, which is correct for
  \ALG. But $\vol_H^*(j)$ is also the same, which is not correct for
  \OPT. Hence, we say that job $j$ \emph{demands} $\tau(j)$ in $\OPT$.
  Note that $\tau(j)=\gamma$   since \OPT did not work on $j$ and \ALG processed it for $\gamma$ units.
\item For each job $j \in A_{\ell}^+$, we have
  $\vol^*_H(j) := \sum_{j'} w(j',j) = r_{j}^*(\ell)$, i.e., it is
  correct for $\OPT$. But $\vol_H(j)$ is the same, when it should be
  $r_j^*(\ell) + \tau^*(j)$, so we say that such a job $j$ \emph{demands} $\tau^*(j)$ in \ALG.  Note that $r_j^*(\ell) =0$
  since \OPT finished $j$
  and $\tau^*(j) = p_j -\gamma$.
\end{itemize}

By the discussion so far, $\sum_{i \in
  O_{\ell}^+}\tau(i) = \sum_{i \in A_{\ell}^+} \tau^*(i)$,
which motivates the fixing step as follows.

\begin{leftbar}
  \textbf{Step 3.}  We sort $O_{\ell}^+$ and $A_{\ell}^+$ in
  non-increasing order of the remaining processing times in $\OPT$ and
  \ALG respectively at time $\ell$. We then greedily assign the
  demands from \OPT to \ALG.
\end{leftbar}

More formally, we start with a bipartite graph
$G = (\alg(\ell),\OPT(\ell),\emptyset,w)$ between the jobs in \ALG and
$\OPT$ at time $\ell$ without any edges. We add the edges to the
bipartite graph using the following greedy assignment: for each job
$j \in O_{\ell}^+$, assign $\tau(j)$ amount from $j$ in $\OPT$ to the
smallest suffix of jobs $i \in A_{\ell}^+$ in \ALG whose cumulative
demand in \ALG is at least $\tau(j)$,
and update the demands after the assignment accordingly. Let $G$ be
the bipartite graph arising from the greedy assignment at the
end. Since
$\sum_{i \in O_{\ell}^+}\tau(i) = \sum_{i \in A_{\ell}^+} \tau^*(i)$,
the construction ensures that $G$ fixes all the deficits.  The final
assignment between \ALG and $\OPT$ at time $\ell$ is defined by the
weighted edges of $G\cup M$. See  \Cref{fig:toy-step-3} for an illustration.

\paragraph{Low Prefix Expansion.} To show that this assignment is
valid, it suffices to prove that $G \cup M$ has low expansion.  That
is, for every prefix $P^* \subseteq \OPT(\ell)$,
$|N_{G \cup M}(P^*)| \leq \factor \cdot |P^*|$ where  $N_H(S)$ is the set of neighbors of $S$ in the graph $H$, excluding $S$. By the definition of \ALG, the
structure of this instance, and the simplifying assumption, we know
that:
\begin{enumerate} [nosep]
\item [(a)] For all jobs $j \in O_{\ell}^+,$ the demand in \OPT is
  $\tau(j) = \gamma.$
\item [(b)] For all jobs $j \in A_{\ell}^+$, the demand in \ALG is
  $\tau^*(j) = p_j - \gamma \geq \threshold\cdot \gamma$.
\end{enumerate}
Now the greedy property of the matching means that for any prefix of
$O_\ell^+$, the number $|N_G(P^*)|$ of nodes in $A_\ell^+$ required to
match the demand of $O_\ell^+$ in $G$ is at most
\begin{align*}
  |N_G(P^*)| \leq  \ceil{ \frac{\gamma |P^*|}{\threshold\cdot \gamma}} \leq \ceil {\frac{1-\eps}{\eps} } \cdot |P^*| \ .
\end{align*}
Since each node in $O_\ell^+$ has unit degree in $M$, the total degree
is at most $
  |P^*| +  \bceil {\frac{1-\eps}{\eps} }\cdot |P^*| = \bfactor \cdot |P^*|$.

\paragraph{Removing the Simplifying Assumption.} If \OPT has a
partially completed job $z$ at time $\ell$, this job may either be in
$O_\ell^+$ or $A_\ell^+$ or neither, depending on whether \OPT finishes less, more or equal than $\gamma$ amount of processing on $z$. The case $z \in O_{\ell}^+$ is
almost identical, where the demand for $z \in O_\ell^+$ now
becomes at most $\gamma$, and hence it is even easier to show small
expansion. In case $z \in A_\ell^+$, the worrying case is when the demand
for $z \in A_\ell^+$ is less than $\threshold \cdot \gamma$---but then
$z$ is the last node of $A_\ell^+$ to be considered in the greedy
assignment, and hence, the same argument goes through. Finally, $\ALG$
and $\OPT$ may have the same elapsed time on job $z$; this
case is handled by Step~1.

\paragraph{Extending to the General Case.}
Things now become more complex when we start with a valid
assignment between \ALG and $\OPT$ at time $s$ (which is typically not
a perfect matching) and update it when the jobs in $J_{new}$
arrive. We deal with this complexity in \Cref{sec:proof-update-lemma}.

\subsection{Related Work}

The lower bound of $\Omega(\log n)$-competitiveness for randomized
algorithms was matched by \textcite{BecchettiL04} who showed that a
randomized variant of multilevel feedback gave
$O(\log n)$-competitiveness; this built upon work by
\textcite{KalyanasundaramP03}. \textcite{MotwaniPT94} also showed that
every deterministic algorithm has a competitive ratio of at least $P$,
where $P$ is the ratio of the largest to smallest processing times;
the algorithm that runs each job to completion achieves this bound.
For scheduling on $m$ parallel identical machines, SRPT is
$\cO(\log(\min\{\nf{n}{m} , P\}))$-competitive, which is
best-possible for online algorithms~\cite{LeonardiR97,LeonardiR07};
the current best non-clairvoyant algorithm is
$\cO(\log(n) \cdot \log(\min\{\nf{n}{m},
P\}))$-competitive~\cite{BecchettiL04}.
We refer to the survey on online scheduling theory by \Textcite{PruhsST04}.

\paragraph{Resource Augmentation.} Another important research direction to sidestep these strong lower
bounds was to give the algorithm more power than the benchmark: this
approach was proposed by \Textcite{KalyanasundaramP00}, and is called
\emph{resource augmentation}.
They showed that the \emph{Shortest
  Elapsed Time First} (SETF) algorithm with $(1+\varepsilon)$ speed
had total flow time at most $(1+ \nicefrac{1}{\varepsilon})$ times that of
the optimal unit-speed solution, for any $\eps > 0$.
Speed augmentation has since been widely studied and results are known
for very general settings, e.g., for weighted flow time, unrelated machines, or abstract polytope
scheduling~\cite{BansalD07,ImKMP14,ImKM18,CIP25}; we refer
to the survey by~\Textcite{ImMP11}.

\paragraph{Semi-Clairvoyant Bounds.}
\Textcite{BenderMR04} studied algorithms in a
``quantized'' model where only the ``class'' $\floor{\log_\mu r_j(t)}$
of the remaining processing times is known at each time. They showed
that a variant of SRPT is $O(1)$-competitive for minimizing total
\emph{stretch}, where the stretch of a job is its flow time normalized
by its length, and also for minimizing total flow time. If only
$\floor{\log_\mu p_j}$ is known, they showed a variant of SPT
algorithm is $O(1)$-competitive for minimizing total stretch. In a
follow-up work, \Textcite{BecchettiLMP04} gave another variant of SPT
which is $O(1)$-competitive for minimizing total flow time.

The recent work of \Textcite{AzarLT21,AzarLT22,AzarPT22,ZhaoLZ22} is
in a similar vein, but now allows the ``predictions'' to differ by a
factor of $\mu$ in either direction. The first work gives an algorithm
that is $O(\mu^2)$-competitive when $\mu$ is known (along with other
results for extensions to multiple machines, and to weights); in the
follow-up work they gave an algorithm with competitive ratios
$O(\mu \log \mu)$ for the distortion-oblivious setting (where the
value of $\mu$ is not known). These models require (coarse) estimates
at the moment the job arrives into the system, which a prediction
mechanism must derive from external attributes of the jobs (rather
than processing the jobs, as in our model). Our algorithms are quite
different from these works, though our lower bounds have a very similar
structure. The notions of $\eps$-clairvoyance
and semi-clairvoyance seem orthogonal, and in future work, it would be
interesting to combine aspects of these two models.

\subsection{Paper Organization}
The rest of the paper is organized as follows. We give some terminology and notation in \Cref{sec:prelim}. In \Cref{sec:alg slf}, we formally define the algorithm and the notion of valid assignment and early-arriving jobs. We also present the Valid Assignment Theorem and the Fast-Forward Lemma whose proofs can be found in \Cref{sec:proof-valid-theorem,sec:proof-update-lemma}, respectively. We present the lower bounds in \Cref{app:lowerbounds}, and the reduction of $\SETF$ to $\SLF$ in \Cref{sec:reduction}. Finally, we discuss the special case of simultaneous release times in \Cref{sec:simultaneous release}.

\section{Preliminaries} \label{sec:prelim}

We consider the problem of scheduling on a single machine with
preemption, which means that at any time $t \in \mathbb{R}$, an
algorithm can process at most one job $\lambda(t)$.
Note that a ``parallel processing'' of jobs like in SETF can be simulated by changing the processed job in sufficiently small time intervals.
Each job $j$ has processing time $p_j > 0$ and release date $q_j \geq 0$.
For simplicity we assume that the processing times are distinct,
i.e., $p_j \neq p_i$ for all $i,j$ with $i \neq j$; this can be done without
loss of generality by breaking ties consistently.

In our online setting, jobs arrive at their release dates, and an
online algorithm $\cA$ needs to produce a schedule over time. In
particular, at time $t$, it cannot change its schedule at time $t' < t$.

Now consider a fixed schedule.
For a time $t$ and a job $j$, the \emph{elapsed time} $e_j(t)$ is the total amount of processing that $j$ has received until time $t$, that is $e_j(t) = \int_{0}^t \mathbf{1}[j = \lambda(t')] \, \mathrm{d}t'$, and $r_j(t) = p_j - e_j(t)$ is the \emph{remaining processing time} of job $j$ at time $t$.
The \emph{completion time} $C_j$ of job $j$ is the first time
$t$ when $e_j(t) = p_j$. The objective is to minimize the total
\emph{flow time}, that is, $\sum_{j \in J} (C_j - q_j)$ for the set of
jobs $J$. We usually use $e_j(t)$, $r_j(t)$, and $C_j$ for the schedule of an algorithm that we consider. For the optimal schedule, we denote these quantities by $e^*_j(t)$, $r^*_j(t)$, and $C^*_j$.

For an algorithm $\calA$, we denote by
$\calA(t) = \{j : q_j \leq t \text{ and } r_j(t) > 0 \}$ the set of
available jobs at time $t$. The total flow time of the algorithm's
schedule equals $\int_{t}^{\infty} |\calA(t)| \, \mathrm{d} t$.  We
use $\opt$ to denote the optimal strategy, which is SRPT. We say that
an algorithm $\calA$ is \emph{locally} $\rho$-competitive if
$|\calA(t)| \leq \rho \cdot |\opt(t)|$ at all times $t$, and it is
$\rho$-competitive if its objective value is at most $\rho$ times the
optimal objective value; the smallest $\rho$ such that $\calA$ is
$\rho$-competitive is called the competitive ratio of $\calA$.
Proving local competitiveness is sufficient to bound the competitive
ratio of $\calA$ by integrating over time.

When proving local competitiveness for a time $t$, we can assume
without loss of generality that the algorithm (and thus, the optimal
solution) does not idle between time $0$ and time $t$.

\section{The Algorithm and Analysis} \label{sec:alg slf}

\paragraph{The \ALG Algorithm.} We formally define the shortest lower-bound first (\ALG) algorithm.  At any time $t$, for each job $j \in \ALG(t)$ in the queue, define an \emph{estimate} $\eta_{j}(t)$
to be
\[
\eta_j(t) =
\begin{cases}
\threshold \cdot e_{j}(t), & \text{if the remaining time is unknown} , \\
r_{j}(t), & \text{else.}
\end{cases}
\]

\begin{defn} [\bf Known/Unknown Jobs]
Let $j \in \ALG(t)$ be an active job in $\ALG$ at time $t$. We say that a job $j$ is \textit{known} if $r_{j}(t) \leq \oneminusA \cdot p_j$. Otherwise, $j$ is \textit{unknown}.
\end{defn}

Let $U(t)$ and $K(t)$ denote the unknown and known jobs in the algorithm's schedule at time $t$.  At any time $t$, \ALG works on a job with the smallest estimate:
\begin{enumerate}
    \item If $\min_{j \in K(t)} \eta_j(t) \le \min_{j \in U(t)} \eta_j(t)$, \ALG processes an arbitrary known job $j$ with minimum $\eta_j(t)$.
    \item Otherwise, \ALG processes \emph{all} unknown jobs $j$ with minimum $\eta_j(t)$ in parallel at an equal pace. We can achieve this by scheduling them in Round-Robin with infinitesimally small steps.
\end{enumerate}
Intuitively, \ALG favors the jobs with unknown processing time until their estimate is larger than the estimate of the smallest known job. The threshold $\threshold e_j(t)$ is the  lower bound of the remaining time of an unknown job $j$ at time $t$. \ALG smoothly interpolates SETF (Shortest Elapsed Time First) and SRPT (Shortest Remaining Time First).

We assume without loss of generality that the value of $\eps$ is known to the algorithm from the start of the instance. Otherwise, the algorithm can compute it from the elapsed times once it is notified of the remaining processing time for the first job that becomes known. Before the first job becomes known, \ALG behaves like $\SETF$; thus, knowing the value of $\eps$ is not necessary.

\subsection{Equivalent Instances and Valid Assignments}

To show the local competitiveness for \ALG,
we iteratively change the original instance $\calJ$ into a new ``nicer'' and easier-to-analyze instance $\calJ'$ by releasing jobs earlier.
This is done carefully so that:
\begin{enumerate}[nosep,label=(\alph*)]
    \item the algorithm has the same state at time $t$, where $t$ is the point in time for which we want to show local competitiveness, when run on $\calJ$ and on $\calJ'$, whereas
    \item SRPT has only fewer jobs remaining at time $t$ in $\calJ'$ than in the original instance $\calJ$.
\end{enumerate}
(Such an instance is said to be $t$-equivalent to $\calJ$.)
Formally, let us emphasize the role of the specific instance $\calJ$ by writing $\calA_{\calJ}(t)$ instead of $\calA(t)$ and $e_{\calJ}(j,t)$ instead of $e_j(t)$.

 \begin{defn}[$t$-\early and $t$-equivalent Instances]
     Let $\calJ$ be a job instance and $t$ be a time. We say that a
     job instance $\calJ'$ is an \emph{$t$-\early} instance of $\calJ$
     if $\calJ'$ is obtained from job instance $\calJ$ by reducing the
     release times of some jobs $j$ in $\calJ$ that satisfy $q_j < t$.
     We say that $\calJ'$ is $t$-\emph{equivalent} to $\calJ$ if
     $\ALG_{\calJ'}(t) = \ALG_{\calJ}(t)$ and for all $j$, $e_{\calJ}(j,t) = e_{\calJ'}(j,t)$.
 \end{defn}

Since every schedule for $\calJ$ is also feasible for $\calJ'$ and every job that is released after $t$ in $\calJ$ is also released after $t$ in $\calJ'$, it follows that
 $|\OPT_{\mathcal{J}'}(t)| \leq |\OPT_{\mathcal{J}}(t)|$
    for any instance $\calJ'$ that is $t$-\early for $\calJ$.
    Henceforth, we find some $t$-equivalent and $t$-\early
    instance $\calJ'$ and prove local competitiveness for it.
\begin{theorem}
  \label{thm:main}
  For every job instance $\calJ$, and for all times $t$, there is an $t$-equivalent $t$-\early instance $\calJ'$ of $\calJ$  such that $|\ALG_{\calJ'}(t)| \leq \bfactor \cdot |\OPT_{\calJ'}(t)|$. Therefore, $|\ALG_{\calJ}(t)| \leq \bfactor \cdot |\OPT_{\calJ}(t)|$.
\end{theorem}

Next, to show local competitiveness, consider jobs in $\ALG$ and $\OPT$ sorted in non-increasing order of
remaining processing times. We will argue that the total remaining time of
the first $k$ jobs in $\OPT$ is at most that of the first $\bfactor k$
jobs in $\ALG$. Formally, our proofs will set up a fractional matching between the
jobs in $\OPT$ and $\ALG$ such that the neighborhood of any
prefix of jobs in $\OPT$ is at most $\bfactor$ times as large.

\begin{defn} [{\bf Assignment}] \label{def:feas assign} An
  \emph{assignment} (between $\ALG$ and $\OPT$) at time $t$ (on job
  instance $\cJ$) is a function
  $\sigma: \ALG_{\cJ}(t) \times \OPT_{\cJ}(t) \rightarrow
  \mathbb{R}_{\geq 0}$ if
  \begin{enumerate} [nosep]
  \item for all $i \in \ALG_{\cJ}(t), \sum_{j}\sigma(i,j) = r_{i}(t)$, and
  \item for all $j \in\OPT_{\cJ}(t), \sum_{i}\sigma(i,j) = r^*_{j}(t)$.
  \end{enumerate}
  Unless stated otherwise, jobs $i \in \ALG(t)$ are ordered by
  non-increasing values of $r_i(t)$, and jobs $i \in \OPT(t)$ are
  ordered by non-increasing values of $r^*_i(t)$; we break ties by the
  unique job identifiers.
\end{defn}

A $k$-\emph{prefix} (respectively, $k$-\emph{suffix}) of $\ALG(t)$ in
an ordering $\sigma$ is the set of the first $k$ jobs (respectively,
last $k$) in this order; the $k$-prefix/suffix of $\OPT(t)$ is defined
analogously. Let $E_{\sigma} = \{ (i,j)\mid \sigma(i,j) > 0\}$ be the
support of $\sigma$. For any subset $S \subseteq \OPT(t)$, let the
\emph{neighbors} of $S$ be
$N_{\sigma}(S) = \{i \in \ALG(t) \mid \exists j \in S, (i,j) \in
E_\sigma\}$.

\begin{defn} [{\bf Prefix Expansion and Valid Assignment}] \label{def:prefix expansion}
    Given an assignment $\sigma$ at time $t$, the \emph{prefix expansion} is \[\phi(\sigma) := \max_{P^* \text{ is a prefix of } \OPT(t) {\text{ in } \sigma} } \frac{|N_{\sigma}(P^*)|}{|P^*|} \ .\]
    An assignment $\sigma$ at time $t$ is called \emph{valid} if $\phi(\sigma) \leq \bfactor$.
\end{defn}

\begin{claim}
  \label{cor:bounded-degree}
  If there is a valid assignment  at time $t$, then
  $ |\ALG(t)| \leq \factor\cdot |\OPT(t)|.$
\end{claim}

\subsection{The Main Technical Statements}
\label{sec:main-techn-lemm}

Armed with these concepts,
\Cref{thm:main} is proven by combining
 \Cref{cor:bounded-degree} with the following result:

 \begin{theorem}[Valid Assignment Theorem]
  \label{thm:there is always a valid assignment}
  For every job instance $\calJ$, and for any time $t$, there is a
  valid assignment between $\ALG_{\calJ'}(t)$ and $\OPT_{\calJ'}(t)$
  at time $t$ for some $t$-equivalent $t$-\early instance $\calJ'$ of
  $\calJ$.
\end{theorem}

The proof of \Cref{thm:there is always a valid assignment} proceeds
inductively: given a valid assignment at time $s$, suppose that a set of
jobs arrives. We argue that the process is very structured, and we can ``fast-forward" from time $s$ to some time in the future $s'$ where we can maintain a valid assignment at time $s'$.
The technical tool that allows us to ``fast-forward" time while maintaining a valid assignment is formalized in the following lemma.

\begin{restatable}[Fast-Forward Lemma]{lemma}{LemmaTwo}
  \label{lem:nontrivial update assignment} Let $J$ be the
  set of jobs released before time $s$, and let $K(s)$ be the set of
  known jobs in $\ALG_{J}(s)$ at time $s$.  Assume that there is a
  valid assignment between $\ALG_{J}(s)$ and $\OPT_{J}(s)$ at time $s$
  right before a batch of jobs $J_{new}$ arrives. Let the \emph{leader}
  $L(s)$ be the job in $J_{new}$ with the highest processing time at
  time $s$.\footnote{$L(s)$ is well defined since we assumed that
    processing times are unique.}
  In addition, assume that the following three properties hold
  during the time interval $[s,\ell]$ for some $\ell > s$:
 \begin{enumerate} [noitemsep,nolistsep]
     \item $\ALG$ only works on the jobs in $J_{new} \cup K(s)$,
     \item At time $\ell$, $\ALG$ touches the leader $L(s)$ and $L(s)$ is still unknown, and
     \item No other job arrives.
 \end{enumerate}
 Then there is a valid assignment between $\ALG_{J'}(\ell)$ and $\OPT_{J'}(\ell)$ at time $\ell$ where $J' = J \cup J_{new}$.
\end{restatable}

At a very high level, if there are multiple batches of arrivals in the interval $[s,t]$, the proof of \Cref{thm:there is always a valid assignment} repeatedly converts the original instance into $t$-equivalent instances until there is a single batch of arrivals at time $s$ satisfying the assumptions above, whereupon it can use the Fast-Forward Lemma.

In the rest of the paper, we first prove \Cref{thm:there is always a valid
  assignment}
assuming the Fast-Forward Lemma
in \Cref{sec:proof-valid-theorem},
and then prove \Cref{lem:nontrivial update assignment} in \Cref{sec:proof-update-lemma}.

\section{Proof of the Valid Assignment Theorem}
\label{sec:proof-valid-theorem}

For this section, fix a job instance $\calJ$ and the target time $t$.
We now show that for any time $t$, we can find a $t$-equivalent $t$-\early
instance $\calJ'$ of $\calJ$, and a valid assignment at that time $t$.
Throughout this section, we assume that the Fast-Forward Lemma
(\Cref{lem:nontrivial update assignment}) holds.

\subsection{Proof Idea}
\label{sec:proof-idea}

The main idea in this section is to maintain a valid assignment
inductively over time. We first observe that if $\ALG$ processes the
jobs with the smallest remaining processing time, any valid assignment
remains valid.
\begin{lemma}[SRPT Maintains Validity] \label{lem:carve min
    assignment} If there is a valid assignment between $\ALG(s)$ and
  $\OPT(s)$ at time $s$ and $\ALG$ works only on known jobs during an
  interval $[s,s']$, then, for all $t \in [s,s']$, there is a valid
  assignment between $\ALG(t)$ and $\OPT(t)$.
\end{lemma}

We defer the proof of \Cref{lem:carve min assignment} to
\Cref{sec:operations} where we introduce operations to modify
assignments. Next, we define the notion of \emph{magical time.}

\begin{defn} \label{def:magical time} Call a time $s$ \emph{magical}
  if $\ALG$ does not touch the jobs that are alive and unknown at time
  $s$ during the interval $[s,t]$.
\end{defn}
Suppose we have a valid assignment at some magical time $s < t$. There
are two cases:
\begin{enumerate} [nosep]
\item If $\ALG$ works on known jobs for some time interval $[s,s']$,
  we can use \Cref{lem:carve min assignment} to get a valid assignment
  at time $s'$; moreover, it is easy to see that $s'$ must be magical
  as well.
\item Else, new jobs must arrive at time $s$, because $s$ is magical.
  We define the ``leader''
  to be the new job $j'$ with the largest processing time. Let $b_s$ be
  the time when the leader becomes known.
  \begin{enumerate}
  \item If $b_s \leq t$, and no new jobs arrive during $[s,b_s]$, then
    we can apply \Cref{lem:nontrivial update assignment} to get a
    valid assignment at time $b_s$. Since the unknown jobs at $b_s$ are
    the same as those at time $s$, it follows that $b_s$ is magical,
    and since $b_s > s$, we made progress.

  \item Else if $b_s \leq t$ and new jobs arrive in $[s,b_s]$, then we
    argue in \Cref{lem:early} that we can let these jobs arrive at
    time $s$ instead, without changing the state of $\ALG$ at time
    $b_s$ (and only helping $\OPT$). This may change the leader, so we
    redefine $b_s$ and repeat this step.

  \item Else if $b_s > t$, then we consider the time $\ell_s$ is the last
    time the leader $j'$ is touched before time $t$. Again, we do the
    same steps: if jobs arrive in $[s,\ell_s]$, we let them arrive at
    time~$s$ and redefine the leader as necessary. Otherwise, we get a
    valid assignment at time $\ell_s$, argue that $\ell_s$ is magical, and
    make progress again.
  \end{enumerate}
\end{enumerate}

In the rest of the section, we give an algorithm that crystallizes the
above argument and show how it proves \Cref{thm:there is always a
  valid assignment}.

\subsection{The Create-Valid-Assignment Algorithm}

To formalize this sketch, consider the $\textsc{MoveJobs}$ operation:
given a job instance $\calJ$, times $x,y$ where $x \leq y$,
$\textsc{MoveJobs}(\calJ,x,y)$ moves the release times of all jobs
$\calJ$ during $(x,y]$ to the time right after~$x$.

\begin{defn}
  $\textsc{MoveJobs}(\calJ,x,y)$ returns a job instance $(\calJ \setminus J') \cup J$ where $J'$ is the set of
  jobs in $\calJ$ that are released during $(x,y]$, and $J$ is the set
  of jobs identical to the jobs in $J'$ but with release times modified to $x^+$ where $x^+$ is the time immediately after $x$.
\end{defn}

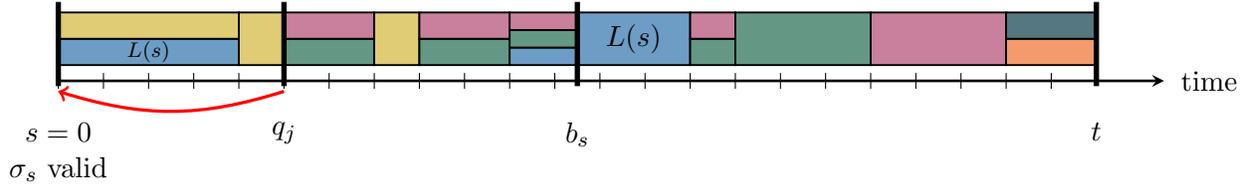
\begin{figure}
    \begin{subfigure}[r]{\textwidth}
    \begin{flushright}
    \begin{tikzpicture}[xscale=0.6, yscale=0.7,tips=on proper draw]

  \rectjobT[job1](0,0)(4,1/2)({\scriptsize$L(s)$});
  \rectjob[job2](0,1/2)(4,1/2);

  \rectjob[job2](4,0)(1,1);

  \rectjob[job3](5,0)(2,1/2);
  \rectjob[job4](5,1/2)(2,1/2);

  \rectjob[job2](7,0)(1,1);

  \rectjob[job3](8,0)(2,1/2);
  \rectjob[job4](8,1/2)(2,1/2);

  \rectjob[job1](10,0)(3/2,1/3);
  \rectjob[job3](10,1/3)(3/2,1/3);
  \rectjob[job4](10,2/3)(3/2,1/3);

  \rectjobT[job1](11.5,0)(2.5,1)({$L(s)$});

  \rectjob[job3](14,0)(1,1/2);
  \rectjob[job4](14,1/2)(1,1/2);

  \rectjob[job3](15,0)(3,1);
  \rectjob[job4](18,0)(3,1);

  \rectjob[job5](21,0)(2,1/2);
  \rectjob[job6](21,1/2)(2,1/2);

  \node at (25.5,-0.3) {time};
  \draw[black,line width=1pt,->,>=stealth] (0,-0.3) -- (24.5,-0.3);
      \foreach \x in {0,1,...,22}{
          \draw[black] (\x,-0.3-0.15) -- (\x,-0.3+0.15);
    }

  \draw[line width=2pt] (0,-0.4) -- node[below=25pt,align=center] (s) {$s=0$\\ $\sigma_s$ valid} (0,1.2);
  \draw[line width=2pt] (23,-0.4) -- node[below=25pt] {$t$} (23,1.2);
  \draw[line width=2pt] (5,-0.4) -- node[below=25pt] (r) {$q_j$} (5,1.2);
  \draw[line width=2pt] (11.5,-0.4) -- node[below=25pt] {$b_s$} (11.5,1.2);

  \draw[red,line width=1.2pt,->] (5,-0.5) edge[bend left=15] (0,-0.5);
\end{tikzpicture}
      \end{flushright}
    \caption{New jobs arrive in $(s,b_s]$. We are in Case~\ref{line:s get bs some jobs} and let these jobs arrive at time $s$ instead.}
    \end{subfigure}
    \par\bigskip
    
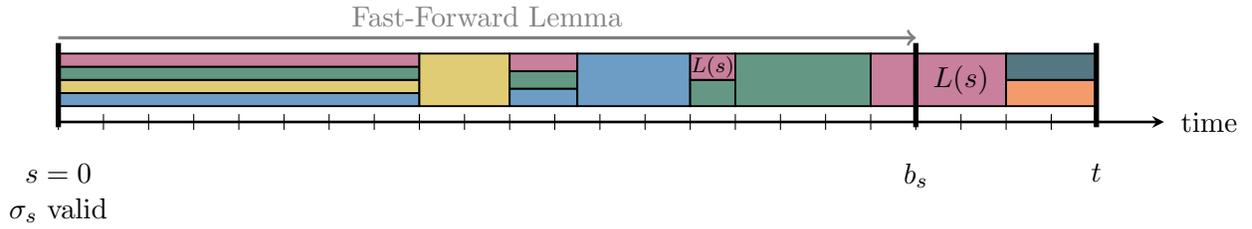
\begin{subfigure}[r]{\textwidth}
    \begin{flushright}
    \begin{tikzpicture}[xscale=0.6, yscale=0.7]

  \rectjob[job1](0,0)(8,1/4);
  \rectjob[job2](0,1/4)(8,1/4);
  \rectjob[job3](0,2/4)(8,1/4);
  \rectjob[job4](0,3/4)(8,1/4);

  \rectjob[job2](8,0)(2,1);

 \rectjob[job1](10,0)(3/2,1/3);
  \rectjob[job3](10,1/3)(3/2,1/3);
  \rectjob[job4](10,2/3)(3/2,1/3);

 \rectjob[job1](11.5,0)(2.5,1);

 \rectjob[job3](14,0)(1,1/2);
  \rectjobT[job4](14,1/2)(1,1/2)({\scriptsize{$L(s)$}});

  \rectjob[job3](15,0)(3,1);
  \rectjob[job4](18,0)(1,1);
  \rectjobT[job4](19,0)(2,1)({$L(s)$});

  \rectjob[job5](21,0)(2,1/2);
  \rectjob[job6](21,1/2)(2,1/2);

  \node at (25.5,-0.3) {time};
  \draw[black,line width=1pt,->,>=stealth] (0,-0.3) -- (24.5,-0.3);
      \foreach \x in {0,1,...,22}{
          \draw[black] (\x,-0.3-0.15) -- (\x,-0.3+0.15);
    }

 \draw[line width=2pt] (0,-0.4) -- node[below=25pt,align=center] (s) {$s=0$\\ $\sigma_s$ valid} (0,1.2);
  \draw[line width=2pt] (23,-0.4) -- node[below=25pt] {$t$} (23,1.2);
  \draw[line width=2pt] (19,-0.4) -- node[below=25pt] {$b_s$} (19,1.2);

  \draw[gray,line width=1.2pt,->] (0,1.3) --node[midway,above] {Fast-Forward Lemma} (19,1.3);

\end{tikzpicture}
      \end{flushright}
      \caption{The schedule for the modified instance $\calJ'$. Note that the schedule after the time when the previous leader (blue) becomes known has not changed. The leader for jobs arriving at time $s$ has changed (blue to pink). We are now in Case~\ref{line:s get bs no jobs} and apply the Fast-Forward Lemma (\Cref{lem:nontrivial update assignment}) to get a valid assignment for time $b_s$.}
     \end{subfigure}
     
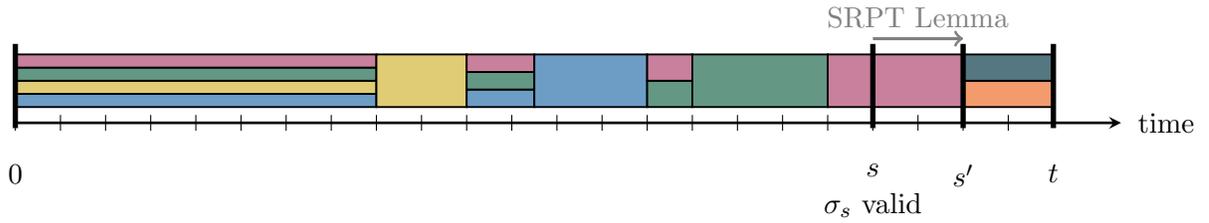
\par\bigskip
    \begin{subfigure}[r]{\textwidth}
    \begin{flushright}
    \begin{tikzpicture}[xscale=0.6, yscale=0.7]

  \rectjob[job1](0,0)(8,1/4);
  \rectjob[job2](0,1/4)(8,1/4);
  \rectjob[job3](0,2/4)(8,1/4);
  \rectjob[job4](0,3/4)(8,1/4);

  \rectjob[job2](8,0)(2,1);

 \rectjob[job1](10,0)(3/2,1/3);
  \rectjob[job3](10,1/3)(3/2,1/3);
  \rectjob[job4](10,2/3)(3/2,1/3);

 \rectjob[job1](11.5,0)(2.5,1);

 \rectjob[job3](14,0)(1,1/2);
  \rectjob[job4](14,1/2)(1,1/2);

  \rectjob[job3](15,0)(3,1);
  \rectjob[job4](18,0)(3,1);

  \rectjob[job5](21,0)(2,1/2);
  \rectjob[job6](21,1/2)(2,1/2);

  \node at (25.5,-0.3) {time};
  \draw[black,line width=1pt,->,>=stealth] (0,-0.3) -- (24.5,-0.3);
      \foreach \x in {0,1,...,22}{
          \draw[black] (\x,-0.3-0.15) -- (\x,-0.3+0.15);
    }

 \draw[line width=2pt] (19,-0.4) -- node[below=25pt,align=center] (s) {$s$\\ $\sigma_s$ valid} (19,1.2);
  \draw[line width=2pt] (23,-0.4) -- node[below=25pt] {$t$} (23,1.2);
  \draw[line width=2pt] (21,-0.4) -- node[below=25pt] {$s'$} (21,1.2);
  \draw[line width=2pt] (0,-0.4) -- node[below=25pt] {$0$} (0,1.2);

  \draw[gray,line width=1.2pt,->] (19,1.3) --node[midway,above] {SRPT Lemma} (21,1.3);

\end{tikzpicture}
    \end{flushright}
    \caption{At time $s$, the algorithm touches a known job (pink). We are in Case~\ref{line:alg touches known} and apply the SRPT Lemma (\Cref{lem:carve min
    assignment}) to get a valid assignment for time $s'$.}
     \end{subfigure}
     
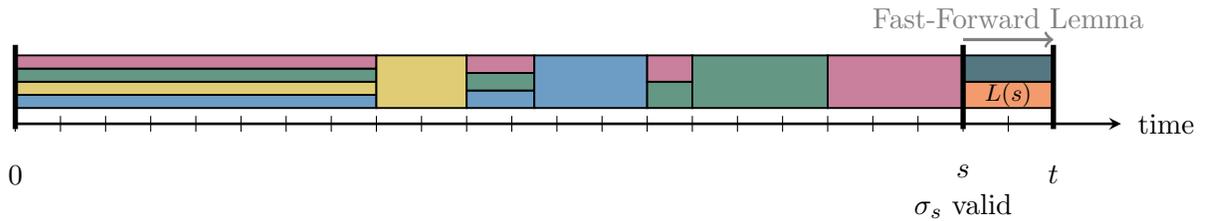
\par\bigskip
    \begin{subfigure}[r]{\textwidth}
    \begin{flushright}
    \begin{tikzpicture}[xscale=0.6, yscale=0.7]

  \rectjob[job1](0,0)(8,1/4);
  \rectjob[job2](0,1/4)(8,1/4);
  \rectjob[job3](0,2/4)(8,1/4);
  \rectjob[job4](0,3/4)(8,1/4);

  \rectjob[job2](8,0)(2,1);

 \rectjob[job1](10,0)(3/2,1/3);
  \rectjob[job3](10,1/3)(3/2,1/3);
  \rectjob[job4](10,2/3)(3/2,1/3);

 \rectjob[job1](11.5,0)(2.5,1);

 \rectjob[job3](14,0)(1,1/2);
  \rectjob[job4](14,1/2)(1,1/2);

  \rectjob[job3](15,0)(3,1);
  \rectjob[job4](18,0)(3,1);

  \rectjobT[job5](21,0)(2,1/2)({\footnotesize$L(s)$});
  \rectjob[job6](21,1/2)(2,1/2);

  \node at (25.5,-0.3) {time};
  \draw[black,line width=1pt,->,>=stealth] (0,-0.3) -- (24.5,-0.3);
      \foreach \x in {0,1,...,22}{
          \draw[black] (\x,-0.3-0.15) -- (\x,-0.3+0.15);
    }

 \draw[line width=2pt] (21,-0.4) -- node[below=25pt,align=center] (s) {$s$\\ $\sigma_s$ valid} (21,1.2);
  \draw[line width=2pt] (23,-0.4) -- node[below=25pt] {$t$} (23,1.2);
  \draw[line width=2pt] (0,-0.4) -- node[below=25pt] {$0$} (0,1.2);

  \draw[gray,line width=1.2pt,->] (21,1.3) --node[midway,above] {Fast-Forward Lemma} (23,1.3);

\end{tikzpicture}
      \end{flushright}
      \caption{New jobs arrive at time $s$. We are in Case~\ref{line:step d} and have $\ell_s = t$, but no new jobs arrive during $(s,t]$. We apply the Fast-Forward Lemma (\Cref{lem:nontrivial update assignment}) to get a valid assignment for time $t$, and are done.}
    \end{subfigure}
    \caption{An example execution of the Create-Valid-Assignment Algorithm (\Cref{alg:create assignment}). We consider an instance with $\eps = \nicefrac{1}{2}$ where the yellow job has length $4$, the blue job has length $5$, the green job has length $6$, and the pink job has length $8$.}
    \label{fig:create-valid-assignment-example}
\end{figure}

The procedure to get the $t$-equivalent (and hence $t$-\early) instance $\calJ'$ appears
in \Cref{alg:create assignment}. As outlined above, we iteratively
decrease the release time of some jobs as we induct over times. An example execution of the algorithm is illustrated in \Cref{fig:create-valid-assignment-example}.

\begin{algorithm}

\caption{Create-Valid-Assignment$(\cJ, t)$}\label{alg:create assignment}

\medskip

\begin{itemize}
\item Start with a (time) counter $s = 0$, and $\calJ' = \calJ$.

\item \textbf{While} $s < t$,
  \begin{enumerate}
  \item The algorithm $\ALG$ has some known jobs $K(s)$ and some unknown
    jobs $U(s)$ at time $s$ for job instance $\calJ'$; either set may be empty. We
    maintain the following invariants at time $s$, denoted by  \textbf{Inv}$(s,\calJ')$:
    \begin{itemize} [noitemsep,nolistsep]
    \item \textbf{Inv1}$(s,\calJ')$: None of the jobs in $U(s)$ is
      processed between time $s$ and $t$. This holds regardless of job
      arrivals during $[s,t).$ Recall: this time $s$ is called \emph{magical}.
    \item $\textbf{Inv2}(s,\calJ')$: $\calJ'$ is $s$-equivalent to $\calJ$.
    \item $\textbf{Inv3}(s,\calJ')$: There is a valid assignment between $\ALG$ and $\OPT$ at time $s$ for job instance $\calJ'$.

    \end{itemize} We prove the loop invariants in \Cref{label:inv maintained}.
  \item \textbf{If} $\ALG$ touches a known job at time $s$, \label{line:alg touches known}
    \textbf{then}
    \begin{enumerate} [nosep]
        \item It runs only known jobs in $K(s)$ until
    some time $s'$.
        \item     Apply \Cref{lem:carve min assignment} and set $s \gets \min\{s',t\}$.
    \end{enumerate}

  \item \label{line:non trivial else} \textbf{Else}, by
    $\textbf{Inv1}(s, \calJ')$, we have the following.
    \begin{claim} \label{claim:new jobs alg}
      A set of jobs from $\calJ'$ arrives at time $s$.
    \end{claim}
    \begin{enumerate}
    \item Let $J_{new}$ be the  set of jobs in $\calJ'$ that arrives at time $s$.
      Call the longest (the one with the largest processing time) job in
      $J_{new}$ the \emph{leader} for time $s$, and denote this job by
      $\leader(s)$.  \label{item:4} Let $b_s$ be the time at which job
      $\leader(s)$ becomes known.

    \item\label{line:s get bs} \textbf{If} $b_s \leq t$:
      \begin{enumerate}
      \item \textbf{If} there are jobs arriving in the interval $(s,b_s]$, then $\calJ' \gets \textsc{MoveJobs}(\calJ', s,b_s)$\label{line:s get bs some jobs}
      \item \textbf{Else} (there are no jobs arriving in $(s,b_s]$),
      apply \Cref{lem:nontrivial update assignment} and set $s \gets b_s$. \label{line:s get bs no jobs}
      \end{enumerate}

    \item\label{line:step d} \textbf{Else} $b_s > t$:
      \begin{enumerate}

      \item \label{line:ltt} Consider the \emph{last touch time} $\ell_s$ of the leader
        $\leader(s)$: this is the last time in $[s,t]$ that the job
        $\leader(s)$ is touched by the algorithm. This time $\ell_s$ exists because any new job is touched when it is released.
        \item $\calJ' \gets \textsc{MoveJobs}(\calJ', s, \ell_s)$
        \item Apply \Cref{lem:nontrivial update assignment} and set $s \gets \ell_s$.
      \end{enumerate}
    \end{enumerate}
  \end{enumerate}
  \item \textbf{Return} a valid assignment between  $\ALG_{\calJ'}(t)$ and $\OPT_{\calJ'}(t)$ at time $t$. Note that $s = t$.
\end{itemize}

\end{algorithm}

\subsection{Analysis}

We want to show that the steps of \Cref{alg:create assignment} are
well-defined, and the invariants are maintained. Let us first record
  a useful observation, whose proof we defer to the appendix.

\begin{observation}
  \label{obs:known-jobs-block-earlier-jobs}
  Let $j$ be a job touched by \ALG at time $t'$ while being
  known at $t'$. Then no job $j' \not= j$ with $q_{j'} < t'$ that is
  unknown at $t'$ is touched by \ALG during $(t',c_j]$, where $c_j$ is
  the time at which $j$ completes.
\end{observation}
 
Next, we prove the
crucial fact that if we are at some time $t' \in [s, t]$ such that
$\ALG$ is working on the leader $L(s)$, then it is safe to move all
jobs arriving during $[s,t']$ to $s$.

\begin{lemma} [Early-Arriving Lemma] \label{lem:early} Let
  $\mathcal{J}$ be a job instance.  Let $J_{new} \subseteq \calJ$ be a
  set of jobs that arrive at time $s$.  Let $L_{\calJ}(s)$ be the job
  with the largest processing time in $J_{new}$.  If $\ALG$ touches
  $L_{\calJ}(s)$ at time $t' \in [s,t]$ and $L_{\calJ}(s)$ is still
  unknown at $t'$, then $\calJ'$ and $\calJ$ are
  $t'$-equivalent
  where $\calJ' = \textsc{MoveJobs}(\calJ,s,t').$
\end{lemma}

\begin{proof}
  Let $\gamma$ be the elapsed time of $L_{\calJ}(s)$ at time $t'$ and let $\bar{J} = \{ j \in \calJ \mid s <  q_j \le t'\}$ denote the set of jobs released during $(s,t']$. Note that $\bar{J}$ contains exactly the jobs whose release times are changed by the $\textsc{MoveJobs}(\calJ,s,t')$-operation. For each job $j \in \bar{J}$, we denote $j'$ as the corresponding job whose release time is changed to $s^+$ and use $q_j'$ to refer to the changed release time.
We also denote $\bar{J}' = \{ j' \colon j \in \bar{J}\}$.

Consider the execution of $\ALG$ on the instance $\calJ'$. Let $t''$
be the first time at which $e_{\calJ'}(L_{\calJ}(s),t'') =
e_{\calJ}(L_{\calJ}(s),t') = \gamma$. To show that $t' = t''$ and the
instances are $t'$-equivalent, it suffices to show that
\begin{enumerate}[nosep]
\item[(A)] $e_{\calJ}(j,t') = e_{\calJ'}(j',t'')$ holds for all jobs
  $j \in \bar{J}$ and
\item[(B)] $e_{\calJ}(j,t') = e_{\calJ}(j,t'')$ holds
  for all jobs $j \in \calJ' \cap \calJ$ with $q_j = q_{j}' \le t'$.
\end{enumerate}

To show~(A),
fix an $j \in \bar{J}$ and let $j' \in \bar{J}'$ be the corresponding job in $\calJ$. By \Cref{lem:new job joins RR}, there are two cases:
\begin{itemize} [nosep]
    \item Case A1: $j$ is completed by time $t'$ on the instance $\calJ$. Thus, $\myAlfa \cdot p_j < \gamma$ since the algorithm processes $j$ at the point in time $d < t'$ when it becomes known, either in parallel with $L_{\calJ}(s)$ or when it preferred $j$ over $L_{\calJ}(s)$.
    We show that $j'$ must be completed by time $t''$. Since $\myAlfa \cdot p_{j'} = \myAlfa \cdot p_j < \gamma$, $j'$ must become known before $t''$. Once $j'$ is known, the algorithm will only touch $L_{\calJ}(s)$ again once $j'$ is completed (cf.~\Cref{obs:known-jobs-block-earlier-jobs}). Thus, $j'$ must be completed  by time $t''$ on the instance $\calJ'$.
    \item Case A2: $e_{\calJ}(j,t') = e_{\calJ}(L_{\calJ}(s),t') = \gamma$. This can only happen if $\myAlfa \cdot p_j \ge \gamma$, as otherwise $j$ would become known before $t'$ and thus the algorithm would complete $j$ before $t'$.
    We show that $e_{\calJ'}(j',t'') = \gamma.$ Indeed, since $\myAlfa \cdot p_{j'} = \myAlfa \cdot p_j \ge \gamma$ and $q'_{j'}$ is immediately after $s = q'_{L_{\calJ}(s)}$, the algorithm on instance $\calJ'$ will only process $j'$ and $L_{\calJ}(s)$ in parallel until their respective elapsed processing times reach $\gamma$.
    Thus, $e_{\calJ'}(j',t'') = e_{\calJ}(L_{\calJ}(s),t'') = \gamma =
   e_{\calJ}(j,t')$.
\end{itemize}
This proves (A). Next, to show (B), we only focus on jobs $j \in
\ALG_{\calJ}(s)\setminus \bar{J}$, since all jobs that are completed by  time $s$ on instance $\calJ$ clearly also complete by  time $s$ on instance $\calJ'$.
Fix an arbitrary job  $j \in \ALG_{\calJ}(s) \setminus \bar{J}$. We distinguish between the following cases:
\begin{itemize}
    \item Case B1: $j$ is not touched during $[s,t']$ on $\calJ$. In
      this case, we either have $e_{\calJ}(j,s) \ge \gamma$ if $j$ is
      still unknown at $s$ or $r_{\calJ}(j,s) \ge \threshold \gamma$
      if $j$ is already known, as otherwise the algorithm should have
      preferred $j$ over $L_{\calJ}(s)$ at some point during
      $[s,t']$. This implies either $e_{\calJ'}(j,s) \ge \gamma$ or
      $r_{\calJ'}(j,s) \ge \threshold \gamma$. In either case, the definition of the algorithm implies that $j$ will not be touched on $\calJ'$ before the elapsed processing time of $L_{\calJ}(s)$ reaches $\gamma$. Thus, $e_{\calJ'}(j,t'') = e_{\calJ'}(j,s) = e_{\calJ}(j,s) = e_{\calJ}(j,t')$.
    \item Case B2: $j$ is touched but not completed during $[s,t']$ on $\calJ$. In this case, we must have $e_{\calJ}(j,t') = e_{\calJ}(L_{\calJ}(s),t')=\gamma$ as otherwise $j$ would have become known before $t'$ and thus the algorithm should have completed $j$ before $t'$. We can now follow the same argument as in Case A2 above.
    \item Case B3: $j$ is touched and completed during $[s,t']$ on $\calJ$. In this case, we must have $\myAlfa \cdot p_j < \gamma$ due to the fact that the algorithm processes $j$ at some time $d < t'$ when it becomes known, either in parallel with $L_{\calJ}(s)$ or when it preferred $j$ over $L_{\calJ}(s)$. We can now follow the same argument as in Case A1.
\end{itemize}
This proves (B) and hence the proof of the lemma.
\end{proof}

The next lemma illustrates the all-important property of the last
touch time of the leader $L(s)$: that this time is magical, and moreover, all
jobs released at time $s$ (and $s^+$) have either finished, or have received the
same processing as $L(s)$.

\begin{lemma} \label{lemma:J_new not touched}
  If \Cref{alg:create assignment} reaches Step~\ref{line:ltt} where $\ell_s$ is defined as the last touch time of the leader,
  then $\ell_s$ is magical, and for all jobs $j \in J_{new}$, $j$ either remains unknown or is completed at time~$\ell_s$.
\end{lemma}
\begin{proof}
  At Step~\ref{line:ltt}, we have $\ell_s \le t < b_s$, and thus the leader $L(s)$ remains unknown at time $\ell_s$. Fix a job $j \in J_{new}$ at time $\ell_s$. By \Cref{lem:new job joins RR}, $j$ is either unknown at $\ell_s$ or completed by time  $\ell_s$. We next prove that $\ell_s$ is magical. That is, we prove that if $j$ is unknown then $j$ will not be touched again during the interval $(\ell_s,t]$. By \Cref{lem:new job joins RR}, we have $e_{j}(\ell_s) = e_{L(s)}(\ell_s)$, i.e., the elapsed time at time $\ell_s$ of $j$ equals that of the leader.
  We claim that, after time $\ell_s$, none of the jobs in $J_{new}$ will be processed. Suppose otherwise. Let $t'$ be the first time in the interval $(\ell_s,t]$ that some job in $J_{new}$ is touched. Since $j'$ has the same elapsed time as the leader $L(s)$ right before time $t'$, the leader $L(s)$ must be touched as well, and thus the last touch time of $L(s)$ is $t'\le \ell_s $, contradicting to $t' > \ell_s$.
\end{proof}

Recall that having a magical time $\ell_s$ (in addition to some
technical conditions) is what allows us to apply \Cref{lem:nontrivial
  update assignment}: this is what drives the next lemma, which shows
that we maintain our invariants in each iteration of the main
\textsw{while} loop of the algorithm.

\begin{lemma} \label{label:inv maintained}
  The loop invariant $\textbf{Inv}(s,\calJ')$ is maintained
  at the beginning of each iteration.
\end{lemma}

\begin{proof}
  Recall that the invariants are:
  \begin{itemize} [nolistsep]
  \item \textbf{Inv1}$(s,\calJ')$: $s$ is magical (\Cref{def:magical time}): None of the jobs in $U(s)$ is processed between time $s$ and $t$. This holds regardless of job arrivals during $[s,t)$.
  \item $\textbf{Inv2}(s,\calJ')$: $\calJ'$ is $s$-equivalent to $\calJ$.
  \item $\textbf{Inv3}(s,\calJ')$: There is a valid assignment between $\ALG$ and $\OPT$ at time $s$ for
    $\calJ'$.
  \end{itemize}
  Initially, $s = 0, \calJ' = \calJ$ and $U(s) = \emptyset$, and thus the invariant is trivially maintained. Assume that
  $\textbf{Inv}(s,\calJ')$
  holds at the beginning of an iteration.  Let $(s',\calJ'')$ be the time and job instance at the end of the iteration. We show that  $\textbf{Inv}(s',\calJ'')$ holds. By the description of the algorithm, either $s' > s$ or $\calJ''$ is a proper ($\calJ''\not= \calJ$)  $t$-\early instance of $\calJ'$.

  There are two cases. The easy case is if \ALG works on a known job
  at time $s$, then we apply \Cref{lem:carve min assignment} using
  time $s'$ to get a valid assignment at time $s'$. Since
  $U(s') = U(s)$ and $\calJ'' = \calJ'$, $s'$ is also
  magical. Therefore, $\textbf{Inv}(s',\calJ'')$ holds in this case.

  Else, we prove \Cref{claim:new jobs alg}.  Indeed, observe that at
  any time, $\ALG$ works on either known jobs or unknown jobs since
  the algorithm is non-idling. Since $\ALG$ does not run on known jobs
  in $K(s)$ at time $s$, $\ALG$ must start working on some unknown
  jobs at time $s$. However, $\ALG$ does not work on $U(s)$ at time
  $s$ by Invariant $\textbf{Inv1}(s, \calJ')$. Therefore, some new
  jobs $J_{new}$ must be released at time $s$ in $\calJ'$. This proves \Cref{claim:new jobs alg}.

  We now resume showing that the invariants hold at time $s$.
  Recall that
  $L(s) \in J_{new}$ is the leader and $b_s$ is the time when the
  leader becomes known.

  \begin{itemize}
  \item If $b_s \leq t$ and there is some job arriving in the interval $(s,b_s]$, then $\calJ'' \gets \textsc{MoveJobs}(\calJ',s, b_s)$. In this case, $s' = s$ but $\calJ''$ is a proper $t$-\early instance of $\calJ'$. We verify that \textbf{Inv}$(s',\calJ'')$ holds:
    \begin{enumerate}[nosep]
    \item \textbf{Inv1}$(s',\calJ'')$. $s'$ is magical because $U(s) = U(s')$.
    \item \textbf{Inv2}$(s',\calJ'')$. By assumption, $\calJ'$ is $s$-equivalent to $\calJ$. Since $\calJ'$ and $\calJ''$ only differ in release dates \emph{after} $s=s'$, we get that $\calJ'$ and $\calJ''$ are $s'$-equivalent. Hence $\calJ$ and $\calJ''$ are $s'$-equivalent.
    \item \textbf{Inv3}$(s',\calJ'')$. This follows because $s = s'$ and the set of jobs in $\calJ''$ that arrive before $s$ is identical to that of $\calJ'$.
    \end{enumerate}

  \item If $b_s \leq t$ and no job arriving during $(s,b_s]$, then  $s' = b_s$ and $\calJ'' = \calJ'$ by Step~\ref{line:s get bs no jobs}.  We verify that \textbf{Inv}$(s',\calJ'')$ holds:
    \begin{enumerate}[nosep]
    \item \textbf{Inv1}$(s',\calJ'')$.  We prove that
      at time $b_s$, every other job in $J_{new} \setminus \{L(s)\}$ is  completed.
      This implies that  $s'$ is magical because the set of unknown jobs remains the same as in time $s$.
      \textbf{Proof.}  Since every job has unique processing time and the leader has the highest processing time then Thus,  every job in $J_{new} \setminus \{L(s)\}$ at time $b_s$ cannot have the same elapsed time as the leader and remains unknown. It follows, by  \Cref{lem:new job joins RR}, every job in $J_{new} \setminus \{L(s)\}$ is completed.

    \item \textbf{Inv2}$(s',\calJ'')$.
      If the algorithm did not modify any release dates after $s$ in previous iterations, i.e., all release dates after $s$ are still the same as in instance $\calJ$, then the invariant follows trivially as the current iteration also does not move release dates after $s$. Hence $\calJ' = \calJ''$ and $\calJ$ being $s$-equivalent implies them being $s'$-equivalent.

      If the algorithm modified release dates after $s$ in previous iterations, then this must have happened in previous executions of Step~\ref{line:s get bs some jobs} as this is the only place in the algorithm where jobs after the current $s$ are moved without increasing $s$. Furthermore, all these executions of Step~\ref{line:s get bs some jobs} must have happened consecutively and directly before the current execution of Step~\ref{line:s get bs no jobs}, as otherwise the changed release dates would be before the current $s$. Note that multiple such executions of Step~\ref{line:s get bs some jobs} can happen, as each of the \textsc{MoveJobs}-operations can change the leader at time $s$. However, each change in the leader only increases $b_s$ and for each increase in $b_s$, \Cref{lem:early} and the second invariant for previous iterations imply $b_s$-equivalence between $\calJ$ and the current modified instance. In particular, this holds for time $s'$, which implies the invariant.
    \item \textbf{Inv3}$(s',\calJ'')$.   Observe that in this case, we have:
      \begin{enumerate} [nosep]
      \item The leader is touched while being unknown at time $b_s$
      \item there is no job arriving between $(s,b_s]$, and
      \item $\ALG$ does not work on $U(s)$ during $(s,t]$, which contains $(s,b_s]$.
      \end{enumerate}
      Therefore, we apply \Cref{lem:nontrivial update assignment} to get a valid assignment at time $b_s$.
    \end{enumerate}

  \item Else ($b_s > t$). In this case, $s' = \ell_s$ and $\calJ''$ is a $t$-\early instance  of $\calJ'$.  We verify that \textbf{Inv}$(s',\calJ'')$ holds:
    \begin{enumerate}[nosep]
    \item \textbf{Inv1}$(s',\calJ'')$. This follows from \Cref{lemma:J_new not touched}. Note that these unknown jobs in $J_{new}$ will not be processed again since $s'$ is magical.
    \item \textbf{Inv2}$(s',\calJ'')$. This follows from \Cref{lem:early} since $\ell_s$ is the time that the leader was touched.
    \item \textbf{Inv3}$(s',\calJ'')$. Observe that in this case, we have:
      \begin{enumerate} [nosep]
      \item The leader is touched while being unknown at time $\ell_s$,
      \item there is no job arriving between $(s,\ell_s]$, and
      \item $\ALG$ does not work on $U(s)$ during $(s,t]$, which contains $(s,\ell_s]$.
      \end{enumerate}
      Therefore, we apply \Cref{lem:nontrivial update assignment} to get a valid assignment at time $\ell_s$.
    \end{enumerate}
  \end{itemize}

  This completes the proof that \textbf{Inv}$(s',\calJ'')$ holds at the end of the iteration.
\end{proof}

\begin{lemma} \label{lem:main alg terminates}
  \Cref{alg:create assignment} terminates and returns a valid
  assignment
  on some $t$-equivalent job instance~$\calJ'$ at time $t$.
\end{lemma}
\begin{proof}
We show that the main while loop terminates. Each job in $\calJ'$ is in one of five states: unreleased, arrived, frozen,  known, and finished. Every job starts in the unreleased state. It is \textit{frozen} if it is unknown and remains untouched until time $t$.
When job is frozen or finished then it cannot change state anymore (before time $t$) so we refer to these two states as a \textit{terminating}. A job goes from unreleased state to the arrived state, and from the arrived state to the either the frozen or the known state, and from the known state to the finished state. During the loop, there are four possible ways to complete the iteration. We argue that in any of these four ways, either
the state of some job moves forward when we move the time, or we identify the release time of a job with the release time of an earlier job.

    \begin{itemize}[nosep]
        \item Case~\ref{line:alg touches known}: $s \gets \min\{s',t\}$ where $s'$ is the time that a job is finished or a new job arrives.
        \item  Case~\ref{line:s get bs some jobs}: $s = s'$, but the release times of some jobs have been moved back and identified with the release times of some other jobs.
        \item  Case~\ref{line:s get bs no jobs}: $s \gets b_s$ and a job becomes known.
        \item  Case~\ref{line:step d}: $s \gets \ell_s$, and a job becomes frozen.
    \end{itemize}

    Since each job can progress at most 4 times before reaching a
    terminating state and every job's released time is modified at
    most once, the loop eventually terminates with some $t$-equivalent
    instance $\calJ'$. A valid assignment can then be computed because of \textbf{Inv3}$(t,\calJ')$.
\end{proof}

\begin{proof}[Proof of \Cref{thm:there is always a valid assignment}]
Fix a time $t$ and a job instance $\cJ$. We run
\textsc{Create-Valid-Assignment}$(\cJ, t)$. By \Cref{lem:main alg
  terminates}, it returns a valid assignment $\sigma'$ between $\ALG$
and $\OPT$ at time $t$ on the $t$-equivalent job instance  $\calJ'$.
In addition, $\cJ'$ is $t$-\early because the operation calls to
$\textsc{MoveJobs}$ only decreased the release times of jobs released
before time $t$.
Therefore, we have shown that $\sigma'$ is a valid
assignment between $\ALG_{\cJ'}(t)$ and $\OPT_{\cJ'}(t)$ at time $t$
and $\cJ'$ is a $t$-equivalent instance of $\cJ$, concluding the proof.
\end{proof}

\section{The Fast-Forward Lemma}
\label{sec:proof-update-lemma}

We now prove \Cref{lem:nontrivial update assignment},
which we used above to prove
\Cref{thm:there is always a valid assignment}, and therefore
\Cref{thm:main1}. Recall its statement:

\LemmaTwo*

 We begin by defining a set of operations to manipulate the fractional matching in \Cref{sec:operations}, and then use these operations to prove  \Cref{lem:nontrivial update assignment} in \Cref{sec:update-proof}.

\subsection{Operations on Assignments}
\label{sec:operations}

We view any assignment as a (fractional) matching,
i.e., an edge-weighted bipartite graph on the jobs of $\OPT$ and $\ALG$
whose marginals are the remaining processing times in the two algorithms.

\begin{defn}
Given a function $\sigma: V \times V^* \rightarrow \mathbb{R}_{\geq 0}$, the edge weighted bipartite graph representation of $\sigma$ is $H = (V,V^*,E,w)$ where $E$ is the support of $\sigma$ and $w = \sigma$. Each node $i \in V$ has weight $\vol_H(i) := \sum_{j \in V^*}w(i,j)$, and similarly, each node $j \in V^*$ has weight $\vol^*_H(j) := \sum_{i \in V}w(i,j)$.  Unless explicitly stated, the ordering is by non-increasing  $\vol_H$ values for $V$ and $\vol_H^*$ values for $V^*$ (breaking ties lexicographically, i.e., by their unique identifiers).
\end{defn}

\begin{defn} [Min-Suffix]
    \label{def:min-suffix}
    Let $H = (V,V^*,E,w)$ be a weighted bipartite graph. We define $\minsuffix_H(\beta)$ where $0 \leq \beta \leq \vol(H)$ as the smallest suffix $S$ of $V$ such that $\vol_H(S) \geq \beta$.
\end{defn}

\begin{defn}
  Let $H = (V,V^*,E,w)$ be a bipartite graph. We say that $P^*$ is $k$-\emph{prefix} of $B^* \subseteq V^*$ in $H$ if $P^* \subseteq B^*$ and $P^*$ contains first $k$ vertices in $B^*$.  We also say that  $P^*$ is a \emph{prefix} of $B^* \subseteq V^*$ in $H$ if $P^*$ is a $k$-prefix of $B^*$ for some $k$.
\end{defn}

A particularly nice ``canonical'' assignment is obtained by sorting jobs in both $\OPT$ and $\ALG$ in non-increasing order of remaining processing times, and greedily matching them, as follows:
\begin{defn}  [Canonical Assignment]
    An assignment at time $t$ is \emph{canonical} if it is obtained by the following
    greedy scheme:
    For each job $j\in \OPT(t)$ in monotonically non-increasing order of remaining time in $\OPT(t)$,  allocate the minimum number of edges of total value $r^*_{j}(t)$ to the jobs of the largest remaining time in $\ALG(t)$ without violating feasibility constraint (2) in \Cref{def:feas assign} for the jobs in $\ALG(t)$.
\end{defn}

\begin{lemma}\label{lem:valid imp canonical}
    If there is a valid assignment at time $t$, then the canonical assignment at time $t$ is valid.
\end{lemma}
\begin{proof}
 It is enough to prove that the canonical assignment $\sigma$
 has the smallest prefix expansion among all
 valid assignments.    Fix $k \leq |\OPT(t)|$.
 Let $S\subseteq V^*_\sigma$ be the $k$-prefix $\OPT(t)$.
 Let $v^*_t(k) := \sum_{i \leq k} r^*_{i}(t)$ be the total remaining time of the jobs in $S$.
    Every feasible assignment must assign at least $v^*_t(k)$ value  to the $k$-prefix of
    $\OPT(t)$. Since the canonical assignment collects this value from the jobs  of largest remaining time in $\ALG(t)$ it minimizes $|N_\sigma(S)|$.
\end{proof}

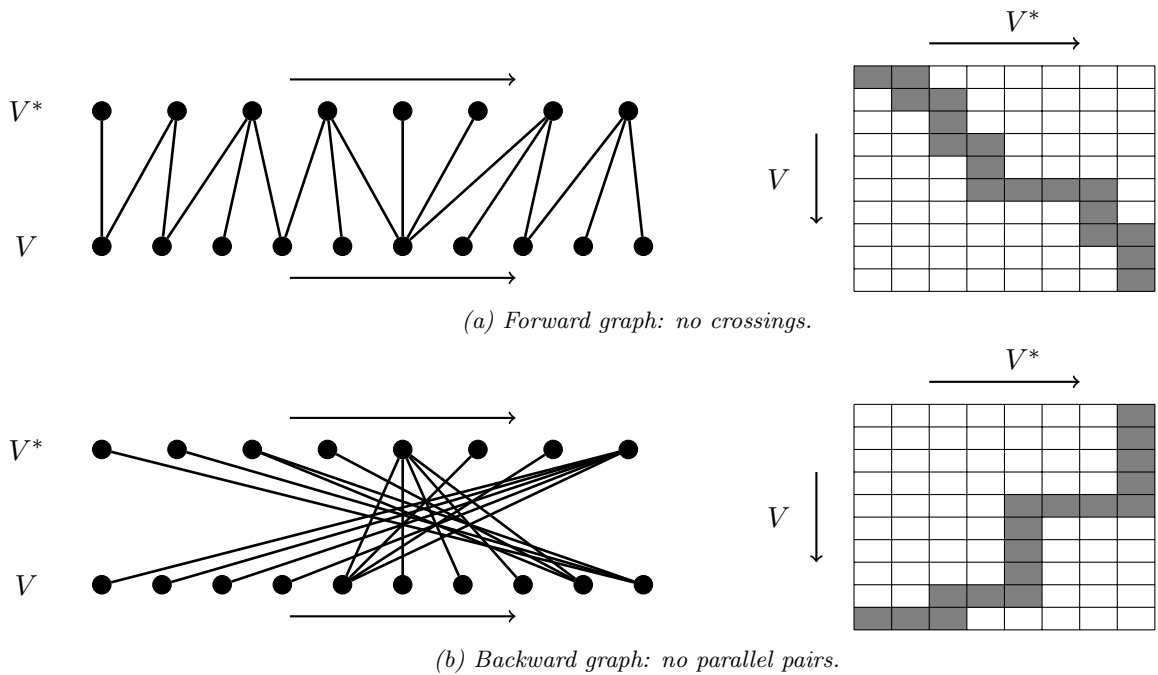
\begin{figure}
    \centering
    \begin{subfigure}[c]{\textwidth}
        \begin{tikzpicture}[yscale=0.6]
    \node at (-1,0) {$V$};
    \node at (-1,3) {$V^*$};

    \foreach \x in {0,...,7} {
    \node[vertex] (o\x) at (\x,3) {};
    }
    \foreach \x in {0,...,9} {
    \node[vertex] (a\x) at (9*0.8 - \x*0.8,0) {};
    }

    \draw[->,thick] (2.5,3.7) -- (5.5,3.7);
    \draw[->,thick] (2.5,-0.7) -- (5.5,-0.7);

    \begin{scope}[shift={(10,-1)}, scale=0.5]

        \foreach \x/\y in {0/9, 1/9, 1/8, 2/8, 2/7, 2/6, 3/6, 3/5, 3/4, 4/4, 5/4, 6/4, 6/3, 6/2, 7/2, 7/1, 7/0} {
            \draw[asg,black] (a\y) -- (o\x);
            \fill[gray] (\x,\y) rectangle++ (1,1);
        }

        \node at (-2,5) {$V$};
        \node at (4.5,12) {$V^*$};

        \draw[->,thick] (-1,7) -- (-1,3);
        \draw[->,thick] (2,11) -- (6,11);

        \draw[step=1cm, black, thin] (0,0) grid (8,10);
    \end{scope}
\end{tikzpicture}
        \caption{Forward graph: no crossings.}
    \end{subfigure}
    \begin{subfigure}[c]{\textwidth}
        \begin{tikzpicture}[yscale=0.6]
    \node at (-1,0) {$V$};
    \node at (-1,3) {$V^*$};

    \foreach \x in {0,...,7} {
    \node[vertex] (o\x) at (\x,3) {};
    }
    \foreach \x in {0,...,9} {
    \node[vertex] (a\x) at (9*0.8 - \x*0.8,0) {};
    }

    \draw[->,thick] (2.5,3.7) -- (5.5,3.7);
    \draw[->,thick] (2.5,-0.7) -- (5.5,-0.7);

    \begin{scope}[shift={(10,-1)}, scale=0.5]

        \foreach \x/\y in {0/0,1/0,2/0,2/1,3/1,4/1,4/2,4/3,4/4,4/5,5/5,6/5,7/5,7/6,7/7,7/8,7/9} {
            \draw[asg,black] (a\y) -- (o\x);
            \fill[gray] (\x,\y) rectangle++ (1,1);
        }

        \node at (-2,5) {$V$};
        \node at (4.5,12) {$V^*$};

        \draw[->,thick] (-1,7) -- (-1,3);
        \draw[->,thick] (2,11) -- (6,11);

        \draw[step=1cm, black, thin] (0,0) grid (8,10);
    \end{scope}
\end{tikzpicture}
        \caption{Backward graph: no parallel pairs.}
    \end{subfigure}
    \caption{Examples of forward and backward bipartite graphs $(V,V^*,E,w)$ and their corresponding matrix representations. Filled boxes correspond to edges. In the graphs, the vertices of $V$ and $V^*$ are ordered from left to right. In the matrix representation, the vertices of $V$ are ordered from
 top to bottom, and the vertices of $V^*$ are ordered from left to right. }
    \label{fig:forward-backward}
\end{figure}

In the proof of \Cref{lem:nontrivial update assignment}, we start with a canonical assignment $\sigma$  at time $s$ right before a batch of jobs arrives, which is valid by \Cref{lem:valid imp canonical} since there is a valid assignment.  Then, we view $\sigma$ as a bipartite graph $H$, and transform  $H$, along with the graph of the jobs in $J_{new}$, so that $\vol_H(i)$ and $\vol^*_H(i)$ correspond to the remaining time of job $i$ in $\ALG(\ell)$ and $\OPT(\ell)$, respectively.  The graph $H$ has a nice structure, which we call \emph{forward}. Sometimes, we might need to reverse the order, and thus, the notion of \emph{backward} graph is also useful. Adjacency matrices of forward and backward graphs are shown in \Cref{fig:forward-backward}.
Flipping the
ordering of $V^*$ changes a forward graph into a backward one, and
vice versa.

\begin{defn}[Forward and Backward Graphs]\label{def:forward backward} We define the following for a weighted bipartite graph $H = (V,V^*,E,w)$.
\begin{enumerate} [nosep]
    \item Let $\vol(H) := \sum_{i \in V} \vol_H(i).$ Note that $\vol^*(H) =  \sum_{i^* \in V^*}\vol^*_H(i^*) = \vol(H)$.
    \item We say that $H$  \emph{contains} a \emph{crossing} if there exist two distinct edges $(u_1,v_1), (u_2,v_2) \in E$ such that $u_1 > u_2$ and $v_1 < v_2$ where
    the ordering relations are defined over $V$ and $V^*$. We say that $H$ is \emph{forward} if $H$ does not contain a crossing.
    \item We say that $H$ \emph{contains} a \emph{parallel pair} if
      there exist two distinct edges $(u_1,v_1), (u_2,v_2) \in E$ such
      that $u_1 < u_2$ and $v_1 < v_2$ where the ordering relations are
      defined over $V$ and $V^*$. We say that $H$ is \emph{backward} if
      it does not contain a parallel pair.
\end{enumerate}
\end{defn}

We describe the following operations intuitively before proceeding with the formal definitions.
\begin{itemize}[nosep]
    \item $\textsc{GreedyMatching}(A,A^*,c,c^*):$ Takes two lists of demands $c$ on $A$ and $c^*$ on $A^*$, and creates a canonical assignment
    between $A^*$ and $A$ reversed.
  This operation will fix the demands  as in Stage 3 of \Cref{sec:toy}.
    \item $\textsc{Split}(H,\beta)$: Given a forward bipartite graph $H$, we split it into two forward subgraphs $H_p$ and $H_s$ where
the edges in $H_p$ ``precede'' the edges in $H_s$ (in the natural ordering induced on the edges of $H$ by the ordering of $V$ and $V^*$). We split such that the total weight of the edges in $H_s$ is $\beta$
    (one edge may split its weight between
    $H_p$ and $H_s$
    to make the volume of $H_s$ exactly $\beta$). This operation will update the assignment when $\OPT$ and $\ALG$ simultaneously process the smallest remaining time job in their respective queue.
    \item $\textsc{Merge}(H_1 = (V_1,V_1^*,E_1,w_1),H_2 = (V_2,V_2^*,E_2,w_2))$: Return $H = (V_1\cup V_2, V_1^*\cup V_2^*, E,w)$ where $w = w_1 + w_2$, and $E$ is obtained by running the $\textsc{GreedyMatching}$.
    This operation will be used to merge the perfect matching given by $J_{new}$ with the canonical assignment after some modifications.
\end{itemize}

First, we formalize the  Greedy assignment operation (c.f. \Cref{sec:toy}).

\begin{lemma} [GreedyMatching] \label{lem:greedy}
Let $A$ and $A^*$ be two sets of vertices with ordering relations $\leq_A$ and
$\leq_{A^*}$, respectively. Let $c: A \rightarrow \mathbb{R}_{\geq 0}$ and
$c^*: A^* \rightarrow \mathbb{R}_{\geq 0}$ be two mappings where $c(A) =
c^*(A^*)$.\footnote{$c(A)=\sum_{x\in A} c(x)$ and similarly for $A^*$.} There is a procedure which we call
$\textsc{GreedyMatching}(A,A^*,c,c^*)$ that returns a bipartite graph $H = (A,A^*,E,w)$ satisfying the following properties:
\begin{enumerate}  [nosep]
\item $A$ is ordered by $\leq_A$ and $A^*$ is ordered by $\leq_{A^*}$,
\item $\forall i \in A, \vol_H(i) = c(i)$ and $\forall i \in A^*,
  \vol_H^*(i) = c^*(i)$, and
\item $H$ is backward.
\end{enumerate}
\end{lemma}
\begin{proof}
The following construction gives all desired properties. We start with a bipartite graph $H = (A,A^*,\emptyset,w)$ where $w$ is a zero function. For each node $u \in A^*$ in $\leq_{A^*}$ ordering, let $S$ be the smallest suffix of $A$ in $\leq_{A}$ ordering of such that $c(S) \geq c^*(u)$: we assign $c^*(u)$ weight from $u$ to $S$ in such a way that every node $v$ in $S$ but the last one receives $c(v)$ weight; Then we update $c(u) \gets c(u) - \gamma_u$ where $\gamma_u$ is the amount of weights it receive in this iteration, and continue with the next node in $A^*$.
\end{proof}

Next, we define the union operation between two graphs.

\begin{defn} [Union]
Given bipartite graphs $H_k = (V_k,V^*_k,E_k,w_k)$ for $k \in
\{1,2\}$, their union $H_1 \cup H_2$ is the bipartite graph $H =
(V,V^*,E,w)$ with
\begin{enumerate} [nosep]
\item  $V = V_1 \cup V_2$ and $V^* = V^*_1 \cup V^*_2$, and
\item $w(i,j) = w_1(i,j) + w_2(i,j)$ for all $(i,j) \in V \times V^*$. (We define $w_k(i,j)=0$ when $(i,j) \not= H_k$.)
We put $(i,j)$ in $E$ iff $w(i,j) > 0$.
\end{enumerate}
\end{defn}

If $H = (V,V^*,E,w)$ is forward, we can define a total order, $\prec$, on its edge set $E$ as follows. For any two edges $e_1=(u_1,v_1),e_2=(u_2,v_2) \in E$  we say that $e_1 \prec e_2$ if $u_1 \leq u_2$ and $v_1 \leq v_2$ and $e_1 \neq e_2$.

Next, we define prefix and suffix graphs.
\begin{defn}[Prefix and Suffix Graphs]
Let $H = (V,V^*,E,w)$ be a forward bipartite graph. We say that $H_p$ is a \emph{prefix subgraph} of $H$ if
it contains exactly a subset
$E_p \subseteq E$ that forms a prefix according to $\prec$. Similarly,  we say that $H_s$ is a \emph{suffix subgraph} of $H$ if
it contains exactly a subset
$E_s \subseteq E$   that forms a suffix according to $\prec$.
\end{defn}

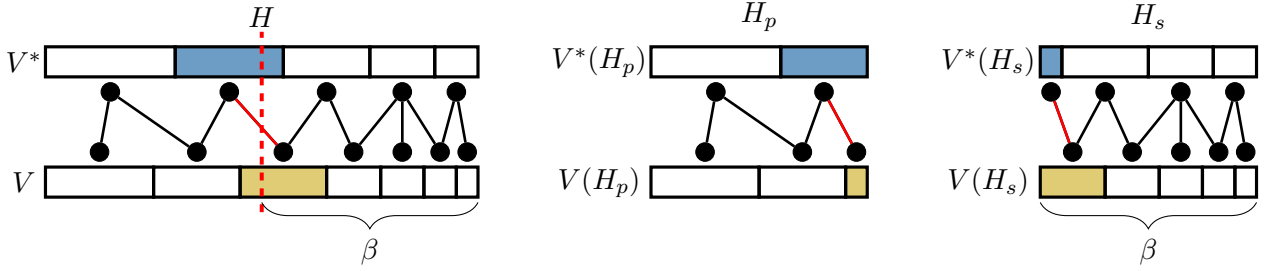
\begin{figure}
    \centering
    \begin{tikzpicture}[yscale=0.4,xscale=0.575]

     \rectjobnob[job1](3,4)(2.5,1);
     \rectjobnob[job2](4.5,0)(2,1);

		\node at (-0.5,4.5) {$V^*$};
		\pgfmathsetmacro\sum{0}
		\pgfmathsetmacro\index{0}
		\pgfplotsforeachungrouped \x in {3,2.5,2,1.5,1} {
			\matchingjob(\sum,4)(\x,1);
			\node[vertex] (o\index) at (\sum + 0.5 * \x ,3.5) {};
			\pgfmathsetmacro\sum{\sum + \x};
			\pgfmathsetmacro\index{int(\index + 1)};
		}

		\node at (-0.5,0.5) {$V$};
		\pgfmathsetmacro\sum{0}
		\pgfmathsetmacro\index{0}
		\pgfplotsforeachungrouped \x in {2.5,2,2,1.25,1,0.75,0.5} {
			\matchingjob(\sum,0)(\x,1);
			\node[vertex] (a\index) at (\sum + 0.5 * \x ,1.5) {};
			\pgfmathsetmacro\sum{\sum + \x};
			\pgfmathsetmacro\index{int(\index + 1)};
		}

        \foreach \x/\y in {0/0,0/1,1/1,1/2,2/2,2/3,3/3,3/4,3/5,4/5,4/6} {
              \draw[asg,black] (a\y) -- (o\x);
        }

        \node at (5,6) {$H$};
        \draw[asg,red] (a2) -- (o1);
        \draw[line width=1.5pt,red,dashed] (5,-0.5) -- (5,5.5);
        \draw [decorate,decoration={brace,mirror,amplitude=10pt,raise=2pt}] (5,0) -- (10,0) node[midway,below=12pt] {$\beta$};

        \begin{scope}[shift={(14,0)}]

            \rectjobnob[job1](3,4)(2,1);
            \rectjobnob[job2](4.5,0)(0.5,1);

            \node at (-1.2,4.5) {$V^*(H_p)$};
		\pgfmathsetmacro\sum{0}
		\pgfmathsetmacro\index{0}
		\pgfplotsforeachungrouped \x in {3,2} {
			\matchingjob(\sum,4)(\x,1);
			\node[vertex] (o\index) at (\sum + 0.5 * \x ,3.5) {};
			\pgfmathsetmacro\sum{\sum + \x};
			\pgfmathsetmacro\index{int(\index + 1)};
		}

		\node at (-1.2,0.5) {$V(H_p)$};
		\pgfmathsetmacro\sum{0}
		\pgfmathsetmacro\index{0}
		\pgfplotsforeachungrouped \x in {2.5,2,0.5} {
			\matchingjob(\sum,0)(\x,1);
			\node[vertex] (a\index) at (\sum + 0.5 * \x ,1.5) {};
			\pgfmathsetmacro\sum{\sum + \x};
			\pgfmathsetmacro\index{int(\index + 1)};
		}

        \foreach \x/\y in {0/0,0/1,1/1,1/2} {
              \draw[asg,black] (a\y) -- (o\x);
        }

        \node at (2.5,6) {$H_p$};
        \draw[asg,red] (a2) -- (o1);

        \end{scope}

    \begin{scope}[shift={(23,0)}]

        \rectjobnob[job1](0,4)(0.5,1);
        \rectjobnob[job2](0,0)(1.5,1);

        \node at (-1.2,4.5) {$V^*(H_s)$};
		\pgfmathsetmacro\sum{0}
		\pgfmathsetmacro\index{1}
		\pgfplotsforeachungrouped \x in {0.5,2,1.5,1} {
			\matchingjob(\sum,4)(\x,1);
			\node[vertex] (o\index) at (\sum + 0.5 * \x ,3.5) {};
			\pgfmathsetmacro\sum{\sum + \x};
			\pgfmathsetmacro\index{int(\index + 1)};
		}

		\node at (-1.2,0.5) {$V(H_s)$};
		\pgfmathsetmacro\sum{0}
		\pgfmathsetmacro\index{2}
		\pgfplotsforeachungrouped \x in {1.5,1.25,1,0.75,0.5} {
			\matchingjob(\sum,0)(\x,1);
			\node[vertex] (a\index) at (\sum + 0.5 * \x ,1.5) {};
			\pgfmathsetmacro\sum{\sum + \x};
			\pgfmathsetmacro\index{int(\index + 1)};
		}

        \foreach \x/\y in {1/2,2/2,2/3,3/3,3/4,3/5,4/5,4/6} {
              \draw[asg,black] (a\y) -- (o\x);
        }

        \node at (2.5,6) {$H_s$};
        \draw[asg,red] (a2) -- (o1);

        \draw [decorate,decoration={brace,mirror,amplitude=10pt,raise=2pt}] (0,0) -- (5,0) node[midway,below=12pt] {$\beta$};

    \end{scope}

\end{tikzpicture}
    \caption{Split operation of $H = (V,V^*,E,w)$ for a volume threshold of $\beta$.
    Both graphs $H_p$ and $H_s$ contain the (red) edge between the two vertices (blue and yellow) whose volume contribute to both graphs. The widths of the rectangles above/below vertices correspond to their weights.}
     \label{fig:split}
\end{figure}

 Next, we define the \emph{prefix expansion} of a bipartite graph. It is similar to the definition of prefix expansion of an assignment (\Cref{def:prefix expansion}).

 \begin{defn}[Prefix Expansion]
Given a bipartite graph $H = (V,V^*, E,w)$, the \emph{prefix expansion} of $H$ is \[\phi(H) := \max_{P^* \text{ is a prefix of } V^*(H) } \frac{|N_{H}(P^*)|}{|P^*|}.\]
 \end{defn}

The following lemma define a \emph{split}
operation illustrated in \Cref{fig:split}.
This operation ``splits" $H$ into  prefix and suffix subgraphs of $H$ where the suffix subgraph has volume $\beta \in [0, \vol(H)]$.

\begin{lemma} [Splitting] \label{lem:spliting}
Given a  forward bipartite graph $H$ and a real $\beta \in [0,\vol(H)]$, there exists a procedure called $\textsc{Split}(H,\beta)$ that returns a pair of bipartite graphs  $(H_p, H_s)$ such that
    \begin{enumerate} [nosep]
        \item $E(H )= E(H_p) \cup E(H_s)$,
        \item $H_p$ is a prefix subgraph of $H$ with ordering induced by $H$,
        \item $H_s$ is a suffix subgraph of $H$ with ordering induced by $H$,
        \item  $H_p$ and $H_s$ are  forward, and
        \item $\vol(H_s) = \beta.$
    \end{enumerate}
In this case, we say that $(H_p,H_s)$ is a \textit{split} of $H$. Furthermore, $\phi(H_p) \leq \phi(H).$
\end{lemma}
\begin{proof}

    Let $E_s \subseteq E$ be the smallest suffix of edges (using the relation $\prec$) in $E$ such that $w(E_s) \geq \beta$. Similarly, let $E_p \subseteq E$ be the smallest prefix of edges in $E$ such that $w(E_p) \geq \vol(H) - \beta$. Observe that $E = E_p \cup E_s$ and $|E_p \cap E_s| \leq 1$. Let
    $H_p$ be the subgraph consisting of the edges in $E_p$, and $H_s$ the subgraph consisting of the edges in $E_s$.
    The weights of edges in $H_p$ and $H_s$ are the same as in $H$ except for the two copies of the edge in $E_p \cap E_s$: We adjust the weights of these copies so that we get $\vol(H_p) = \vol(H) - \beta$ and $\vol(H_s) = \beta$. We return $(H_p,H_s)$. It is easy to check that $H_p$ is a prefix subgraph of $H$ and $H_s$ is a suffix subgraph of $H$ and that both graphs are  forward.

    We prove that $\phi(H_p) \leq \phi(H).$  Suppose by contradiction that $\phi(H_p) > \phi(H)$. Since $H_p$ is a prefix subgraph of $H$, every prefix $P$ of $H_p$ is also a prefix  of $H$ and $N_{H_p}(P) \subseteq N_{H}(P)$. Since $\phi(H_p) > \phi(H)$, there is a prefix $P'$ of $H_p$ such that $|N_{H_p}(P')| > \phi(H) \cdot |P'|.$ Therefore, $H$ has a prefix $P'$ such that $|N_H(P')| \ge |N_{H_p}(P')| > \phi(H) \cdot |P'|,$ a contradiction.
\end{proof}

As a simple application of the split operation, we now prove \Cref{lem:carve min assignment}.
\begin{proof} [Proof of \Cref{lem:carve min assignment}]
Let $\sigma$ be the canonical assignment between $\ALG$ and $\OPT$ at time $s$. Let $H = (V,V^*,E,\sigma)$ be the corresponding bipartite graph where $V = \ALG(s), V^* = \OPT(s)$.  Let $\beta = s' - s \geq 0$. By \Cref{lem:spliting}, $\textsc{Split}(H,\beta)$ returns a pair of bipartite graphs  $(H_p, H_s)$ that is a split of $H$.  By \Cref{lem:spliting},  the assignment obtained from $H_p$ is a valid assignment between $\ALG$ and $\OPT$ at time~$s'$.
\end{proof}

Finally, we define a merge operation, which
unites the vertex sets of two disjoint graphs and then computes a canonical assignment.

\begin{lemma} [Merging] \label{lem:merge}
Let $H_1 = (V_1,V^*_1,E_1,w_1)$ and $H_2 = (V_2,V^*_2,E_2,w_2)$ be disjoint  bipartite graphs, that is, $V_1 \cap V_2 = \emptyset$ and $V^*_1 \cap V^*_2 = \emptyset$.
Then, there exists a procedure called \textsc{Merge}($H_1,H_2$) that returns bipartite graph $H$ such that
\begin{enumerate} [nosep]
\item The volume of each node is the sum of its volumes in two graphs. More precisely,
\begin{enumerate} [nosep]
\item $V = V_1 \cup V_2$ and $V^* = V^*_1 \cup V^*_2$,
\item for all $i \in V$, $\vol_H(i)  = \vol_{H_1}(i)$
if $i\in V_1$ and
$\vol_H(i)  = \vol_{H_2}(i)$
if $i\in V_2$, and
\item for all $i \in V^*$, $\vol_H(i)  = \vol_{H_1}(i)$
if $i\in V^*_1$ and
$\vol_H(i)  = \vol_{H_2}(i)$
if $i\in V^*_2$.
\end{enumerate}

\item $H$ is forward.
\end{enumerate}
\end{lemma}
\begin{proof}
    Given $H_1$ and $H_2$, define the following:
    \begin{itemize} [nosep]
        \item $A  := V_1 \cup V_2$ and $A^* :=  V_1^* \cup V_2^*$,
        \item a function $c: A \rightarrow \mathbb{R}_{\geq 0}$
          where
          $c(a) := \vol_{H_1}(a)$ if $a\in V_1$ and
        $c(a) := \vol_{H_2}(a)$ if $a\in V_2$,
        \item a function $c^*: A^* \rightarrow \mathbb{R}_{\geq 0}$
          where
          $c^*(a) := \vol^*_{H_1}(a)$ if $a\in V^*_1$ and
        $c^*(a) := \vol^*_{H_2}(a)$ if $a\in V^*_2$,
        \item $A^*$ is ordered by non-increasing values of $c^*$, and $A$ is ordered by \emph{non-decreasing values} of $c$.
    \end{itemize}
    Let $H \gets \textsc{GreedyMatching}(A,A^*,c,c^*)$. By
    \Cref{lem:greedy}, the volume of each node
    in $H$ is
    as required.
 Moreover, since $A^*$ is ordered by non-increasing values of $c^*$ and $A$ is
 ordered by   \emph{non-decreasing} values of $c$, $\textsc{GreedyMatching}$
 in fact produces a forward graph (with respect to a non-increasing order of~$A$).
\end{proof}

Having defined useful operations and properties of weighted bipartite
graphs representing the fractional matching between $\ALG$ and $\OPT$, we are ready to dive into the proof of \Cref{lem:nontrivial update assignment}.

\subsection{Proof of \Cref{lem:nontrivial update assignment}}
\label{sec:update-proof}

 Let $J' := J \cup J_{new}$. Our proof strategy starts with the
 canonical assignment $\sigma$ between $\ALG_J(s)$ and $\OPT_J(s)$ right before a batch of jobs $J_{new}$ arrives.
 We aim to transform the assignment $\sigma$ to get a valid assignment at time $\ell$.
The transformations are based on the operations defined in
\Cref{sec:operations}.

\begin{leftbar}
  Throughout the analysis, we denote by $\gamma$ the elapsed time of
  the leader $L(s)$ at time $\ell$.
\end{leftbar}

\subsubsection{The Structure of the Instances}

Our goal is to transform a valid assignment for time $s$ into a valid assignment for time $\ell$. As part of this process, we formalize how the remaining instance changes during $(s,\ell]$ for $\ALG$ and $\OPT$, respectively. Since no new jobs are released during that interval by assumption, all changes are caused only by the work executed during $(s,\ell]$. Similarly to the example given in~\Cref{sec:toy}, formalizing the work done by $\ALG$ and $\OPT$ during $(s,\ell]$ will help us to understand the demands at time $\ell$ that must be fixed to create an assignment and to give volume bounds that imply the bounded prefix expansion. In contrast to the example in~\Cref{sec:toy}, we need to keep track of not only the work on jobs in $J_{new}$, but also the work on ``old jobs" that have been released before time $s$. To this end, we first characterize the changes in  $\ALG$ and afterwards track the difference of the work on jobs in $J_{new}$ between $\ALG$ and $\OPT$ by using the functions $\Delta$, $\tau$ and $\tau^*$ as in the illustrative example.

\paragraph{Structure of $\ALG$.}
Let us define
\begin{itemize}[nosep]
    \item $A_{\ell} := \ALG_{J'}(\ell) \cap J_{new}$, and
    \item $O_{\ell} := \OPT_{J'}(\ell) \cap J_{new}$.
\end{itemize}
We start by stating properties of jobs in $\ALG$ at time $\ell$.

\begin{lemma}[$\ALG$ at time $\ell$] \label{lem:Ks Kl Al in alg} The following properties hold at time $\ell$:
\begin{enumerate} [noitemsep,nolistsep]
    \item For every job $j \in K(s)$, $j \in \ALG_{J'}(\ell)$ if and only if $r_{j}(s) \geq  \threshold\cdot \gamma$.
    \item If $ j\in K(\ell)$, then $r_{j}(s) = r_{j}(\ell).$
    Thus, $K(\ell) \subseteq K(s)$ and $\ALG$ does not touch $j$ during $(s,\ell]$.
    \item For every job $j \in A_{\ell}$, $e_{j}(\ell) = \gamma$.
    \item For every job $j \in J_{new} \setminus A_{\ell}$, $e_{j}(\ell) \leq \ainv\cdot  \gamma$.
\end{enumerate}
\end{lemma}

\begin{proof}
Recall that $L(s)$ is the leader among $J_{new}$. We prove the first two items. Fix a job $j \in K(s)$. By the description of $\ALG$, it switches from processing $L(s)$ to processing $j$ at some point if there is a time $t' \in [s,\ell]$ such that $\threshold \cdot e_{L(s)}(t')  \geq r_j(s)$. Once it does, it will finish job $j$ before coming back to $L(s)$. On the other hand, if there is no such time $t'$, then $\ALG$ never works on $j$ during $[s,\ell]$. Therefore, the first two items follow.

We next prove the last two items.  By the description of $\ALG$ and the fact that every job in $J_{new}$ arrives at time $s$, the set of jobs in $J_{new}$ is processed in round-robin manner starting at time $s$. Once one of the jobs in $J_{new}$ becomes known, $\ALG$  completes this job before returning to round-robin on the remaining unknown jobs. Since $\ALG$ touches the leader at time $\ell$, any job in $J_{new}$ that becomes known must be completed by time $\ell$.
Furthermore, each such job  $j$ becomes known before \ALG processes it for $\gamma$ time, and  therefore $e_j(\ell)=p_j \le \ainv\gamma$ and item (4) follows.
On the other hand, any job in $J_{new}$ that remains unknown at time $\ell$ must have the same elapsed time as the leader and therefore item (3) follows.
\end{proof}

On the other hand, $\OPT$'s state at time $\ell$ is simple to infer from its state at $s$: $\OPT$  spends the entire time from $s$ to $\ell$ working on jobs in its queue ordered by increasing remaining time.

\paragraph{Tracking the differences in $J_{new}$.} As in the special case of  \Cref{sec:toy}, we consider the amount of work done by $\ALG$ and $\OPT$ on the jobs in $J_{new}$. Recall that, by our assumption that there is no idle time in $(s,\ell]$, so the total work of $\ALG$ and $\OPT$, respectively, on all jobs during $(s,\ell]$ is exactly $\ell-s$.

\begin{defn}\label{def:delta tau tau*} [$\Delta,\tau,\tau^*$] We classify the elapsed times of jobs done by $\ALG$ and $\OPT$ during $(s,\ell]$ as follows.

\begin{enumerate}[nosep]
    \item For each job $i \in J_{new}$, let $\Delta(i) := \min\{e_i(\ell),e^*_i(\ell)\}$  be the common amount of work that both algorithms perform on job $i$. Let $\tau(i) := \max\{e_i(\ell) - e^*_i(\ell),0\}$ and $\tau^*(i) := \max\{e^*_i(\ell) - e_i(\ell),0\}$  be the amount of work done on $i$ by either $\ALG$ or $\OPT$ but not the other.
    \item Let $\Delta := \sum_{i \in J_{new}} \Delta(i)$.
    \item Let $\nu^* := \ell - s - \Delta - \tau^*(J_{new})$ and $\nu := \ell - s - \Delta - \tau(J_{new})$ be the time that $\OPT$ and $\ALG$ spend on the old jobs, respectively, during the interval $(s, \ell]$.
\end{enumerate}
\end{defn}

The following fact follows directly from \Cref{def:delta tau tau*}.
\begin{fact}  \label{fact:total work ALG OPT}
$\ell - s = \Delta + \tau(J_{new})  + \nu = \Delta + \tau^*(J_{new}) + \nu^*$.
\end{fact}

For the next fact, recall that the algorithm does not work on unknown old jobs during $(s,\ell]$ by assumption of~\Cref{lem:nontrivial update assignment} and that all jobs in $K(s) \setminus K(\ell)$ are completed during $(s,\ell]$ by~\Cref{lem:Ks Kl Al in alg}.

\begin{fact} \label{fact:volumes nu}
    $\nu = \sum_{j \in K(s) \setminus K(\ell)} r_j(s).$
\end{fact}

As in the illustrative example, we partition the jobs in $J_{new}$ depending on whether $\ALG$ or $\OPT$ works more on them during $(s,\ell]$.

\begin{defn} \label{def:O ell plus} We define $O^{+}_{\ell}, A^{+}_{\ell}$, and $D_{\ell}$ as follows.
\begin{itemize}[nosep]
    \item   $O^{+}_{\ell} := \{ j \in O_{\ell} \colon \tau(j) > 0 \}$, the set of new jobs in $\OPT(\ell)$ that $\ALG$ did more work on.
    \item   $A^{+}_{\ell} := \{ j \in A_{\ell} \colon \tau^*(j) > 0\}$,  the set of new jobs in $\ALG(\ell)$ that $\OPT$ did more work on.
    \item $D_{\ell} := \{ j \in J_{new} \colon \tau(j) = \tau^*(j) = 0 \}$.
\end{itemize}
\end{defn}

Observe that $J_{new} = O^{+}_{\ell} \cup A^{+}_{\ell} \cup D_{\ell}$.
The following lemma allows us to focus on the excess mass of $\ALG$ on $O_{\ell}^+$ and the excess mass of $\OPT$ on $A_{\ell}^+$.
\begin{lemma} \label{lem:sum tau i and sum tau * i}
\[\sum_{i \in O_{\ell}^+} \tau(i) = \sum_{i \in J_{new}}\tau (i) \mbox{ and } \sum_{i \in A_{\ell}^+} \tau^*(i) = \sum_{i \in J_{new}}\tau^*(i) \ . \]
\end{lemma}
\begin{proof}
It is enough to prove that $\forall i \in J_{new} \setminus O_{\ell}, \tau(i) = 0$ and $\forall i \in J_{new} \setminus A_{\ell}, \tau^*(i) = 0$. To see that $\forall i \in J_{new} \setminus O_{\ell}, \tau(i) = 0$, observe that each $i \in J_{new} \setminus O_{\ell}$ is a job whose elapsed time in $\OPT$ at time $\ell$ is equal to its processing time, and thus $\tau(i) = 0$. The argument that for $\forall i \in J_{new} \setminus A_{\ell}$ we have that  $\tau^*(i) = 0$ is similar.
\end{proof}

\paragraph{Structure of $\Delta, \tau,\tau^*$.} The following \Cref{lem:characterize tau delta} specifies the values of $\Delta(i)$, $\tau^*(i)$, and $\tau^*(i)$ for jobs in $A_\ell \cup O_\ell$.  Let $z$ be the last job that $\OPT$ touched at time $\ell$. Recall that $\gamma := e_{L(s)}(\ell)$ is the elapsed processing time of the leader $L(s)$ of $J_{new}$ at  time $\ell$.

To put \Cref{lem:characterize tau delta} into perspective of~\Cref{def:O ell plus}, note that $O_\ell \setminus \{z\} = O_\ell^+ \setminus \{z\}$ due to the fact that $\OPT$ (SRPT) finishes all jobs it touches during $(s,\ell]$ except for $z$. Hence, the jobs in $O_\ell \setminus \{z\}$ remain untouched by $\OPT$ but are touched by $\ALG$. The first two statements of the lemma concern the jobs in $O_\ell^+\setminus \{z\}$. Similarly, $A_\ell \setminus O_\ell = A_\ell^+ \setminus \{z\}$ as $\OPT$ clearly works more on these jobs than $\ALG$ and does not work at all on jobs in $O_\ell\setminus\{z\} $. Thus, the third property of the lemma holds for jobs in $A_\ell^+ \setminus \{z\}$. Finally, the last property of the lemma concerns the jobs in $D_\ell \setminus \{z\}$. Note that the job $z$ can potentially belong to any of the sets $A_\ell^+,O_\ell^+$ and $D_\ell$.

\begin{lemma} \label{lem:characterize tau delta}
Let $j \in J_{new} \setminus \{z\}$. Then,
\begin{itemize} [noitemsep, nolistsep]
\item If $j \in O_{\ell} \cap A_\ell$, then $\tau(j) = \gamma$ and $\tau^*(j) = \Delta(j) = 0$.
\item If $j \in O_{\ell} \setminus A_{\ell}$, then $\tau(j) \leq \ainv \cdot \gamma$ and $\tau^*(j) = \Delta(j) = 0$.
\item If $j \in A_{\ell} \setminus O_{\ell}$, then $\Delta(j) = \gamma$, $\tau(j) = 0$  and $\tau^*(j) \geq \threshold \cdot \gamma$.
\item Else ($j \not\in A_\ell \cup O_\ell$), $\Delta(j) = p_j$ (so $\tau(j) = \tau^*(j) = 0$).
\end{itemize}
\end{lemma}

\begin{proof}
We use the fact that $\OPT$ is SRPT. Since no jobs are released during $(s,\ell]$ by assumption, the fact that $\OPT$ runs SRPT implies that jobs  $j\in O_\ell \setminus \{z\}$ are not touched during $[s,\ell]$. In particular,  $\tau^*(j)=\Delta(j)=0$.

By \Cref{lem:Ks Kl Al in alg}, if $j \in A_\ell$, then $\ALG$ worked on $j$ for $\gamma$ time and $j$ is still unknown at $\ell$ in $\ALG$.

Let $j \in J_{new} \setminus \{z\}$:

\begin{itemize} [nosep]
    \item If $j \in O_{\ell} \cap A_{\ell}$, then $\tau^*(j)=\Delta(j)=0$
    and $\tau(j) = \gamma$ as
     argued above.
    \item If $j \in O_{\ell} \setminus A_{\ell}$, then $j$ was finished by $\ALG$, but  $\OPT$ did not touch it. Since the leader $L(s)$ has elapsed time $\gamma$ at time $\ell$, $j$ must have become known when its elapsed time was $q \in (0,\gamma)$, and thus its processing time is $p_j = \ainv q \leq \ainv  \gamma.$ Therefore, $\tau(j) = p_j \leq \ainv  \gamma$.
    \item  If $j \in A_{\ell} \setminus O_{\ell}$, then $j$ was finished by $\OPT$, but unfinished by $\ALG$, and thus $\Delta(i) = \gamma$ and $\tau^*(j) = p_j - \gamma$, which is also the remaining time of $J$ in $\ALG(\ell)$. Since the elapsed time of $j$ at time $\ell$ is $\gamma$ and $j$ is still unknown at $\ell$ in $\ALG$, by the definition of $\ALG$ the  remaining time of $j$ in $\ALG$ is at least $\threshold\gamma$. It follows that $\tau^*(j) = p_j - \gamma \geq \threshold\gamma$.
    \item  Finally, if $j$ was finished by both $\ALG$ and $\OPT$, then $\Delta(j) = p_j$. \qedhere
\end{itemize}
\end{proof}

We are ready to state the algorithm transforming the canonical assignment at time $s$ to an assignment at time $\ell$.

\subsubsection{Algorithm Description}

\begin{figure}[t]
    \centering
    \begin{tikzpicture}[scale=0.7]

\foreach \x in {0,...,7} {
    \node[vertex] (o\x) at (\x,3) {};

    \draw[asg] (o\x) -- (\x,2);
    \node at (\x+0.25,2.2) {\tiny $\ldots$};
    \draw[asg] (o\x) -- (\x + 0.7,2);

    \node[vertex] (a\x) at (\x+0.7,0) {};

    \draw[asg] (a\x) -- (\x+0.7,1);
    \node at (\x+0.4,0.8) {\tiny $\ldots$};
    \draw[asg] (a\x) -- (\x,1);
}

\node at (-1.5,3) {$\OPT(s)$};
\node at (-1.5,0) {$\ALG(s)$};
\node at (-4.5,1.5) {$(a)$};
\node at (-4.5,1.5) {\phantom{$(b)$}};

\node at (3.5,4) {$H^{(1)}$};

\draw (-0.25,2) rectangle (8,1);
\node at (3.5,1.5) {$\sigma$};

\begin{scope}[shift={(11,0)}]

\foreach \x in {0,...,7} {
    \node[vertex] (o\x) at (\x,3) {};
    \node[vertex] (a\x) at (\x,0) {};
    \draw[asg,job3] (a\x) -- (o\x);
}

\node at (3.5,4) {$M^{(1)}$};

\node at (-1,3) {$J_{new}$};
\node at (-1,0) {$J_{new}$};

\end{scope}

\end{tikzpicture}
    \begin{tikzpicture}[scale=0.7]

\foreach \x in {0,...,7} {
    \node[vertex] (o\x) at (\x,3) {};

    \draw[asg] (o\x) -- (\x,2);
    \node at (\x+0.25,2.2) {\tiny $\ldots$};
    \draw[asg] (o\x) -- (\x + 0.7,2);

    \node[vertex] (a\x) at (\x+0.7,0) {};

    \draw[asg] (a\x) -- (\x+0.7,1);
    \node at (\x+0.4,0.8) {\tiny $\ldots$};
    \draw[asg] (a\x) -- (\x,1);
}

\coordinate (oleft) at (0,3);
\coordinate (aleft) at (0,0);
\coordinate (oright) at (7+0.7,3);
\coordinate (aright) at (7+0.7,0);

\draw[line width=1.5pt, dashed, black] \convexpath{oleft,o3,a4,aleft}{10pt};
\draw[line width=1.5pt, dashed, red!90!black] \convexpath{o4,oright,aright,a5}{10pt};

\node at (-1.5,3) {$\OPT(s)$};
\node at (-1.5,0) {$\ALG(s)$};

\node at (-4.5,1.5) {$(b)$};
\node at (-4.5,1.5) {\phantom{$(a)$}};

\node at (2,4) {$H_p^{(1)}$};
\node[red!90!black] at (6,4) {$H_s^{(1)}$};

\draw (-0.25,2) rectangle (8,1);
\node at (3.5,1.5) {$\sigma$};

\begin{scope}[shift={(11,0)}]

\foreach \x in {0,...,7} {
    \node[vertex] (o\x) at (\x,3) {};
    \node[vertex] (a\x) at (\x,0) {};
    \draw[asg,job3] (a\x) -- (o\x);
}

\node at (3.5,4) {$M^{(1)}$};

\node at (-1,3) {$J_{new}$};
\node at (-1,0) {$J_{new}$};

\end{scope}

\end{tikzpicture}
    \caption{Illustration of $H^{(1)},H_p^{(1)},H_s^{(1)}$ and $M^{(1)}$ as defined in~\Cref{alg:update assignment}.}
    \label{fig:update-ALG-1}
\end{figure}
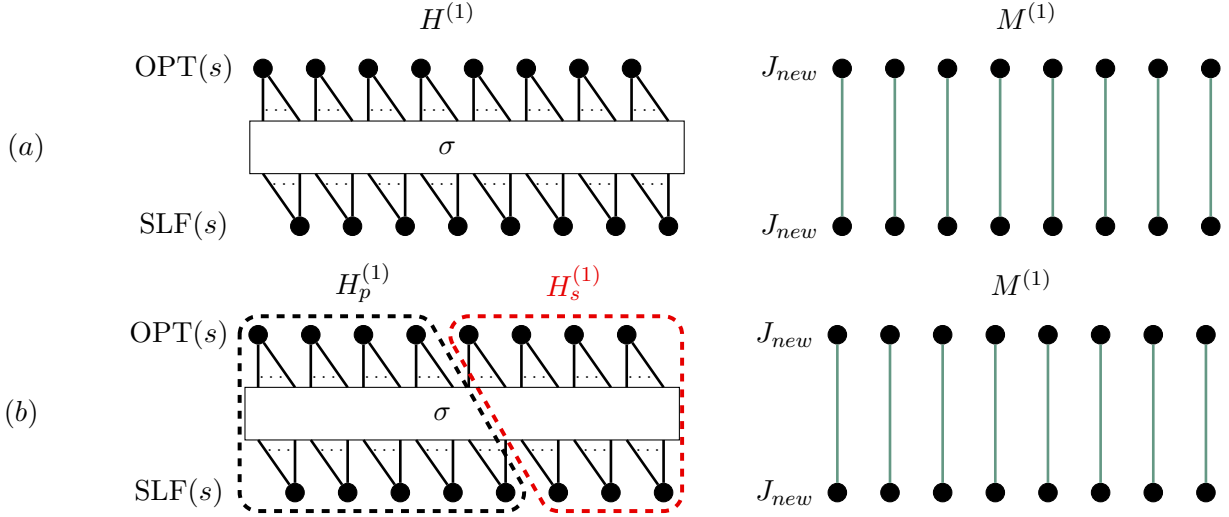

The algorithm is given in  \Cref{alg:update assignment}. We first take $\sigma$ the canonical assignment for job instance $J$ at time $s$ right before all jobs in $J_{new}$ arrive. Let $H^{(1)}$ be the bipartite graph representation of $\sigma$. Since $\sigma$ is the canonical assignment, $H^{(1)}$ is  forward (\Cref{def:forward backward}).
Let $M^{(1)}$ be a perfect matching between the  jobs of $J_{new}$  in $\ALG$ and $\OPT$.
\Cref{fig:update-ALG-1}(a) illustrates this starting state for our algorithm.

At time $s$, the graph $H^{(1)} \cup M^{(1)}$ corresponds to a valid assignment after $J_{new}$ arrives. We  update $H^{(1)}$ and $M^{(1)}$ so that the marginals of every job $j$ in $\ALG$ and in $\OPT$ are equal to the remaining time of $j$ at time $\ell$.  The transformation is based on updating the graph using $\Delta, \tau^*,\tau$ as in the toy example of \Cref{sec:toy}, but now we deal with the general case.  For every job $j \in J_{new}$, we remove $\Delta(j)$ weight from the edge $(j,j^*)\in M^{(1)}$. After this step
we can focus on the symmetric difference of the work of $\ALG$ and $\OPT$ on  $J_{new}$, i.e., on $\tau^*$ and $\tau$.  In the next step, we obtain $(H_p^{(1)},H_s^{(1)})$ from the $\textsc{Split}(H^{(1)}, \min\{\nu,\nu^*\})$ operation (recall that $H^{(1)}$ is   forward). Then, we set $H^{(2)} \gets H_p^{(1)}$.

\begin{algorithm}
  \caption{Update-Valid-Assignment$(J,J_{new},[s,\ell])$
}

    \medskip

  Let $\sigma$ be the canonical assignment for $J$ at the
  time $s$ right before jobs $J_{new}$ arrive.
  \begin{enumerate}
    \item Define the following:
    \begin{enumerate} [nosep]
        \item Let $H^{(1)}$ be the  representation of $\sigma$ as a bipartite graph.
        \item \label{line:1b} Let $M^{(1)}$ be the perfect describing $J_{new}$ at time $s$.
            \begin{itemize} [nosep]
                \item That is, $M^{(1)} := (J_{new},J_{new},E,w_M)$ where $w_M(i,i^*) = p_i$.
            \end{itemize}
    \end{enumerate}
    \item \label{line:2} Perform the following:
    \begin{enumerate} [nosep]
        \item Let $M^{(2)} \gets M^{(1)}$
        \item \label{line:2a} $\forall i \in J_{new}$ do $w_{M^{(2)}}(i,i^*) \gets w_{M^{(2)}}(i,i^*) - \Delta(i)$
        \item \label{line:2b} $(H^{(1)}_p,H^{(1)}_s) \gets \textsc{Split}(H^{(1)},\min\{\nu,\nu^*\})$
        \item \label{line:2c} $H^{(2)} \gets H^{(1)}_p$
    \end{enumerate}

    \item  \label{line:3} \textbf{If} $\nu \leq \nu^*$, \textbf{Return} $\textsc{Update1}(H^{(2)},M^{(2)})$.
     \textbf{Else,} \textbf{Return} $\textsc{Update2}(H^{(2)},M^{(2)})$.
\end{enumerate}
\label{alg:update assignment}
\end{algorithm}

We explain the intuition behind splitting $\min\{\nu,\nu^*\}$ units.
 Recall that $\nu$ and $\nu^*$ are the amounts of work done on jobs not in $J_{new}$ by $\ALG$ and $\OPT$, respectively. By~\Cref{lem:nontrivial update assignment}, the only old jobs that \ALG works on during $(s,\ell]$ are known and, thus, processed in SRPT manner.
This means that $\min\{\nu,\nu^*\}$ represents ``the common amount of volume" that both algorithms spend on running old jobs in SRPT manner during $(s,\ell]$.
In terms of the bipartite graph, since $H^{(1)}$ is   forward, we can remove the suffix of volume $\min\{\nu,\nu^*\}$ from both $V(H^{(1)})$ and $V^*(H^{(1)})$ using the $\textsc{Split}(H^{(1)}, \min\{\nu,\nu^*\})$ operation to obtain $(H_p^{(1)},H_s^{(1)})$ (cf.~\Cref{fig:update-ALG-1}(b) for an illustration). Since all jobs in $V^*(H_s^{(1)})$ (resp.~$V(H_s^{(1)})$) with possibly a single exception are completed by $\OPT$ (resp.~$\ALG$) at time $\ell$ by the definition of SRPT, and the exception also appears in $H_p^{(1)}$ by definition of the split operation, the graph $H_s^{(1)}$ is not relevant anymore for creating an assignment for time $\ell$ and we can just remove it.
After this, either $\ALG$ or $\OPT$ may still have some volume left to run SRPT on the old jobs that appear in $H_p^{(1)}$. We distinguish the two cases $\nu \leq \nu^*$ and $\nu^* < \nu$, and depending on this, call
either \textsc{Update1} or \textsc{Update2}, which are given in \Cref{fig:update1} and \Cref{fig:update2}.

We proceed as follows. In the next section (\Cref{sec:update-alg-correctness}), we show that the algorithm computes an assignment $\sigma'$ between $\ALG$ and $\OPT$ at time $\ell$. Afterwards, we prove that $\sigma'$ is valid by separately analyzing \Cref{fig:update1} and \Cref{fig:update2} in \Cref{sec:nulenu*,sec:nu*lenu}, respectively, concluding the proof of \Cref{lem:nontrivial update assignment}.

\begin{algorithm}
        \caption{$\textsc{Update1}(H^{(2)},M^{(2)})$}
        \label{fig:update1}

    \medskip
    \begin{enumerate}
    \item Define the following: \label{line:3a}
    \begin{enumerate}  [noitemsep,nolistsep]
    \item Split $M^{(2)} := M^{(2)}_p \cup M^{(2)}_s \cup  M^{(2)}_d$ where
        \begin{itemize} [nosep]
            \item $M^{(2)}_p$ is the perfect matching in $M^{(2)}$ corresponding to the jobs in $O_{\ell}^+$,
            \item $M^{(2)}_s$ is the perfect matching in $M^{(2)}$ corresponding to the jobs in $A_{\ell}^+$, and
            \item $M^{(2)}_d$ is the perfect matching in $M^{(2)}$ corresponding to the jobs in $D_{\ell}$.
        \end{itemize}
    \item $H^{(3)} \gets \textsc{Merge}(H^{(2)},M_s^{(2)})$. \label{line:merge}
    \item \label{line:X case 1}Let $X$ be the smallest suffix in $V(H^{(3)})$ such that $\vol_{H^{(3)}}(X) \geq T,$ where $T =  \tau(O_{\ell}^+)$.
    \end{enumerate}
     \begin{claim} \label{claim:X exists}
    The set $X$ exists.
   \end{claim}
    \item Perform the following:   \label{line:3b}
    \begin{enumerate} [noitemsep,nolistsep]
        \item \label{line:3b0} $M^{(3)} \gets M^{(2)}_p$
        \item   \label{line:3bi}$\forall j \in O_{\ell}^+$ do $w_{M^{(3)}}(j,j^*) \gets  w_{M^{(3)}}(j,j^*) - \tau(j)$
        \item \label{line:3bii} $(H^{(3)}_p, H^{(3)}_s) \gets \textsc{Split}(H^{(3)},T)$
        \item \label{line:3biii} $G \gets \greedy(X:=V(H_s^{(3)}),O_{\ell}^+, \vol_{ H^{(3)}_s}, \tau)$
    \end{enumerate}
    \item \textbf{Return} $\sigma' =$ the weight function of $G \cup  M^{(3)} \cup H^{(3)}_p \cup M_d^{(2)}$ after removing isolating vertices.
    \end{enumerate}
\end{algorithm}

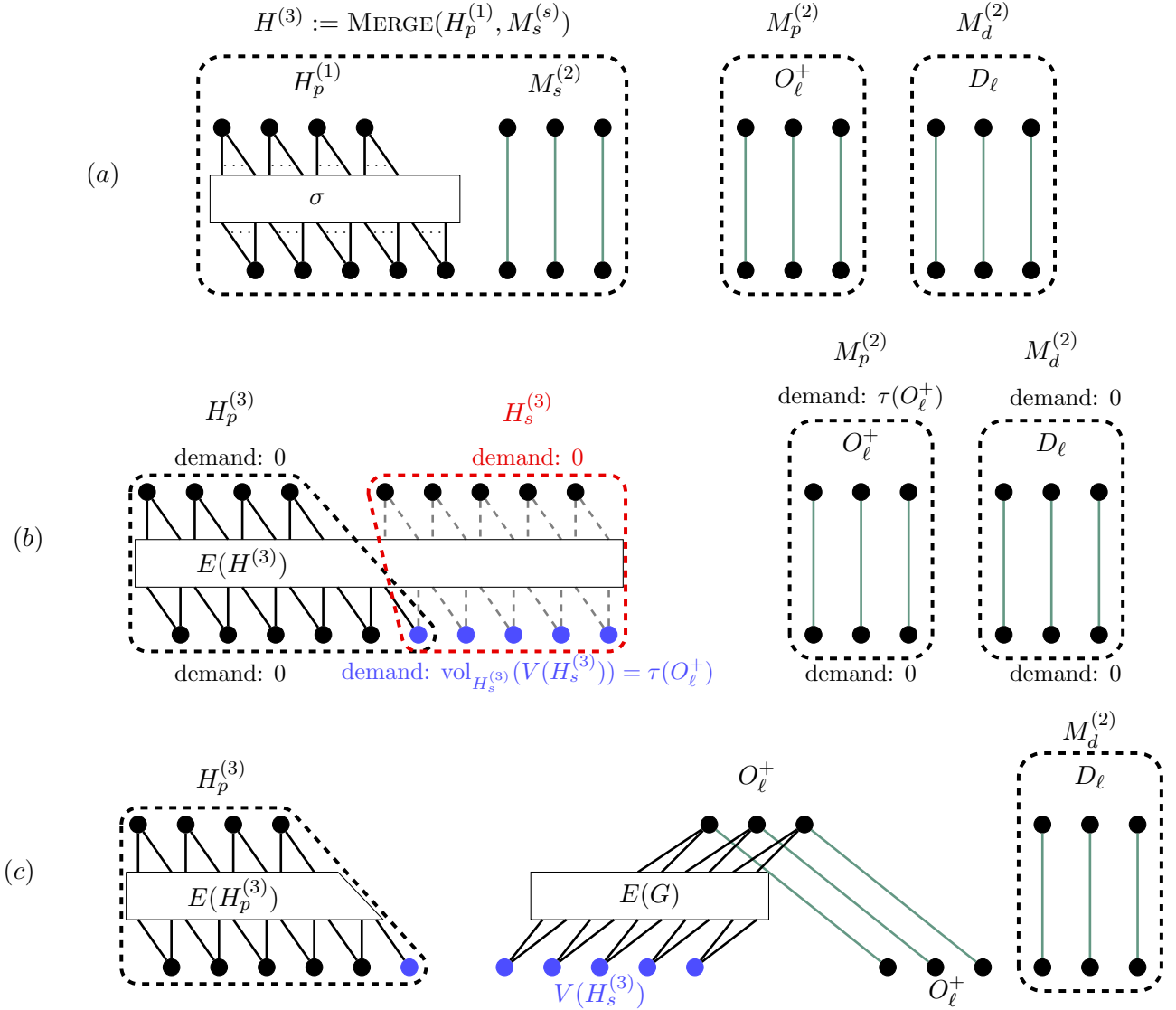
\begin{figure}[th]
    \centering
    \begin{tikzpicture}[scale=0.7]

\foreach \x in {0,...,3} {
    \node[vertex] (o\x) at (\x,3) {};

    \draw[asg] (o\x) -- (\x,2);
    \node at (\x+0.25,2.2) {\tiny $\ldots$};
    \draw[asg] (o\x) -- (\x + 0.7,2);
}
\foreach \x in {0,...,4} {
    \node[vertex] (a\x) at (\x+0.7,0) {};

    \draw[asg] (a\x) -- (\x+0.7,1);
    \node at (\x+0.4,0.8) {\tiny $\ldots$};
    \draw[asg] (a\x) -- (\x,1);
}

\node at (2,4) {$H_p^{(1)}$};
\draw (-0.25,2) rectangle (5,1);
\node at (2,1.5) {$\sigma$};

\begin{scope}[shift={(6,0)}]

\foreach \x in {0,...,2} {
    \node[vertex] (o\x) at (\x,3) {};
    \node[vertex] (a\x) at (\x,0) {};
    \draw[asg,job3] (a\x) -- (o\x);
}

\node at (1,4) {$M_s^{(2)}$};
\end{scope}

\draw[rounded corners=10pt,line width=1.5pt,dashed] (-0.5,-0.5) rectangle ++ (9,5);
\node at (4,5.25) {$H^{(3)} := \textsc{Merge}(H_p^{(1)},M_s^{(s)})$};

\begin{scope}[shift={(11,0)}]

\foreach \x in {0,...,2} {
    \node[vertex] (o\x) at (\x,3) {};
    \node[vertex] (a\x) at (\x,0) {};
    \draw[asg,job3] (a\x) -- (o\x);
}

\node at (1,4) {$O_\ell^+$};
\draw[rounded corners=10pt,line width=1.5pt,dashed] (-0.5,-0.5) rectangle ++ (3,5);
\node at (1,5.25) {$M_p^{(2)}$};
\end{scope}

\begin{scope}[shift={(15,0)}]

\foreach \x in {0,...,2} {
    \node[vertex] (o\x) at (\x,3) {};
    \node[vertex] (a\x) at (\x,0) {};
    \draw[asg,job3] (a\x) -- (o\x);
}

\node at (1,4) {$D_\ell$};
\draw[rounded corners=10pt,line width=1.5pt,dashed] (-0.5,-0.5) rectangle ++ (3,5);
\node at (1,5.25) {$M_d^{(2)}$};
\end{scope}

\node at (-2.5,2) {$(a)$};

\end{tikzpicture}

    \vspace*{10pt}
    \begin{tikzpicture}[scale=0.7]

\foreach \x in {0,...,3} {
    \node[vertex] (o\x) at (\x,3) {};

    \draw[asg] (o\x) -- (\x,2);
    \draw[asg] (o\x) -- (\x + 0.7,2);
}
\foreach \x in {0,...,4} {
    \node[vertex] (a\x) at (\x+0.7,0) {};

    \draw[asg] (a\x) -- (\x+0.7,1);
    \draw[asg] (a\x) -- (\x,1);
}

\node at (1.75,4.75) {$H_p^{(3)}$};
\node at (1.75,3.75) {\small demand: 0};
\node at (1.75,-0.8) {\small demand: 0};

\foreach \x in {5,...,9} {
    \node[vertex] (o\x) at (\x,3) {};

    \draw[asg,dashed, gray] (o\x) -- (\x,2);
    \draw[asg,dashed, gray] (o\x) -- (\x + 0.7,2);
}

\foreach \x in {6} {
    \node[vertex,blue!70!white] (a\x) at (\x-0.3,0) {};

    \draw[asg,dashed, gray] (a\x) -- (\x-0.3,1);
    \draw[asg] (a\x) -- (\x-1,1);
}

\foreach \x in {7,...,10} {
    \node[vertex,blue!70!white] (a\x) at (\x-0.3,0) {};

    \draw[asg,dashed, gray] (a\x) -- (\x-0.3,1);
    \draw[asg,dashed, gray] (a\x) -- (\x-1,1);
}

 \node[blue!70!white] at (8,-0.8) {\small demand: $\vol_{H_s^{(3)}}(V(H_s^{(3)})) = \tau(O_\ell^+)$};

\node[red!90!black] at (8,4.75) {$H_s^{(3)}$};
\node[red!90!black] at (8,3.75) {\small demand: 0};

\draw (-0.25,2) rectangle (10,1);
\node at (2,1.5) {$E(H^{(3)})$};

\coordinate (oleft) at (0,3);
\coordinate (aleft) at (0,0);
\coordinate (oright) at (9+0.7,3);
\coordinate (aright) at (9+0.7,0);

\draw[line width=1.5pt, dashed, black] \convexpath{oleft,o3,a6,aleft}{10pt};
\draw[line width=1.5pt, dashed, red!90!black] \convexpath{o5,oright,aright,a6}{10pt};

\begin{scope}[shift={(14,0)}]

\foreach \x in {0,...,2} {
    \node[vertex] (o\x) at (\x,3) {};
    \node[vertex] (a\x) at (\x,0) {};
    \draw[asg,job3] (a\x) -- (o\x);
}

\node at (1,4) {$O_\ell^+$};
\draw[rounded corners=10pt,line width=1.5pt,dashed] (-0.5,-0.5) rectangle ++ (3,5);
\node at (1,6) {$M_p^{(2)}$};
\node at (1,5) {\small demand: $\tau(O_\ell^+)$};
\node at (1,-0.8) {\small demand: $0$};

\end{scope}

\begin{scope}[shift={(18,0)}]

\foreach \x in {0,...,2} {
    \node[vertex] (o\x) at (\x,3) {};
    \node[vertex] (a\x) at (\x,0) {};
    \draw[asg,job3] (a\x) -- (o\x);
}

\node at (1,4) {$D_\ell$};
\draw[rounded corners=10pt,line width=1.5pt,dashed] (-0.5,-0.5) rectangle ++ (3,5);
\node at (1,6) {$M_d^{(2)}$};
\node at (1.3,5) {\small demand: $0$};
\node at (1.3,-0.8) {\small demand: $0$};

\end{scope}

\node at (-2.5,2) {$(b)$};

\end{tikzpicture}

    \begin{tikzpicture}[scale=0.7]

\foreach \x in {0,...,3} {
    \node[vertex] (o\x) at (\x,3) {};

    \draw[asg] (o\x) -- (\x,2);
    \draw[asg] (o\x) -- (\x + 0.7,2);
}
\foreach \x in {0,...,4} {
    \node[vertex] (a\x) at (\x+0.7,0) {};

    \draw[asg] (a\x) -- (\x+0.7,1);
    \draw[asg] (a\x) -- (\x,1);
}

\node at (1.75,4) {$H_p^{(3)}$};

\foreach \x in {6} {
    \node[vertex,blue!70!white] (a\x) at (\x-0.3,0) {};
    \draw[asg] (a\x) -- (\x-1,1);
}

\path
(-0.25,2) edge (4.2,2)
(-0.25,2) edge (-0.25,1)
(-0.25,1) edge (5.15,1)
 (4.2,2) edge (5.15,1)
;

\node at (2,1.5) {$E(H_p^{(3)})$};

\coordinate (oleft) at (0,3);
\coordinate (aleft) at (0,0);
\coordinate (oright) at (9+0.7,3);
\coordinate (aright) at (9+0.7,0);

\draw[line width=1.5pt, dashed, black] \convexpath{oleft,o3,a6,aleft}{10pt};

\node at (-2.5,2) {$(c)$};

\foreach \x in {6,...,10} {
    \node[vertex,blue!70!white] (a\x) at (\x+2-0.3,0) {};
    \draw[asg] (a\x) -- (\x+3,1);
    \draw[asg] (a\x) -- (\x+2.5,1);
}

\draw (8+0.25,2) rectangle (10+3+0.25,1);
\node at (10.75,1.5) {$E(G)$};
\node[blue!70!white] at (10-0.3,-0.5) {$V(H_s^{(3)})$};

\begin{scope}[shift={(12,0)}]

\foreach \x in {0,...,2} {
    \node[vertex] (o\x) at (\x,3) {};
    \node[vertex] (a\x) at (\x+3.75,0) {};
    \draw[asg,job3] (a\x) -- (o\x);

    \draw[asg] (o\x) -- (\x-1.5,2);
    \draw[asg] (o\x) -- (\x-1,2);

}

\node at (1,4) {$O_\ell^+$};
\node at (5,-0.5) {$O_\ell^+$};

\end{scope}

\begin{scope}[shift={(19,0)}]

\foreach \x in {0,...,2} {
    \node[vertex] (o\x) at (\x,3) {};
    \node[vertex] (a\x) at (\x,0) {};
    \draw[asg,job3] (a\x) -- (o\x);
}

\node at (1,4) {$D_\ell$};
\draw[rounded corners=10pt,line width=1.5pt,dashed] (-0.5,-0.5) rectangle ++ (3,5);
\node at (1,5) {$M_d^{(2)}$};

\end{scope}

\end{tikzpicture}

    \vspace*{-20pt}
    \caption{Illustration of the different operations executed by~\Cref{fig:update1}. The dashed edges in part (b) of the figure indicate that the corresponding edges will be discarded and not be part of the constructed assignment.}
    \label{fig:update-ALG-4}
\end{figure}

\subsubsection{Correctness}\label{sec:update-alg-correctness}

In this section, we show that \Cref{alg:update assignment} returns an assignment $\sigma'$ between $\ALG$ and $\OPT$ on job instance $J'$ at time $\ell$.
Let $H' = (V,V^*,E,\sigma')$ be the bipartite graph for the weight function $\sigma'$ returned by \Cref{alg:update assignment}.
Let $H^{(1)}$ be the bipartite graph corresponding to the assignment $\sigma$ at time $s$ right before $J_{new}$ arrives and let $M^{(1)}$ denote the perfect matching defined by $J_{new}$ (cf. Step~\ref{line:1b} in \Cref{alg:update assignment}). We prove the following properties.

\begin{restatable}{lemma}{lemPropHPrime}
    \label{lem:prop H'} The graph $H'$ satisfies the following properties:
   \begin{enumerate} [noitemsep,nolistsep]
       \item \label{item:1 of h'} For all $i \in J_{new}$, $\vol_{M^{(1)}}(i) - \vol_{H'}(i) = \Delta(i) + \tau(i).$
       \item \label{item:2 of h'}For all $i \in K(s) \setminus K(\ell)$, $\vol_{H^{(1)}}(i) - \vol_{H'}(i) = r_i(s)$.
       \item \label{item:3 of h'}For all $i \in \ALG_{J}(s) \setminus (K(s) \setminus K(\ell)), \vol_{H'}(i) = \vol_{H^{(1)}}(i)$.
       \item \label{item:4 of h'}For all $i \in J_{new}$,  $\vol_{M^{(1)}}^*(i) - \vol_{H'}^*(i) = \Delta(i) + \tau^*(i).$
       \item \label{item:5 of h'}For all $i \in \OPT_{J}(s)$,  $\vol_{H^{(1)}}^*(i) - \vol_{H'}^*(i) = r_i^*(s) - r_i^*(\ell)$.
   \end{enumerate}
\end{restatable}
The proof of \Cref{lem:prop H'} is a straightforward verification of the algorithms given in \Cref{alg:update assignment}, \Cref{fig:update1}, and \Cref{fig:update2}, and we defer it to~\Cref{sec:omitted}. With these properties in hand, we are ready to prove that $\sigma'$ is an assignment.

\begin{lemma} \label{lem:sigma' assignment}
 At time $\ell$, $\sigma'$ is an assignment between $\ALG$ and $\OPT$ on job instance $J'$.
\end{lemma}

\begin{proof}
To show that $\sigma'$ is an assignment at time $\ell$, it suffices to show that $\vol_{H'}^*(j) = r_j^*(\ell)$ for all $j \in \OPT_{J}(s) \cup J_{new}$ and that  $\vol_{H'}(j) = r_j(\ell)$ for all $j \in \ALG_{J}(s) \cup J_{new}$.

First, consider an arbitrary $j\in \OPT_{J}(s) \cup J_{new}$. If $j \in J_{new}$, then~\Cref{lem:prop H'}(\ref{item:4 of h'}) implies  $\vol_{H'}^*(j) = \vol_{M^{(1)}}^*(j) - \Delta(j) - \tau^*(j) = p_j - \Delta(j) - \tau^*(j) = r_j^*(\ell)$, where the last two equalities use the definition of $M^{(1)}$ and the fact that $\OPT$ processes jobs $j$ for $\Delta(j) + \tau^*(j)$ units during $(s,\ell]$.
If $j \not\in J_{new}$, then $j \in \OPT_J(s)$ and~\Cref{lem:prop H'}(\ref{item:5 of h'}) implies  $\vol_{H'}^*(j) = \vol_{H^{(1)}}^*(j) -r_j^*(s) + r_j^*(\ell) =  r_j^*(s) -r_j^*(s) + r_j^*(\ell) = r_j^*(\ell)$, where we use that $\vol_{H^{(1)}}^*(j) = r_j^*(s)$  since $H^{(1)}$ is induced by the valid assignment $\sigma$ for time $s$.

Next, consider an arbitrary $j\in \ALG_{J}(s) \cup J_{new}$. If $j \in J_{new}$, then~\Cref{lem:prop H'}(\ref{item:1 of h'}) implies $\vol_{H'}(j) = \vol_{M^{(1)}}(j) - \Delta(j) - \tau(j) = p_j - \Delta(j) - \tau(j) = r_j(\ell)$, where the last two equalities use the definition of $M^{(1)}$ and the fact that $\ALG$ processes jobs $j$ for $\Delta(j) + \tau(j)$ units during $(s,\ell]$. If $j \in \ALG_{J}(s)$, then we distinguish between $j \in \ALG_{J'}(\ell)$ and $j \not\in \ALG_{J'}(\ell)$.

In case that $j \in \ALG_{J}(s) \cap \ALG_{J'}(\ell)$, we have $j\in \ALG_J(s)\setminus (K(s)\setminus K(\ell))$ since all jobs in $K(s)$ that are not in $K(\ell)$ must have been completed by time $\ell$ and thus are not part of $\ALG_{J'}(\ell)$. Then,~\Cref{lem:prop H'}(\ref{item:3 of h'}) implies $\vol_{H'}(j) = \vol_{H^{(1)}}(j) = r_j(s)$, where the last equality uses that $H^{(1)}$ is induced by the valid assignment $\sigma$ for time $s$. We claim that $r_j(s)=r_j(\ell)$.  Indeed, because of \Cref{lem:Ks Kl Al in alg}(2) and the assumption that $\ALG$ does not work on unknown old jobs during $(s,\ell]$, $\ALG$ does not work on any job in  $\ALG_J(s) \setminus (K(s) \setminus K(\ell))$ during $[s,\ell]$. Hence, $r_j(s)=r_j(\ell)$.

In case that $j \in  \ALG_{J}(s)\setminus \ALG_{J'}(\ell)$, we have $j \in K(s) \setminus K(\ell)$ by  \Cref{lem:Ks Kl Al in alg}(2) and by the assumption that $\ALG$ does not work on unknown old jobs during $(s,\ell]$. Then,~\Cref{lem:prop H'}(\ref{item:2 of h'}) implies  $\vol_{H'}(j) = \vol_{H^{(1)}}(j) - r_j(s) = r_j(s) - r_j(s) = 0= r_j(\ell)$, where the last two equalities use that $H^{(1)}$ is induced by the valid assignment $\sigma$ for time $s$ and that $j \not\in \ALG_{J'}(\ell)$.
\end{proof}

\subsubsection{Analysis of $\textsc{Update1}$ for the Case $\nu \leq \nu^*$}
\label{sec:nulenu*}

We describe the \textsc{Update1} algorithm as shown in \Cref{fig:update1}. If $\nu \leq \nu^*$, then $\ALG$ spends less (or the same) time running SRPT on the old jobs than $\OPT$. We split the matching $M^{(2)}$ into three parts, the matching $M_p^{(2)}$ corresponding to the jobs in $O_\ell^+$, the matching $M_s^{(2)}$ corresponding to the jobs in $A_\ell^+$, and the matching $M_d^{(2)}$ corresponding to the remaining jobs. At this point, each edge $(j,j^*) \in M^{(2)}$ has weight $w_{M^{(2)}}(j,j^*) = p_j - \Delta(j)$ (cf.~Step~\ref{line:2b} of~\Cref{alg:update assignment}). Consequently, each job $j \in J_{new}$ has a weight of $\vol_{M^{(2)}}(j) = \vol^*_{M^{(2)}}(j) = p_j - \Delta(j)$.

For the jobs $j$ appearing in $M_d^{(2)}$, we have $r_j(\ell) = r^*_j(\ell) = p_j - \Delta(j)$ because $\ALG$ and $\OPT$ work on these jobs for the same amount of time during $(s,\ell]$ by definition of $D_\ell$. Hence, we already have $\vol_{M^{(2)}}(j) = r_j(\ell)$ and $\vol^*_{M^{(2)}}(j) = r^*_j(\ell)$ for all such jobs, which means that matching $M_d^{(2)}$ already implies a valid assignment for these jobs at $\ell$. This allows us to ignore the jobs in $M_d^{(2)}$ and focus on $M_p^{(2)}$ and $M_s^{(2)}$. Note that the jobs appearing in $M_d^{(2)}$ are exactly the jobs in $D_\ell$, which means that \ALG and \OPT work the exact same time on these jobs during $(s,\ell]$. By the fact that \OPT executes SRPT, all jobs in $D_\ell$ with one possible exception are completed at time $\ell$ by both \OPT and \ALG.  So after the subtraction of the common work in step 2b of \Cref{alg:update assignment}, $M_d^{(2)}$ contains at most one edge -- the others have weight $0$.

For the jobs $j$  in $M_s^{(2)}$, we have $j \in A_\ell^+$ which means that $\OPT$ works more on $j$ during $(s,\ell]$ than $\ALG$. If we would reduce the weight of the edges $(j,j^*) \in M_s^{(2)}$ by $\tau^*(j)$, then this would create a demand of $\tau^*(j)$ for each $j \in V(M_s^{(2)})$ in $\ALG$, while no demands would be created for $\OPT$. Instead of directly creating and fixing these demands, we first merge $M_s^{(2)}$ and $H^{(2)}$ (which is the same as  $H_p^{(1)}$)
to obtain $H^{(3)}$ (cf.~\Cref{fig:update-ALG-4}(a)). We do this because $\OPT$ spends more time working on jobs in $H^{(2)}$ and $M_s^{(2)}$ than $\ALG$ ($\nu\le \nu^*$), and $\OPT$ \emph{only} works on these jobs (apart from the already removed common work). In total, \OPT works $\tau^*(A_\ell^+)$ additional time units on jobs in $M_s^{(2)}$ and $\nu^* - \nu$ additional time units on jobs in $H^{(2)}$. By~\Cref{fact:total work ALG OPT}, we have $\tau^*(A_\ell^+) + \nu^* - \nu = \tau(O_\ell^+)$. Hence, $\OPT$ works $\tau(O_\ell^+)$ more time units on jobs in $V^*(H^{(3)})$ than $\ALG$ on jobs in $V(H^{(3)})$.

To handle this discrepancy
we use the \textsc{Split}-operation in Step~\ref{line:3bii} to shave off $\tau(O_\ell^+)$ units of work from $H^{(3)}$ and create $(H_p^{(3)},H_s^{(3)})$. The algorithm keeps $H_p^{(3)}$ as part of the constructed assignment, but discards $H_s^{(3)}$. Since all extra work of $\OPT$ is on jobs in $H_s^{(3)}$ (using the fact that $\OPT$ executes SRPT), there are no demands for $\OPT$ or $\ALG$ for the jobs in $V^*(H_p^{(3)})$ and  $V(H_p^{(3)})$ (with possibly one exception for the partial job in $V(H_p^{(3)}) \cap V(H_s^{(3)})$). On the other hand, discarding $H_s^{(3)}$ creates a demand of $\vol_{H_s^{(3)}}(j)$ for each job $j \in V(H_s^{(3)})$ in $\ALG$ leading to a total demand of $\vol_{H_s^{(3)}}(V(H_s^{(3)})) = \tau(O_\ell^+)$ for these jobs in $\ALG$, while creating no demand for $\OPT$ as $\OPT$ finishes all jobs in $V^*(H_s^{(3)}) \setminus V^*(H_p^{(3)})$ by time $\ell$. \Cref{fig:update-ALG-4}(b) illustrates the created demands.

To fix the demands of the jobs in $X:= V(H_s^{(3)})$ in $\ALG$, we observe that the jobs in $O_\ell^+$ have the same total demand of $\tau(O_\ell^+)$ after updating the edge weights of $M_p^{(2)}$ in Step~\ref{line:3bi}. Hence, we can use $\greedy(X:=V(H_s^{(3)}),O_{\ell}^+, \vol_{ H^{(3)}_s}, \tau)$ to fix all remaining deficits on the marginal volume of jobs in $\ALG$ and $\OPT$ (cf.~\Cref{fig:update-ALG-4}(c)).

The proof of \Cref{claim:X exists} stated in the algorithm is now straightforward:

\begin{proof} [Proof of \Cref{claim:X exists}]
Recall that we are in the case $\nu \leq \nu^*$. By definition of $O_{\ell}^+$ and $\tau$,  during $(s,\ell]$,  $\ALG$ works on jobs in $O_{\ell}^+$ for total $\tau(O_{\ell}^+) = T$ time whereas $\OPT$ spent $T$ time on jobs that are not in $O_{\ell}^+$. Thus, $\OPT$  must work $T$ time on the suffix of $V^*(H^{(3)})$. It follows that there must be a suffix $X$ in $V(H^{(3)})$ such that $\vol_{H^{(3)}}(X) \geq T$.
\end{proof}

For the remainder of this section, we prove that the assignment $\sigma'$ which we return is a valid assignment at time $\ell$. Since~\Cref{lem:sigma' assignment} already shows that $\sigma'$ is an assignment, it only remains to prove bounded prefix expansion. To this end, we first establish the following two auxiliary lemmas.

\begin{lemma} \label{lem:forall r alg' ge e leader} Let $X$ be the smallest suffix in $V(H^{(3)})$ such that $\vol_{H^{(3)}}(X) \geq T$ (defined in Step~\ref{line:X case 1} of \Cref{fig:update1}). Let $j'$ be the first element  in $X$ in the ordering of $V(H^{(3)})$.
Let $G$ be the output of \\ $\greedy(X,O_{\ell}^+, \vol_{ H^{(3)}_s}, \tau)$ in Step~\ref{line:3biii} in \Cref{fig:update1}. We have
\[ \forall j \in X \setminus \{j'\}, \vol_G(j) \geq \threshold \cdot \gamma \
 .\]
\end{lemma}
\begin{proof}

Fix $j \in X \setminus \{j'\}$.
Since $j \neq j'$, i.e., $j$ is not the job that appears in both $H_p^{(3)}$ and $H_s^{(3)}$,
so we have $\vol_{H_p^{(3)}}(j) = 0$. Hence, $\vol_{H'}(j) =  \vol_{G}(j) + \vol_{H_p^{(3)}}(j) =  \vol_{G}(j)$, where $H'$ is the graph for the weight function $\sigma'$ returned by the algorithm. \Cref{lem:sigma' assignment} implies that $ \vol_{H'}(j) = r_{j}(\ell)$, and therefore
\begin{align}  \label{eq:volj eq rj}
\vol_{G}(j) = r_{j}(\ell).
\end{align}

Furthermore, observe that each $j \in X$ is still alive at time $\ell$ in $\ALG$. This is, because $X \subseteq V(H^{(2)}=H_p^{(1)}) \cup V(M_s^{2})$ either implies $j \in V(M_s^{2}) = A_\ell^+ \subseteq \ALG(\ell)$ or $j \in V(H_p^{(1)}) \subseteq  \ALG(\ell)$. Note that $H_p^{(1)} \subseteq  \ALG(\ell)$ holds because the \textsc{split}-operation in Step~\ref{line:2b} of~\Cref{alg:update assignment} selects $H_p^{(1)}$ as the prefix of old jobs that are not touched by $\ALG$ during $(s,\ell]$.

By definition, $j \in K(s) \cup U(s) \cup J_{new}$.  Recall that $K(s)$ and $U(s)$ are the sets of known and unknown jobs, respectively, at time $s$ in $\ALG(s)$ before $J_{new}$ arrive.
We consider each case as follows.
\begin{itemize} [nosep]
    \item  If $j \in K(s)$, then $j \in K(\ell)$ since $j$ is not completed by $\ALG$ at time $\ell$, and thus  $\vol_G(j) \overset{(\ref{eq:volj eq rj})}{=} r_j(\ell) = r_j(s) \geq \threshold \cdot \gamma$,
    by  \Cref{lem:Ks Kl Al in alg}.
    \item If $j \in U(s)$, then $\ALG$ never worked on $j$ during $(s,\ell]$ by the assumption of~\Cref{lem:nontrivial update assignment}. This means $e_j(s) = e_j(\ell) > e_{L(s)}(\ell) = \gamma$ using the fact that $\ALG$ touches the leader $L(s)$ at time $\ell$ by the assumption of~\Cref{lem:nontrivial update assignment}.
    Thus, $j$'s remaining time is $r_j(\ell) \geq \threshold \cdot \gamma$ and, therefore, $ \vol_G(j) \overset{(\ref{eq:volj eq rj})}{=} r_j(\ell) \geq \threshold \cdot \gamma$.
    \item If $j \in J_{new}$, then $j \in A_{\ell}^+$ by the definition of the perfect matching $M_s^{(2)}$. Thus, $\ALG$ does not finish $j$ by time $\ell$ but $j$'s elapsed time in $\ALG$ is $\gamma$ (\Cref{lem:Ks Kl Al in alg}(3)). Hence, $\vol_G(j) \overset{(\ref{eq:volj eq rj})}{=} r_j(\ell) \geq \threshold \cdot \gamma.$ \qedhere
\end{itemize}
\end{proof}

\begin{restatable}{lemma}{NGUMez}
\label{lem:nulenu* NGUM}
If $P^*$ is a prefix of $V^*(G \cup  M^{(3)})$, then
\[|N_{G\cup  M^{(3)}}(P^*)| \leq \bfactor\cdot |P^*|.\]
\end{restatable}

 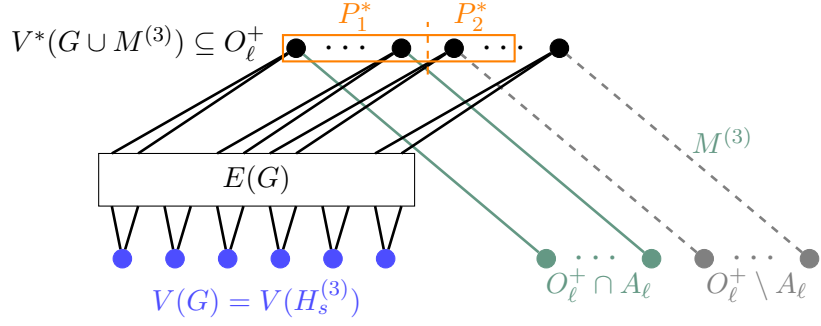
\begin{figure}[t]
     \centering
     \begin{tikzpicture}[scale=0.7]

\foreach \x in {6,...,11} {
    \node[vertex,blue!70!white] (a\x) at (\x+2-0.3,0) {};
    \draw[asg] (a\x) -- (\x+2,1);
    \draw[asg] (a\x) -- (\x+1.5,1);
}

\draw (7+0.25,2) rectangle (10+3+0.25,1);
\node at (10.25,1.5) {$E(G)$};
\node[blue!70!white] at (10.25,-0.75) {$V(G) = V(H_s^{(3)})$};
\node at (8,4.1) {$V^*(G \cup M^{(3)}) \subseteq O_\ell^+$};

\begin{scope}[shift={(11,0)}]

    \node[vertex] (o0) at (0,4) {};
    \node[vertex,job3] (a0) at (0+4.75,0) {};
    \draw[asg,job3] (a0) -- (o0);

    \draw[asg] (o0) -- (0-3.5,2);
    \draw[asg] (o0) -- (0-3,2);
    \node at (1,4) {\Large$\cdots$};

    \node[vertex] (o2) at (2,4) {};
    \node[vertex,job3] (a2) at (2+4.75,0) {};
    \draw[asg,job3] (a2) -- (o2);

    \draw[asg] (o2) -- (2-3.5,2);
    \draw[asg] (o2) -- (2-3,2);
    \node[job3] at (1+4.75,0) {\Large$\cdots$};

    \node[vertex] (o3) at (3,4) {};
    \node[vertex,gray] (a3) at (3+4.75,0) {};
    \draw[asg,gray,dashed] (a3) -- (o3);

    \draw[asg] (o3) -- (3-3.5,2);
    \draw[asg] (o3) -- (3-3,2);
    \node at (4,4) {\Large$\cdots$};

    \node[vertex] (o5) at (5,4) {};
    \node[vertex,gray] (a5) at (5+4.75,0) {};
    \draw[asg,gray,dashed] (a5) -- (o5);

    \draw[asg] (o5) -- (5-3.5,2);
    \draw[asg] (o5) -- (5-3,2);
    \node[gray] at (4+4.75,0) {\Large$\cdots$};

    \node[job3] at (1+4.75,-0.5) {$O_\ell^+ \cap A_\ell$};
    \node[gray] at (4+4.75,-0.5) {$O_\ell^+ \setminus A_\ell$};

    \draw[orange,thick] (-0.25,3.75) rectangle (4.15,4.25);
    \draw[orange,thick,dashed] (2.5,3.5) -- (2.5,4.5);
    \node[orange] at (1.125,4.6) {$P_1^*$};
    \node[orange] at (3.325,4.6) {$P_2^*$};

    \node[job3] at (8.1,2.25) {$M^{(3)}$};

\end{scope}

\end{tikzpicture}
     \caption{Illustration of the situation considered in the proof of~\Cref{lem:nulenu* NGUM}. Dashed edges represent edges of weight zero.}
     \label{fig:NGUMez}
 \end{figure}

\begin{proof}
 First, observe that $V^*(G \cup M^{(3)}) \subseteq O_\ell^+$. This is because $V^*(M^{(3)}) = V^*(M_p^{(2)}) \subseteq O_\ell^+$ by Steps~\ref{line:3b0} and~\ref{line:2a} of~\Cref{fig:update1}  and $V^*(G) \subseteq O_\ell^+$ by Step~\ref{line:3biii} of~\Cref{fig:update1}. Hence, $P^* \subseteq O_\ell^+$.

 Next, we split $P^*$ into $P^*_1$ and $P^*_2$ as follows.  The set $P^*_1$ contains jobs in $P^* \subseteq O_{\ell}^+$ such that the corresponding job is not completed at time $\ell$ by $\ALG$, while the set $P^*_2$ contains jobs in $P^* \subseteq O_{\ell}^+$ such that $\ALG$ finished the corresponding job by time $\ell$.
 Formally, we define $P^*_1 := P^* \cap A_{\ell}$,
 and $P^*_1 := P^* \cap O_{\ell}^+ \setminus A_{\ell}$.
 Note that $P^* = P^*_1 \cup P^*_2$ and $P^*_1 \cap P^*_2 = \emptyset$. See~\Cref{fig:NGUMez} for an illustration.
We have
 \begin{align}
        |N_{G \cup M^{(3)}}(P^*)| &= |P^*_1| + \left |\minsuffix_{G}\big( \tau(P^*)\big)\right | \label{eq:NGUMez1} \\
      &\leq |P^*_1| + \ceil {\frac{ \tau(P^*)}{\threshold\cdot \gamma}} \label{eq:NGUMez2} \\
            &\leq |P^*_1| + \ceil { \frac{\gamma\cdot |P^*_1| + \ainv  \cdot |P^*_2| \cdot \gamma}{\threshold\cdot \gamma}}\label{eq:NGUMez3}\\
      &= |P^*_1| + \ceil {\frac{\myAlfa \cdot |P^*_1|}{\oneminusA} + \frac{|P^*_2|}{\oneminusA }} \nonumber\\
      &= \ceil{\frac{|P^*|}{\oneminusA }} \nonumber \\
      &\leq \bfactor\cdot |P^*| \label{eq:NGUMez4} \ .
 \end{align}
where we justify \Cref{eq:NGUMez1,eq:NGUMez2,eq:NGUMez3,eq:NGUMez4} as follows.
\begin{itemize}
    \item [(\ref{eq:NGUMez1})]  Observe that $|N_{G \cup M^{(3)}}(P^*)| = |N_{M^{(3)}}(P_1^*)| + |N_{G}(P^*)|$ as the jobs in $P_2^*$ do not have neighbors in $M^{(3)}$.   We first show that $|N_{M^{(3)}}(P_1^*)| = |P^*_1|$. For every job $j \in P^*$, if $j \in P^*_1$, then $j \in A_{\ell}$, which has an extra neighbor from the matching $M^{(3)}$.  We next show the second term. Observe that $G$ is the output of $\greedy(X,O_{\ell}^+, \vol_{ H^{(3)}_s}, \tau)$ in Step~\ref{line:3biii} of \Cref{fig:update1}. By \Cref{lem:greedy}, we have $G = (X,O^+_{\ell}, E_G, w_G)$ with $\vol^*_G(j) = \tau(j)$ for all $j \in O^+_{\ell}$. Therefore, \[|N_{G}(P^*)| = |\minsuffix_{G}\big(\vol^*_G(P^*) \big)| = |\minsuffix_{G}\big( \tau(P^*)\big)|\]
    holds by definition of the construction given in the proof of~\Cref{lem:greedy} and by~\Cref{def:min-suffix}.

    \item [(\ref{eq:NGUMez2})]  By~\Cref{lem:forall r alg' ge e leader}, each element $j$ of $X = V(G)$, except possibly the first, satisfies $\vol_G(j) \ge \threshold \cdot \gamma$. Hence, the size of the smallest suffix of $X=V(G)$ with volume at least $\tau(P^*)$ is at most $\ceil {\frac{ \tau(P^*)}{\threshold\cdot \gamma}} $.
    \item [(\ref{eq:NGUMez3})]  Observe that
    \begin{align*}
    \tau(P^*) &= \tau(P^*_1) + \tau(P^*_2)
    = \sum_{j \in P^* \cap A_{\ell}} \tau(j) +  \sum_{j \in P^* \cap     (O^+_{\ell} \setminus  A_{\ell})} \tau (j)
    \leq \gamma |P^*_1| + \ainv  |P_2^*| \cdot \gamma,
    \end{align*}
    where last inequality follows from \Cref{lem:characterize tau delta}.
    \item [(\ref{eq:NGUMez4})]  This follows because $|P^*|$ is a positive integer. \qedhere
\end{itemize}
\end{proof}

\begin{observation}
\label{obs:disjointness}
    The sets $V^*(G \cup M^{(3)})$, $V^*(H_p^{(3)})$ and $V^*(M_d^{(2)})$ are pairwise disjoint.
\end{observation}

\begin{proof}
    First, observe that the sets $O_\ell^+$, $D_\ell$ and $A_\ell^+$ are pairwise disjoint by~\Cref{def:O ell plus}. Furthermore, $V^*(H^{(1)})$ satisfies  $V^*(H^{(1)}) \cap J_{new} = \emptyset$ and, thus, is disjoint to any of the sets $O_\ell^+$, $D_\ell$ and $A_\ell^+$.

    To finish the proof, note that (i) $V^*(G) \subseteq O_\ell^+$ by Step~\ref{line:3biii} of~\Cref{fig:update1}, (ii) $V^+(M^{(3)}) \subseteq V^*(M_p^{(2)}) \subseteq O_{\ell}^+$ by Step~\ref{line:3a} and~\ref{line:3b0} of~\Cref{fig:update1}, (iii) $V^*(H_p^{(3)}) \subseteq V^*(H^{(1)}) \cup V^*(M_s^{(2)}) \subseteq V^*(H^{(1)}) \cup A_\ell^{+}$ by Steps~\ref{line:3a} and~\ref{line:merge} of~\Cref{fig:update1} and Step~\ref{line:2c} of~\Cref{alg:update assignment}, and (iv) $V^*(M_d^{(2)}) \subseteq D_\ell$ by Step~\ref{line:3a} of~\Cref{fig:update1}.
    Hence $V^*(G \cup M^{(3)}) \subseteq O_\ell^+$, $V^*(H_p^{(3)}) \subseteq  V^*(H^{(1)}) \cup A_\ell^{+}$ and $V^*(M_d^{(2)}) \subseteq D_\ell$, which implies that the sets are pairwise disjoint by the observation at the beginning of the proof.
\end{proof}

With the two auxiliary lemmas in place, we are ready to prove that the prefix expansion of $\sigma'$ is bounded. Since we already showed that $\sigma'$ is an assignment with~\Cref{lem:sigma' assignment}, this implies that $\sigma'$ is a valid assignment and, thus, finishes the proof of~\Cref{lem:nontrivial update assignment} for the case that $\nu \le \nu^*$.

\begin{lemma} \label{lem:expansion new case 1}
If  $\nu \leq \nu^*$, then $\phi(\sigma') \leq \bfactor.$
\end{lemma}

\begin{proof}
Recall that $\sigma'$ is the weight function for the graph $H' = G \cup M^{(3)} \cup  H^{(3)}_p \cup  M_d^{(2)}$ returned by~\Cref{fig:update1}.
To prove the lemma, we fix an integer $k$, let $P^*$ be a $k$-prefix of $V^*(H')$ and show that $ |N_{H'}(P^*)| \leq \bfactor \cdot |P^*|.$

 Note that the sets $V^*(G \cup M^{(3)})$, $V^*(H^{(3)}_p)$ and $V^*( M_d^{(2)})$ are pairwise disjoint by~\Cref{obs:disjointness}. We use the pairwise disjointness to partition $P^*$ in the following way:
Let $P^* = P^*_0 \cup  P^*_1 \cup P^*_2$
where $P^*_0 := P^* \cap M_d^{(2)}$ is the prefix of $V^*( M_d^{(2)})$, $P^*_1 := P^* \cap V^*(G \cup M^{(3)})$ is a $k_1$-prefix of $V^*(G \cup M^{(3)})$ in $G \cup M^{(3)}$ and $P^*_2 := P^* \cap V^*(H^{(3)}_p)$ is a $k_2$-prefix of $V^*(H^{(3)}_p)$ such that $k = |P^*_0| + k_1 + k_2$.

Recall that $M_d^{(2)}$ contains at most a single edge, so $P^*_0$ is either empty or a singleton.
We have
     \begin{align}
     |N_{H'}(P^*)| &\leq |N_{M^{(2)}_d}(P^*_0)| +|N_{G \cup M^{(3)}}(P^*_1)| + |N_{H^{(3)}_p}(P^*_2)|  \label{eq:NH'P1}  \\
     &\leq |P^*_0| + \bfactor \cdot |P^*_1|  + |N_{H^{(3)}_p}(P^*_2)| \label{eq:NH'P2}\\
     &\leq \bfactor \cdot |P^*_0| + \bfactor \cdot |P^*_1|  + \bfactor \cdot |P^*_2| \label{eq:NH'P3}\\
     &= \bfactor \cdot |P^*| \ , \nonumber
     \end{align}
where we justify \Cref{eq:NH'P1,eq:NH'P2,eq:NH'P3} as follows.
\begin{itemize}
    \item [(\ref{eq:NH'P1})] This follows since $H' = G \cup M^{(3)} \cup H^{(3)}_p \cup M_d^{(2)}$,
    the edge of $M_d^{(2)}$ if it exists, is incident to  $P^*_0$, the edges of
    $G \cup M^{(3)}$ are incident to $P_1^*$ and the edges of $H^{(3)}_p$
     are incident to $P_2^*$.
    \item [(\ref{eq:NH'P2})] This follows from \Cref{lem:nulenu* NGUM}.
    \item [(\ref{eq:NH'P3})] It is enough to prove $\phi(H^{(3)}_p) \leq \bfactor$. To see this, first note that $\phi(H^{(1)}) \le  \bfactor$ since $H^{(1)}$ corresponds to the assignment $\sigma$, which is valid by \Cref{lem:valid imp canonical} and the fact that there is a valid assignment at time $s$, right before $J_{new}$ arrives.  \Cref{lem:spliting} implies that
    $\phi(H^{(2)}) \le  \bfactor$ since $H^{(2)} = H_p^{(1)}$ and  $(H_p^{(1)},H_s^{(1)})$ is a split of $H^{(1)}$.
    Then, since $\phi(H^{(2)}) \le  \bfactor$, $M_s^{(2)}$ is a matching disjoint from $H^{(2)}$, and $H^{(3)} = \textsc{Merge}(H^{(2)}, M_s^{(2)})$ it follows that  $\phi(H^{(3)}) \leq \bfactor$.
    Finally, we  conclude that $\phi(H^{(3)}_p) \leq \bfactor$
    by using \Cref{lem:spliting} again, since
     $\phi(H^{(3)}) \leq \bfactor$ and $(H^{(3)}_p,H^{(3)}_s)$ is a split of $H^{(3)}$, and~\Cref{lem:spliting}.\qedhere
\end{itemize}
\end{proof}

\begin{algorithm}
\caption{$\textsc{Update2}(H^{(2)},M^{(2)})$} \label{fig:update2}

\medskip

    \begin{enumerate}
\item Define the following:

    \begin{enumerate}  [noitemsep,nolistsep]
     \item Let $d := \nu - \nu^*$.
    \item  Let $X$ be the suffix  in $V(H^{(2)})$ such that $\vol_{H^{(2)}}(X) = d.$
    \begin{claim} \label{claim:X exsits 4}
       The set $X$ exists.
    \end{claim}
    \item  Let $Y$ be the smallest suffix  in $V^*(H^{(2)})$ such that $\vol_{H^{(2)}}^*(Y) \geq d.$
    \begin{claim}\label{claim:Y exsits 4}
       The set $Y$ exists.
    \end{claim}
    \end{enumerate}
 \item  \label{line:4b}Perform the following:
    \begin{enumerate} [noitemsep,nolistsep]
        \item $M^{(3)} \gets M^{(2)}$
        \item \label{line:4bi} $\forall j \in O_{\ell}^+$ do $w_{M^{(3)}}(j,j^*) \gets w_{M^{(3)}}(j,j^*) - \tau(j)$
        \item \label{line:4bii} $\forall j \in A_{\ell}^+$ do $w_{M^{(3)}}(j,j^*) \gets w_{M^{(3)}}(j,j^*) - \tau^*(j)$
        \item  \label{line:4biii} $(H^{(2)}_p, H^{(2)}_s) \gets \textsc{Split}(H^{(2)},d)$
        \item \label{line:4biv} $G \gets \greedy( A^+_{\ell}, O_{\ell}^+\cup Y,  \tau^*, \tau + \vol^*_{H^{(2)}_s})$
    \end{enumerate}
        \item \textbf{Return} $\sigma' =$ the weight function of $G \cup  {M^{(3)}} \cup H^{(2)}_p$ after removing isolating vertices.
    \end{enumerate}
\end{algorithm}

\subsubsection{Analysis of $\textsc{Update2}$ for the Case $\nu^* < \nu$}
\label{sec:nu*lenu}

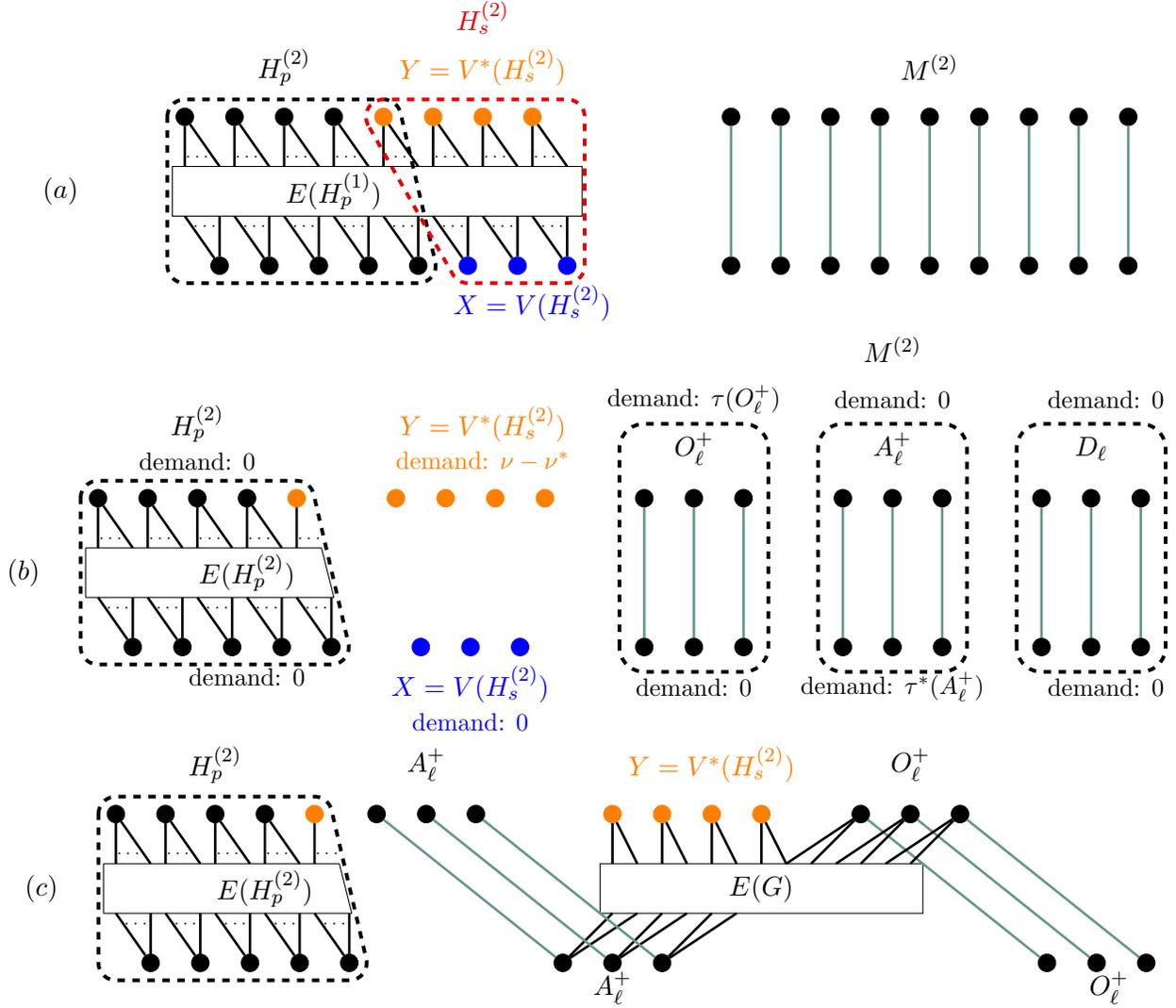
\begin{figure}[th]
    \centering
    \begin{tikzpicture}[scale=0.7]

\foreach \x in {0,...,3} {
    \node[vertex] (o\x) at (\x,3) {};

    \draw[asg] (o\x) -- (\x,2);
    \node at (\x+0.25,2.2) {\tiny $\ldots$};
    \draw[asg] (o\x) -- (\x + 0.7,2);

    \node[vertex] (a\x) at (\x+0.7,0) {};

    \draw[asg] (a\x) -- (\x+0.7,1);
    \node at (\x+0.4,0.8) {\tiny $\ldots$};
    \draw[asg] (a\x) -- (\x,1);
}

\foreach \x in {4} {
    \node[vertex,orange] (o\x) at (\x,3) {};

    \draw[asg] (o\x) -- (\x,2);
    \node at (\x+0.25,2.2) {\tiny $\ldots$};
    \draw[asg] (o\x) -- (\x + 0.7,2);

    \node[vertex] (a\x) at (\x+0.7,0) {};

    \draw[asg] (a\x) -- (\x+0.7,1);
    \node at (\x+0.4,0.8) {\tiny $\ldots$};
    \draw[asg] (a\x) -- (\x,1);
}

\foreach \x in {5,...,7} {
    \node[vertex,orange] (o\x) at (\x,3) {};

    \draw[asg] (o\x) -- (\x,2);
    \node at (\x+0.25,2.2) {\tiny $\ldots$};
    \draw[asg] (o\x) -- (\x + 0.7,2);

    \node[vertex,blue] (a\x) at (\x+0.7,0) {};

    \draw[asg] (a\x) -- (\x+0.7,1);
    \node at (\x+0.4,0.8) {\tiny $\ldots$};
    \draw[asg] (a\x) -- (\x,1);
}

\coordinate (oleft) at (0,3);
\coordinate (aleft) at (0,0);
\coordinate (oright) at (7+0.7,3);
\coordinate (aright) at (7+0.7,0);

\draw[line width=1.5pt, dashed, black] \convexpath{oleft,o4,a4,aleft}{10pt};
\draw[line width=1.5pt, dashed, red!90!black] \convexpath{o4,oright,aright,a5}{10pt};

\node at (-2.5,1.5) {$(a)$};

\node at (2,4) {$H_p^{(2)}$};
\node[red!90!black] at (6,5) {$H_s^{(2)}$};

\node[orange] at (6,4) {$Y= V^*(H_s^{(2)})$};
\node[blue] at (7,-0.75) {$X = V(H_s^{(2)})$};

\draw (-0.25,2) rectangle (8,1);
\node at (3,1.5) {$E(H_p^{(1)})$};

\begin{scope}[shift={(11,0)}]

\foreach \x in {0,...,8} {
    \node[vertex] (o\x) at (\x,3) {};
    \node[vertex] (a\x) at (\x,0) {};
    \draw[asg,job3] (a\x) -- (o\x);
}

\node at (4,4) {$M^{(2)}$};

\end{scope}

\end{tikzpicture}

    \vspace*{-1.5cm}

    \begin{tikzpicture}[scale=0.7]

\foreach \x in {0,...,3} {
    \node[vertex] (o\x) at (\x,3) {};

    \draw[asg] (o\x) -- (\x,2);
    \node at (\x+0.25,2.2) {\tiny $\ldots$};
    \draw[asg] (o\x) -- (\x + 0.7,2);

    \node[vertex] (a\x) at (\x+0.7,0) {};

    \draw[asg] (a\x) -- (\x+0.7,1);
    \node at (\x+0.4,0.8) {\tiny $\ldots$};
    \draw[asg] (a\x) -- (\x,1);
}

\foreach \x in {4} {
    \node[vertex,orange] (o\x) at (\x,3) {};

    \draw[asg] (o\x) -- (\x,2);
    \node at (\x+0.25,2.2) {\tiny $\ldots$};

    \node[vertex] (a\x) at (\x+0.7,0) {};

    \draw[asg] (a\x) -- (\x+0.7,1);
    \node at (\x+0.4,0.8) {\tiny $\ldots$};
    \draw[asg] (a\x) -- (\x,1);
}

\coordinate (oleft) at (0,3);
\coordinate (aleft) at (0,0);
\coordinate (oright) at (7+0.7,3);
\coordinate (aright) at (7+0.7,0);

\draw[line width=1.5pt, dashed, black] \convexpath{oleft,o4,a4,aleft}{10pt};

\node at (-1.5,1.5) {$(b)$};

\node at (2,4.5) {$H_p^{(2)}$};
\node at (2,3.75) {\small demand: $0$};
\node at (3,-0.6) {\small demand: $0$};

\path
(-0.25,2) edge (4.5,2)
(-0.25,2) edge (-0.25,1)
(-0.25,1) edge (4.75,1)
(4.5,2) edge (4.75,1)
;

\node at (3,1.5) {$E(H_p^{(2)})$};

\begin{scope}[shift={(2,0)}]
\foreach \x in {4,...,7} {
    \node[vertex,orange] (o\x) at (\x,3) {};
}
\node[orange] at (5.75,4.5) {$Y= V^*(H_s^{(2)})$};
\node[orange] at (5.75,3.75) {\small demand: $\nu - \nu^*$};

\end{scope}

\begin{scope}[shift={(2,0)}]
\foreach \x in {5,...,7} {
    \node[vertex,blue] (o\x) at (\x-0.5,0) {};
}
\node[blue] at (5.5,-0.75) {$X = V(H_s^{(2)})$};
\node[blue] at (5.5,-1.5) {\small demand: $0$};
\end{scope}

\begin{scope}[shift={(11,0)}]

\foreach \x in {0,...,2} {
    \node[vertex] (o\x) at (\x,3) {};
    \node[vertex] (a\x) at (\x,0) {};
    \draw[asg,job3] (a\x) -- (o\x);
}

\node at (1,4) {$O_\ell^+$};
\draw[rounded corners=10pt,line width=1.5pt,dashed] (-0.5,-0.5) rectangle ++ (3,5);
\node at (1,5) {\small demand: $\tau(O_\ell^+)$};
\node at (1,-0.8) {\small demand: $0$};

\end{scope}

\begin{scope}[shift={(15,0)}]

\foreach \x in {0,...,2} {
    \node[vertex] (o\x) at (\x,3) {};
    \node[vertex] (a\x) at (\x,0) {};
    \draw[asg,job3] (a\x) -- (o\x);
}

\node at (1,4) {$A_\ell^+$};
\draw[rounded corners=10pt,line width=1.5pt,dashed] (-0.5,-0.5) rectangle ++ (3,5);
\node at (1,6) {$M^{(2)}$};
\node at (1,5) {\small demand: $0$};
\node at (1,-0.8) {\small demand: $\tau^*(A_\ell^+)$};

\end{scope}

\begin{scope}[shift={(19,0)}]

\foreach \x in {0,...,2} {
    \node[vertex] (o\x) at (\x,3) {};
    \node[vertex] (a\x) at (\x,0) {};
    \draw[asg,job3] (a\x) -- (o\x);
}

\node at (1,4) {$D_\ell$};
\draw[rounded corners=10pt,line width=1.5pt,dashed] (-0.5,-0.5) rectangle ++ (3,5);
\node at (1.3,5) {\small demand: $0$};
\node at (1.3,-0.8) {\small demand: $0$};

\end{scope}

\end{tikzpicture}

    \vspace*{-1.5cm}

    \begin{tikzpicture}[scale=0.7]

\begin{scope}[shift={(-3,0)}]
\foreach \x in {0,...,3} {
    \node[vertex] (o\x) at (\x,3) {};

    \draw[asg] (o\x) -- (\x,2);
    \node at (\x+0.25,2.2) {\tiny $\ldots$};
    \draw[asg] (o\x) -- (\x + 0.7,2);

    \node[vertex] (a\x) at (\x+0.7,0) {};

    \draw[asg] (a\x) -- (\x+0.7,1);
    \node at (\x+0.4,0.8) {\tiny $\ldots$};
    \draw[asg] (a\x) -- (\x,1);
}

\foreach \x in {4} {
    \node[vertex,orange] (o\x) at (\x,3) {};

    \draw[asg] (o\x) -- (\x,2);
    \node at (\x+0.25,2.2) {\tiny $\ldots$};

    \node[vertex] (a\x) at (\x+0.7,0) {};

    \draw[asg] (a\x) -- (\x+0.7,1);
    \node at (\x+0.4,0.8) {\tiny $\ldots$};
    \draw[asg] (a\x) -- (\x,1);
}

\coordinate (oleft) at (0,3);
\coordinate (aleft) at (0,0);
\coordinate (oright) at (7+0.7,3);
\coordinate (aright) at (7+0.7,0);

\draw[line width=1.5pt, dashed, black] \convexpath{oleft,o4,a4,aleft}{10pt};

\node at (-1.5,1.5) {$(c)$};

\node at (2,4) {$H_p^{(2)}$};

\path
(-0.25,2) edge (4.5,2)
(-0.25,2) edge (-0.25,1)
(-0.25,1) edge (4.75,1)
(4.5,2) edge (4.75,1)
;

\node at (3,1.5) {$E(H_p^{(2)})$};

\end{scope}

\foreach \x in {4,...,7} {
    \node[vertex,orange] (o\x) at (\x+3,3) {};

    \draw[asg] (o\x) -- (\x+3,2);
    \draw[asg] (o\x) -- (\x + 3 + 0.5,2);

}

\node[orange] at (9,4) {$Y= V^*(H_s^{(2)})$};

\draw (6+0.75,2) rectangle (10+3+0.25,1);
\node at (10,1.5) {$E(G)$};

\foreach \x in {0,...,2} {
    \node[vertex] (o\x) at (\x+6-3.75,3) {};
    \node[vertex] (a\x) at (\x+6,0) {};
    \draw[asg,job3] (a\x) -- (o\x);

    \draw[asg] (a\x) -- (\x+6+1.5,1);
    \draw[asg] (a\x) -- (\x+6+1,1);
}

\node at (1+6-3.75,4) {$A_\ell^+$};
\node at (7,-0.5) {$A_\ell^+$};

\begin{scope}[shift={(12,0)}]

\foreach \x in {0,...,2} {
    \node[vertex] (o\x) at (\x,3) {};
    \node[vertex] (a\x) at (\x+3.75,0) {};
    \draw[asg,job3] (a\x) -- (o\x);

    \draw[asg] (o\x) -- (\x-1.5,2);
    \draw[asg] (o\x) -- (\x-1,2);

}

\node at (1,4) {$O_\ell^+$};
\node at (5,-0.5) {$O_\ell^+$};

\end{scope}

\end{tikzpicture}

    \vspace*{-1cm}

    \caption{Illustration of the operations executed by~\Cref{fig:update2} in the case that $\nu^* < \nu$.}
    \label{fig:update-ALG-7}
\end{figure}

We assume $\nu^* < \nu$ throughout this section.
We describe the algorithm for $\textsc{Update2}$ as shown in \Cref{fig:update2}. Since $\nu^* < \nu$, $\OPT$ spends less time running SRPT on the old jobs than $\ALG$. Let $d := \nu - \nu^*$ be the additional time that $\ALG$ works on old jobs compared to $\OPT$.  We claim that there is a suffix $X$ of $V(H^{(2)})$ such that $\vol_{H^{(2)}}(X) = d$ and there is a suffix $Y$ of $V^*(H^{(2)})$ such that $\vol^*_{H^{(2)}}(Y) \geq d$.
Since $H^{(2)} = H_p^{(1)}$, it follows that $H^{(2)}$ is   forward and the  split operation in Step \ref{line:4biii} of \Cref{fig:update2} gives the subgraph  $H^{(2)}_s$ with
$V(H^{(2)}_s)=X$ and  $V^*(H^{(2)}_s)=Y$ (see~\Cref{fig:update-ALG-7}(a)). Removing this subgraph creates a demand of $\vol^*_{H_s^{(2)}}(V^*(H_s^{(2)})) = \nu-\nu^*$ for $Y$ in $\OPT$, without
creating a demand for $\ALG$.
Then, in Steps~\ref{line:4bi} and~\ref{line:4bii} of~\Cref{fig:update2},  we remove the mass on the matching to account for the excess that $\ALG$ and $\OPT$ have according to $\tau$ and $\tau^*$. This creates a demand of $\tau(O_\ell^+)$ for the jobs in $O_\ell^+$ in $\OPT$ and a demand of $\tau^*(A_\ell^+)$ for the jobs in $A_\ell^+$ in $\ALG$. \Cref{fig:update-ALG-7}(b) shows all generated demands.

Since $\nu-\nu^* + \tau(O_\ell^+) = \vol^*(H_s^{(2)}) +  \tau(O_\ell^+) = \tau^*(A_\ell^+)$ by~\Cref{fact:total work ALG OPT}, $\greedy( A^+_{\ell}, O_{\ell}^+\cup Y,  \tau^*, \tau + \vol^*_{H^{(2)}_s})$ fixes the demands on the corresponding jobs between $\ALG$ and $\OPT$ (see~\Cref{fig:update-ALG-7}(c)).

We first prove the two claims stated in~\Cref{fig:update2}.

\begin{proof} [Proof of \Cref{claim:X exsits 4}]
Recall that we are in the case $\nu^* < \nu$, and let $d := \nu - \nu^*$.
By \Cref{lem:Ks Kl Al in alg}, $\ALG$ finished all the jobs in $K(s) \setminus K(\ell)$ by time $\ell$. Since the total volume on $K(s) \setminus K(\ell)$ is $\nu$, and $\vol(H_s^{(1)}) = \min\{\nu,\nu^*\} = \nu^*$, there is exactly $\nu - \nu^* = d$ volume of the set of jobs $\subseteq K(s) \setminus K(\ell)$ remaining in $H^{(2)} = H_p^{(1)}$.
\end{proof}

\begin{proof} [Proof of \Cref{claim:Y exsits 4}]
This follows because $X$ (defined in \Cref{claim:X exsits 4}) exists and $\vol(H^{(2)}) = \vol^*(H^{(2)})$.
\end{proof}

For the remainder of this section, we prove that constructed function $\sigma'$ is a valid assignment at time $\ell$. Since~\Cref{lem:sigma' assignment} already shows that $\sigma'$is an assignment, it only remains to prove bounded prefix expansion. To this end, we first establish a series of auxiliary lemmas that will be useful for proving \Cref{lem:nu*lenu NGUM}.

\begin{lemma} \label{lem:N Y H2_s = X}
$N_{H^{(2)}_{s}}(Y) = X$.
\end{lemma}
\begin{proof}
    This follows because $X$ is the suffix of $V(H^{(2)})$ in $H^{(2)}$ such that $\vol_{H^{(2)}}(X) = d$, $Y$ is the smallest suffix of $V^*(H^{(2)})$ in $H^{(2)}$ such that $\vol_{H^{(2)}}^*(Y) \geq d$, and $(H_p^{(2)},H_s^{(2)})$ is a split of $H^{(2)}$ with threshold $d$.
\end{proof}

\begin{lemma} \label{lem:forall r alg' le e leader}
    For all  $j \in X$, $\vol_{H^{(2)}_s}(j) \leq \threshold \cdot \gamma.$
\end{lemma}
\begin{proof}
Note that, by the assumption that $\ALG$ does not work on old unknown jobs during $(s,\ell]$ and by the choice of $X$, the set $X$ satisfies $X \subseteq K(s)$ and contains  old jobs that $\ALG$ touches during $(s,\ell]$. By~\Cref{lem:Ks Kl Al in alg}, the fact that $\ALG$ touches the jobs in $X$ during $(s,\ell]$ implies  $X \subseteq K(s) \setminus K(\ell)$. Hence, $X$ contains  old jobs that $\ALG$ finishes by time $\ell$.
Fix $j \in X$. Since $\ALG$ finishes $j$ by time $\ell$, \Cref{lem:Ks Kl Al in alg}(1) implies that $r_j(s) \leq \threshold \cdot \gamma.$ Therefore, $\vol_{H^{(2)}_s}(j) = r_j(s) \leq \threshold \cdot \gamma$.
\end{proof}

\begin{lemma} \label{lem:tau o le tau* a}
  $\tau (J_{new}) < \tau^*(J_{new}).$
\end{lemma}
\begin{proof}
    By \Cref{fact:total work ALG OPT},
    $\Delta(J_{new}) + \tau(J_{new}) + \nu = \Delta(J_{new}) + \tau^*(J_{new}) + \nu^* .$
  That is, $\tau(J_{new}) + d = \tau^*(J_{new})$ where $d = \nu - \nu^* > 0$. Therefore, $\tau(J_{new}) < \tau^*(J_{new})$.
\end{proof}

\begin{corollary} \label{cor:O subseteq A}
  $O_{\ell} \subseteq A_{\ell}$.
\end{corollary}
\begin{proof}
For the sake of contradiction, suppose that $O_{\ell} \setminus A_{\ell}\not=\emptyset$.
By definition of $\ALG$ and $\OPT$ (SRPT), both algorithms complete some suffix of $J_{new}$ (ordered by non-increasing processing time) by time $\ell$. Since we assume $O_{\ell} \setminus A_{\ell}\not=\emptyset$, the suffix completed by $\ALG$ must contain more jobs than the suffix completed by $\OPT$. However, this implies that  the total amount of work on $J_{new}$ in $\ALG$ dominates that in $\OPT$, i.e., $ \Delta (J_{new}) + \tau (J_{new}) >  \Delta (J_{new}) + \tau^*(J_{new})$. Therefore,  $\tau (J_{new}) > \tau^*(J_{new})$, which contradicts \Cref{lem:tau o le tau* a}.
\end{proof}
\begin{lemma} \label{lem:j in matching tau j le e1}
If $j \in O_{\ell}^+$, then $\tau(j) \leq \gamma$.
\end{lemma}
\begin{proof}
By \Cref{cor:O subseteq A}, we have $O_{\ell} \subseteq A_{\ell}$. Since $j \in O_{\ell}^+ \subseteq O_\ell$, we conclude that $j \in A_{\ell}$. By \Cref{lem:Ks Kl Al in alg}(3),  the elapsed time of $j$ at time $\ell$ in $\ALG$ is $\gamma$ . Therefore, $\tau(j) \leq e_{j}(\ell) \leq \gamma.$
\end{proof}

With the next lemma, we give a bound on the neighborhood size for a prefix $P^*$ of $V^*(G \cup M^{(3)})$. Before we prove the lemma, we recall some properties of the set $V^*(G \cup M^{(3)})$. First, recall that $V^*(G) \subseteq O_\ell^+ \cup Y$ by Step~\ref{line:4biv} of~\Cref{fig:update2}. Furthermore, note that $O_\ell^+ \subseteq V^*(M^{(3)})$ since the fact that $\ALG$ works more on the jobs in $O_\ell^+$ than $\OPT$ implies that $\OPT$ cannot finish jobs in $O_\ell^+$. Finally, note that $|V^*(M^{(3)}) \setminus O_\ell^+| \le 1$. This is, because $\OPT$ executes SRPT and, therefore, will have at most one job that it touches during $(s,\ell]$ and does not finish by time $\ell$. Since all jobs in $V^*(M^{(3)}) \setminus O_\ell^+$ are touched by $\OPT$ during $(s,\ell]$, at most a single one of them can appear in $V^*(M^{(3)})$ (the rest are isolated vertices that are removed by the algorithm). Note that this single job is the only job in $V^*(G \cup M^{(3)})$ that does not participate in the \greedy \ of Step~\ref{line:4biv} in~\Cref{fig:update2}, i.e., is the only job that is not part of $V^*(G)$.

\begin{restatable}{lemma}{NGUMhard}
\label{lem:nu*lenu NGUM}
Let $H' = G \cup  M^{(3)} \cup H^{(2)}_p$ be the bipartite graph with weight function $\sigma'$ that is returned by \Cref{fig:update2}.
Let $P^* \subseteq V^*(G\cup M^{(3)})$ be a prefix of $V^*(G\cup M^{(3)})$. Then
  \[ |N_{G \cup M^{(3)}}(P^*)| \leq \bfactor\cdot |O^*|+ |N_{H^{(2)}_s}(Y^*)|\] where $Y^* := Y \cap P^*$ and $O^* := P^* \setminus Y$ (so $P^* := O^* \cup Y^*$).
\end{restatable}

\begin{proof}
Define $\bar{O}^* = O^* \cap O_{\ell}^+$. As discussed above, $\bar{O}^*$ contains all elements of $O^*$ with possibly a single exception, and $V^*(G) \cap O^* = \bar{O}^*$.
We first argue that
    \begin{align}
         |N_{G \cup M^{(3)}}(P^*)| &= |O^*| + \bigg |\minsuffix_G\big (\tau(\bar{O}^*) +\vol^*_{H_s^{(2)}}(Y^*)\big ) \bigg| \label{eq:NGUM1}\\
         &\leq |O^*| + \bigg | \minsuffix_G \bigg ( \gamma\cdot |\bar{O}^*| + \gamma\cdot \threshold\cdot |N_{H_{s}^{(2)}}(Y^*)| \bigg ) \bigg| \label{eq:NGUM2} \\
               &\leq |O^*| + \ceil{\frac{ \gamma\cdot |\bar{O}^*| + \gamma\cdot \threshold\cdot |N_{H_{s}^{(2)}}(Y^*)|}{\threshold\cdot \gamma}} \label{eq:NGUM3} \\
                     &= |O^*| + \ceil{\frac{|\bar{O}^*|}{\threshold}} + |N_{H_{s}^{(2)}}(Y^*)|   \nonumber \\
      &=   \ceil{|\bar{O}^*|+\frac{\myAlfa \cdot |\bar{O}^*|}{\oneminusA}} +|O^* \setminus \bar{O}^*|        \nonumber \\
          &=   \ceil{\frac{|\bar{O}^*|}{\oneminusA }} + |N_{H_{s}^{(2)}}(Y^*)| +|O^* \setminus \bar{O}^*|  \nonumber \\
      &\leq \bfactor \cdot |\bar{O}^*| + |N_{H_{s}^{(2)}}(Y^*)|  +|O^* \setminus \bar{O}^*|         \label{eq:NGUM4}\\
        &= \bfactor \cdot |O^*| + |N_{H_{s}^{(2)}}(Y^*)|\nonumber.
    \end{align}
where the proofs of \Cref{eq:NGUM1,eq:NGUM2,eq:NGUM3,eq:NGUM4} follows:
\begin{itemize}
    \item [(\ref{eq:NGUM1})]    The first term, $|O^*|$, accounts for
    the neighbor of each vertex in $O^*$ in the matching
  $M^{(3)}$.
  The second term follows from
   \Cref{lem:greedy}
   since  $G$ is the output of $\greedy( A^+_{\ell}, O_{\ell}^+\cup Y,  \tau^*, \tau + \vol^*_{H^{(2)}_s})$ in Step~\ref{line:4biv} of \Cref{fig:update2}.
   Indeed, we have that if $j \in \bar{O}^+_{\ell}$, then $\vol_G^*(j) = \tau(j)$ and if $j \in Y$, then $\vol_G^*(j) = \vol^*_{H_s^{(2)}}(j)$ so by the definition of $\minsuffix$ (cf.~\Cref{def:min-suffix}) the second term counts the neighbors of $P^*$ in $G$. Note that $V^*(G) = O_\ell^+ \cup Y$, so if the job in $O^*\setminus \bar{O}^*$ exists, it does not appear in $V^*(G)$ and has no effect on $N_{G}(P^*)$.
    \item [(\ref{eq:NGUM2})]
    This follows from the following two statements.
    \begin{itemize}
        \item [(a)] For all $j \in \bar{O}^*$, we have $\tau(j) \leq \gamma$.
            \begin{itemize} [nosep]
            \item [\textbf{Proof.}] Since $j \in \bar{O}^* \subseteq O_{\ell}^+$, \Cref{lem:j in matching tau j le e1}  implies $\tau(j) \leq \gamma$.
            \end{itemize}
        \item [(b)] $\vol_{H_s^{(2)}}^*(Y^*) \leq \gamma \cdot \threshold \cdot |N_{H_s^{(2)}}(Y^*)|.$
         \begin{itemize} [nosep]
            \item [\textbf{Proof.}]
            Since
            $Y^* \subseteq Y$, \Cref{lem:N Y H2_s = X} implies $N_{H_s^{(2)}}(Y^*) \subseteq X$.
           Thus, by \Cref{lem:forall r alg' le e leader},  for all $x \in N_{H_s^{(2)}}(Y^*) \subseteq X$ we have that $\vol_{H_s^{(2)}}(x)\leq \gamma \cdot \threshold$. Therefore,
        \begin{align*}
        \vol_{H_s^{(2)}}^*(Y^*) = \sum_{x \in N_{H_s^{(2)}}(Y^*)}\vol_{H_s^{(2)}}(x)\leq  \gamma \cdot \threshold \cdot |N_{H_s^{(2)}}(Y^*)|.
        \end{align*}
      \end{itemize}

    \end{itemize}
    \item [(\ref{eq:NGUM3})]
         Let $z$ be the last job touched by $\OPT$ before time $\ell$. First note that if $z \in V(G) = A^+_{\ell}$, then $z$ is the only job in $A^+_{\ell}$ that is touched by $\OPT$ during $(s,\ell]$ but not finished. Since $\OPT$ executes SRPT, this implies that $z$ is the largest job in $A^+_{\ell}$. Furthermore, since $\ALG$ processes all jobs in $A^+_{\ell}$ for the same amount of time during $(s,\ell]$, we get that $z$ has the largest remaining processing time among the jobs in $A^+_{\ell}$ at time $\ell$ for $\ALG$. In particular, this means that $z$ is the first element in the order of $A_{\ell}^+$ of $\ALG$ and therefore it belongs to the $\minsuffix$ only if all elements of $A^+_{\ell}$ are in the $\minsuffix$.

        For all $j \in V(G) \setminus \{z\} = A^+_{\ell} \setminus \{z\}$, we argue that $\tau^*(j) \geq \threshold\cdot \gamma$. Since $z$ is the last job touched by $\OPT$ at time $\ell$, and every job in $A^+_{\ell} \setminus \{z\}$ was touched before $\ell$ by $\OPT$, we conclude that $j \not \in O_{\ell}$ holds for all $j \in A^+_{\ell} \setminus \{z\}$. Therefore~\Cref{lem:characterize tau delta} implies $\tau^*(j) \geq \threshold\cdot \gamma$ for all $j \in V(G) \setminus \{z\} = A^+_{\ell} \setminus \{z\}$. If
        $$|V(G)\setminus \{z\}| \ge S := \ceil{\frac{ \gamma\cdot |\bar{O}^*| + \gamma\cdot \threshold\cdot |N_{H_{s}^{(2)}}(Y^*)|}{\threshold\cdot \gamma}},$$
        then a simple volume argument implies that the size of the suffix is at most $S$ and $z$ is not part of the suffix. If on the other hand, $|V(G)\setminus \{z\}| < S$, then $|V(G)| \le S$. As $|V(G)|$ is an upper bound on the suffix size, this implies the target bound of~$\left| \minsuffix_G \left( \gamma\cdot |\bar{O}^*| + \gamma\cdot \threshold\cdot |N_{H_{s}^{(2)}}(Y^*)| \right) \right| \le |V(G)| \le S$

    \item [(\ref{eq:NGUM4})] The inequality $\ceil{\frac{|\bar{O}^*|}{\oneminusA }} \le \bfactor \cdot |\bar{O}^*|$ follows because $|\bar{O}^*|$ is an integer.\qedhere
\end{itemize}
\end{proof}

With the auxiliary lemmas in place, we are ready to prove that the prefix expansion of $\sigma'$ is bounded. Since \Cref{lem:sigma' assignment} showed that
 $\sigma'$ is an assignment,  this implies that $\sigma'$ is a valid assignment and, thus, finishes the proof of~\Cref{lem:nontrivial update assignment} for the case that $\nu > \nu^*$.
Together with the case $\nu \le \nu^*$ that has been established in \Cref{lem:expansion new case 1} of the previous section, this completes the proof of~\Cref{lem:nontrivial update assignment}.

Let $H' = (V,V^*,E,\sigma')$ be the bipartite graph for the weight function $\sigma'$ returned by \Cref{alg:update assignment}.

\begin{lemma} \label{lem:expansion new case 2}
If $\nu^* < \nu$, then $\phi(\sigma') \leq \bfactor.$
\end{lemma}
\begin{proof}
Fix an integer $k$ and let $P^*$ be a $k$-prefix of $V^*(H')$.  We show that $ |N_{H'}(P^*)| \leq \bfactor \cdot |P^*|.$

Recall that $H' = G \cup M^{(3)} \cup H^{(2)}_p$. Let $P^* = P^*_1 \cup P^*_2$ where $P^*_1 := P^* \cap V^*(G \cup  M^{(3)})$ is a prefix of $V^*(G \cup  M^{(3)})$  and $P^*_2 := P^* \cap V^*(H^{(2)}_p)$ is a prefix of $V^*(H^{(2)}_p)$.

We split $P^*_1 := O^* \cup Y^*$ where $Y^* := Y \cap P^*_1$ and $O^* := P^*_1 \setminus Y$.
Therefore, we have
    \begin{align}
     |N_{H'}(P^*)| & \leq |N_{G \cup M^{(3)}}(P^*_1)| + |N_{H^{(2)}_p}(P^*_2)| \label{eq:l1}  \\
     &\leq \bfactor\cdot |O^*|+ |N_{H^{(2)}_s}(Y^*)|  + |N_{H^{(2)}_p}(P^*_2)|  \label{eq:l2} \\
     &=  \bfactor \cdot |O^*| + |N_{H^{(2)}_s}(Y^*)\cup N_{H^{(2)}_p}(P^*_2)|\label{eq:l3}  \\
     &=  \bfactor \cdot |O^*| + |N_{H^{(2)}}\big (Y^* \cup P^*_2\big )| \label{eq:l4} \\
    &\leq  \bfactor \cdot  |O^*| + \bfactor \cdot | Y^* \cup P^*_2| \label{eq:l5} \\
     &= \bfactor \cdot |P^*| \ , \label{eq:l6}
     \end{align}
where we justify \Cref{eq:l1,eq:l2,eq:l3,eq:l4,eq:l5,eq:l6} as follows.
\begin{itemize}
    \item [(\ref{eq:l1})] This follows since $H' = G \cup M^{(3)} \cup H^{(2)}_p$,
    edges of $G \cup M^{(3)}$ are incident to
    $P^*_1$
    and edges of $H^{(2)}_p$ are incident to $P^*_2$.
    \item [(\ref{eq:l2})] This follows from \Cref{lem:nu*lenu NGUM}.
    \item [(\ref{eq:l3})] It is enough to prove disjointness between $N_{H^{(2)}_s}(Y^*)$ and $N_{H^{(2)}_p}(P^*_2)$.

    To this end, we argue that  $N_{H^{(2)}_s}(Y^*) \subseteq X$ and $N_{H^{(2)}_p}(P^*_2) \cap X = \emptyset$, which implies disjointness.
   To see this, observe that $\vol_{H^{(2)}}(X) = d$ by~\Cref{claim:X exsits 4}. Since $X$ is a suffix of $V(H^{(2)})$, the split with threshold $d$ in Step~\ref{line:4biii} of~\Cref{fig:update2} creates $(H^{(2)}_p, H^{(2)}_s)$ with $V(H_s^{(2)}) = X$ and $V(H_p^{(2)}) \cap X = \emptyset$. Hence $N_{H^{(2)}_s}(Y^*) \subseteq V(H_s^{(2)}) = X$ and $N_{H^{(2)}_p}(P^*_2) \cap X \subseteq  V(H_p^{(2)}) \cap X = \emptyset$, which implies that $N_{H^{(2)}_s}(Y^*)$ and $N_{H^{(2)}_p}(P^*_2)$ are disjoint.

    \item [(\ref{eq:l4})] This follows since $(H^{(2)}_p,H^{(2)}_s)$ is a split of $H$.
    \item [(\ref{eq:l5})] It is enough to prove $\phi(H^{(2)}_p) \leq \bfactor$. This follows since $(H^{(2)}_p,H^{(2)}_s)$ is a split of $H^{(2)}$ where  $H^{(2)} = H_p^{(1)}$ and $(H_p^{(1)},H_s^{(1)})$ is a split of $H^{(1)}$ and $H^{(1)}$ is obtained from the canonical assignment $\sigma$, which is valid by \Cref{lem:valid imp canonical} and the fact that there is a valid assignment at time $s$ right before $J_{new}$ arrives.
    \item [(\ref{eq:l6})] This follows from $O^*\cup Y^* = P^*_1$ and $P^* =
      P^*_1 \cup P^*_2.$
      \qedhere
\end{itemize}
\end{proof}

\section{Lower Bounds}
\label{app:lowerbounds}

In this section, we present our lower bounds for $\eps$-clairvoyant algorithms for the problem of minimizing the total flow time on a single machine.

For a fixed algorithm and a fixed instance, we denote by $\delta(t)$ and $\delta(t,1)$ the number of unfinished (active) jobs and the number of unfinished (active) jobs with a remaining processing time of at least $1$ at time $t$ in the algorithm's schedule, respectively, and by $\delta^*(t)$ the number of unfinished (active) jobs at time $t$ in an optimal solution. The following technique is well-known in literature~\cite{MotwaniPT94,AzarLT21,AzarLT22}.

\begin{proposition}\label{prop:dos}
    If there exists an instance such that $\delta(t,1) \geq \rho \cdot \delta^*(t)$ for some constant $\rho$ at some time $t$, then the algorithm has a competitive ratio of at least $\Omega(\rho)$. Moreover, if for every sufficiently large integer $k$ there exists an instance such that $\delta^*(t) = k$ and $\delta(t,1) \geq \rho \cdot \delta^*(t)$ for some constant $\rho$ at some time $t$, then the algorithm has a competitive ratio of at least $\rho$.
\end{proposition}

\begin{proof}
  After time $t$, the adversary releases at each of the next $M$ integer times a job of length $1$. Thus, the best strategy for the algorithm is to run such jobs to completion before any other job with remaining processing time at least $1$ at time $t$.
  Therefore, the algorithm has between time $t$ and $t+M$ at least $\delta(t) + 1$ unfinished jobs. Similarly, we can assume that an optimal solution has at most $\delta^*(t) + 1$ pending jobs during this time. Therefore, as $M$ tends to $\infty$, the competitive ratio of the algorithm is at least $\frac{\delta(t) + 1}{\delta^*(t) + 1} \geq \frac{\rho}{2}$, proving the first part of the statement.

  For the second part, we can analogously conclude that the algorithm's competitive ratio for every instance of the family is at least $\frac{\delta(t) + 1}{\delta^*(t) + 1} \geq \frac{\rho k + 1}{k + 1}$, which tends to $\rho$ as $k \to \infty$.
\end{proof}

\subsection{The Optimal Lower Bound for Deterministic Algorithms}\label{sec:det-lower-bound}

We start with a deterministic lower bound of $\factor$, which is matched by our algorithm.
The construction is based on common arguments used for deterministic lower bounds for the flow time objective~\cite{MotwaniPT94,AzarLT21,AzarLT22}, which applied in a straightforward manner gives a lower bound of $\Omega(\nicefrac{1}{\eps})$. To achieve the tight bound of $\factor$, we analyze the algorithm's state more carefully and apply a delicate scaling argument.

\begin{theorem}
For every $\eps \in (0,1)$, every deterministic $\eps$-clairvoyant algorithm has a competitive ratio of at least $\factor$ for minimizing the total flow time on a single machine.
\end{theorem}
\begin{proof}
Let $k := \ceil{\frac{1-\eps}{\eps}}$. Hence $k+1 = \ceil{\frac{1-\eps}{\oneminusA}} + 1 =  \factor$.
The adversary releases jobs in several rounds. A batch of $k+1$ jobs is released for each round, and the job sizes are declared on the fly based on the algorithm's behavior. The goal for the adversary is to show that at the end of round $i$, the optimal algorithm has $i$ jobs, whereas the algorithm has $i\cdot (k+1)$ jobs. We show that the construction can be done for all $i$. If the goal is achieved, then \Cref{prop:dos} implies that the algorithm has a competitive ratio of $k+1 = \factor$.

\paragraph{First Round.}  We release $J_1$, a set of $k+1$ jobs at time $0$.  Let $t'_1$ denote the first time when the elapsed time of a job in $J_1$ is equal to $1$.
Our goal is to prove that the adversary always ``wins"; that is, there is a time such that $\OPT$ has $1$ job but $A$ has $1+k$ jobs no matter what the algorithm does.  Now, the adversary declares that for all jobs $j \in J_1$,
\[   p_j :=
\begin{cases}
  e_j(t'_1) \cdot (1+\frac{\eps}{1-\eps}) = e_j(t'_1) \cdot \frac{1}{1-\eps} & \text{if } e_j(t'_1) > 0 , \\
\frac{\eps}{1-\eps} & \text{else.}
\end{cases}
\]
At time $t'_1$, every job $j$ in $J_1$ with non-zero elapsed time becomes known, and its remaining time is $r_j = \frac{\eps}{1-\eps} \cdot e_j(t'_1)$. Every job in $J_1$ with zero elapsed time at time $t'_1$ needs $\frac{\eps}{1-\eps}$ more processing units to complete. Let $j_{\max}, j_{\min}$ be the job with the largest and smallest remaining times at time $t'_1$, respectively. Since $k \geq 1$, we can break ties such that $j_{\max} \neq j_{\min}$. Note that $r_{j_{\max}} = \frac{\eps}{1-\eps}$, because there is at least one job with elapsed time equal to $1$. We denote $r(J'_1) = \sum_{i \in J'}r_j$ for all subsets $J'_1 \subseteq J_1$.

\begin{claim} \label{lem:nice eq}
$r(J_1 \setminus \{j_{\max}\}) < 1 + r_{j_{\min}}$.
\end{claim}
\begin{proof}
   This follows because $(\sum_{j \in J_1} r_j) - r_{j_{\max}} - r_{j_{\min}} \leq (k-1)\cdot r_{j_{\max}}= (k-1)\cdot \frac{\eps}{1-\eps} < \frac{1-\eps}{\eps} \cdot \frac{\eps}{1-\eps} = 1.$
\end{proof}

Imagine a smart algorithm $A'$ that uses the 1 unit of time $A$ worked on $j_{\max}$ to work on $J_1 \setminus \{j_{\max}\}$ to fill the remaining time $r_j$ for all $j \in J_1 \setminus \{j_{\max}\}$ and continues to work until all but $j_{\max}$ are completed at time $t''_1$.   On the other hand, the algorithm $A$ would finish the first job at the earliest at time $t'_1 + r_{j_{\min}}$.  By \Cref{lem:nice eq}, $t''_1 < t'_1 + r_{j_{\min}}$. Therefore, at time $t''_1$, $A'$ has $1$ job but $A$ has $k+1 = \factor$ jobs available. The adversary declares victory, and the round ends at time $t''_1$.

\paragraph{Subsequent Rounds.} Round $c \geq 2$ starts at time $t_c = t''_{c-1}$ defined as the time at which the previous round ended. We maintain the following invariant: At the beginning of the round (at time $t_c$), $A'$ has $c-1$ active jobs, whereas the algorithm $A$ has $(c-1)\cdot \factor$ active jobs.
Let $Q_c$ denote the set of active jobs of the previous batches $J_1,\ldots,J_{c-1}$ at time $t_c$ in $A$'s schedule, and let $r^*(t_c)$ be the smallest remaining time of a job in $Q_c$. At time $t_c$, we release a batch $J_c$ of $k+1$ jobs. Let $t_c'$ be the first time after $t_c$ at which $A$ worked on a job $j_c \in Q_c \cup J_c$ during $[t_c,t_c']$ for $\gamma := \frac{r^*(t_c)}{k+1}\cdot \frac{1-\eps}{\eps}$ units.
Note that $\gamma < r^*(t_c)$ since $\frac{1-\eps}{\eps} \leq \ceil{\frac{1-\eps}{\eps}}= k$, which means that every job in $Q_c$ is active at time $t'_c$.

At time $t'_c$, the adversary declares that for all jobs $j \in J_c$,
\[
 p_j :=
\begin{cases}
e_j(t'_c) \cdot (1+\frac{\eps}{1-\eps}) = e_j(t'_c) \cdot \frac{1}{1-\eps} & \text{if } e_j(t'_c) > 0  \\
\frac{\eps}{1-\eps} \cdot \gamma = \frac{r^*(t_c)}{k+1} & \text{else.}
\end{cases}
\]

For all jobs $j \in J_c \cup Q_c$, let $r_j$ be the remaining processing time of job $j$ in $A$ at time $t'_c$. Let $j_{\max} = \argmax_{j \in J_c} r_j$ and note that $j_c \notin J_c \setminus \{j_{\max}\}$.
Note that $r_{j_{\max}} \leq \frac{\eps}{1-\eps} \cdot \gamma \leq r^*(t_c)$.
Thus, there is a job $j_{\min} \in J_c \setminus\{j_{\max}\}$ such that $r_{j_{\min}} \leq r_j$ for all $j \in Q_c \cup J_c$. Analogously to \Cref{lem:nice eq}, we have the following inequality.

\begin{claim} \label{claim:eJ eps lt plus}
    $r(J_c \setminus \{j_{\max}\})  <  \gamma + r_{j_{\min}}.$
\end{claim}
\begin{proof}
       We have $(\sum_{j \in J_c} r_j) - r_{j_{\max}} - r_{j_{\min}} \leq (k-1)\cdot r_{j_{\max}}= (k-1)\cdot \frac{\eps}{1-\eps} \cdot\gamma  < \frac{1-\eps}{\eps} \cdot \frac{\eps}{1-\eps} \cdot \gamma = \gamma.$
\end{proof}

Again, the smart algorithm $A'$ would use $\gamma$ units of time that
$A$ did on $j_c$ to work on all jobs in $J_c
\setminus \{j_{\max}\}$ and continued to work until all are
completed at time $t''_c$. On the other hand, the algorithm $A$ would
complete the first job in this round at the earliest at time $t'_c+r_{j_{\min}}$. By \Cref{claim:eJ eps lt plus}, $t''_c < t'_c +
r_{j_{\min}}$. Therefore, at time  $t''_c$, the algorithm $A'$ has
$(c-1) + 1 = c$ active jobs, whereas $A$ has $(c-1) \cdot \factor +
\factor = c \cdot \factor$ active jobs, establishing the invariant, and
the round ends at time~$t''_c$.
\end{proof}

\subsection{Randomized Lower Bounds}

We continue with a lower bound for randomized $\eps$-clairvoyant algorithms. It matches our deterministic lower bound up to constants, and is based on a commonly used construction for randomized lower bounds for non-clairvoyant flow time minimization~\cite{MotwaniPT94,AzarLT22}. Compared to these previous constructions, we have the additional challenge that an algorithm receives additional information during processing, which makes the analysis of an algorithm's schedule harder and requires additional arguments.

\begin{theorem}\label{theorem:randomized-lb}
  For every $\eps \in (0,1/2)$, every $\eps$-clairvoyant
  randomized algorithm has a competitive ratio of at least
  $\Omega(\factorWithoutCeil)$ for minimizing the total flow time on a
  single machine.
\end{theorem}

We present a randomized instance on which every deterministic
algorithm has an expected competitive ratio of at least
$\Omega(\factorWithoutCeil)$. The theorem then follows using Yao's
principle. For simplicity, assume that $\factorWithoutCeil = 2k$ for
some integer $k \gg 1$.

Fix any deterministic algorithm. We consider a distribution of
instances where at time $0$ we release $n := 2^k$ jobs with integer
processing times $P_j := Y_j + 1$, where the random variables
$Y_j \sim \text{Geom}(\nf12)$ are drawn independently. Note that
$\EX[P_j] = 1 + \EX[Y_j] = 3$. Let $\tau := \floor{3(n-n^{3/4})}$. We
first bound the number of jobs remaining in the optimal solution at
time $\tau$.

\begin{lemma}\label{lemma:randomized-lb-opt}
  $\EX[\delta^*(\tau)] \leq O(\frac{n^{3/4}}{\log n})$.
\end{lemma}

\begin{proof}
  The total processing time $P := \sum_j P_j$ of all jobs is a sum of
  $n$ independent random variables, each with mean $3$ and variance
  $2$. Thus, Chebyshev's inequality gives
  \[
    \pr \left( P \geq 3n + n^{3/4} \right) = \pr \left( P \geq 3n + n^{1/2} \cdot n^{1/4} \right) \leq O \left( \frac{1}{n^{1/2}} \right) \ .
  \]
  Moreover, let $b := \frac{\log n}{4}$ and let $B$ be the number of
  jobs with processing time more than $b$.  Note that $B$ is
  binomially distributed, because it can be written as the sum of $n$
  binary random variables $Y_j = \ind[P_j > b]$ with equal success
  probabilities
  $\pr(P_j > b) = \pr(Y_j > b - 1) = (\frac{1}{2})^{b-1} = 2n^{-1/4}$.
  Therefore, $\EX[B] = n^{-1/4} \cdot n = n^{3/4}$ and
  $\VAR[B] = n^{-1/4} (1-n^{-1/4}) n = O(n^{3/4})$.  Thus, Chebyshev's
  inequality gives
  \[
    \pr \left( B \leq n^{3/4} \right) = \pr \left( B \leq 2 n^{3/8}
      \cdot n^{3/8} - n^{3/4} \right) \leq O \left(
      \frac{1}{n^{3/4}} \right) \ .
  \]
  Therefore, with probability of at least $1 - O(\frac{1}{n^{1/2}})$
  it holds that $P < 3n + n^{3/4}$ and $B > n^{3/4}$.  In this case,
  we compute the maximum number of jobs of size at least $b$ that an
  optimum solution cannot complete until time $\tau$ as follows
  \begin{align*}
    \frac{1}{b} \left(P - \tau \right) \leq \frac{1}{b} \left(3n + n^{3/4} - 3(n-n^{3/4}) + 1 \right)
    \leq O\left( \frac{n^{3/4}}{\log n} \right) \ .
  \end{align*}
  Note that $O(\frac{n^{3/4}}{\log n}) \leq n^{3/4} \leq B$, i.e.,
  there are many long jobs. Since the optimal schedule will process
  these jobs at the end, we have a small number of remaining jobs.
  Specifically, the expected number of alive jobs in an optimum
  solution at time $\tau$ is at most
  \[
    \EX[\delta^*(\tau)] \leq \left(1-O\left(\frac{1}{n^{1/2}}
      \right) \right) \cdot O\left( \frac{n^{3/4}}{\log n} \right) +
    O \left( \frac{1}{n^{1/2}} \right) \cdot n = O\left(
      \frac{n^{3/4}}{\log n} \right). \qedhere
  \]
\end{proof}

Next we want to show that any deterministic algorithm has a large
number of jobs remaining.

\begin{lemma}\label{lemma:randomized-lb-alg}
   $\EX[\delta(\tau,1)] \geq \Omega(n^{3/4})$.
\end{lemma}

\begin{proof}
We first consider the following simplification that can only improve the quality of the algorithm.
We assume that the job becomes known at time $\floor{\myAlfa P_j}$; this gives only more power to the algorithm.

We call a job $j$ \emph{short} if it has
$P_j \leq 2k = \factorWithoutCeil$; otherwise \emph{long}. Observe that for any short job $j$
we have $\floor{\myAlfa P_j} = Y_j$. Let $\cE_1$ be the event that
all $n$ jobs are short.
A simple calculation shows that the probability of a job $j$ being
long is $\pr(P_j > 2k) = \pr(Y_j \geq 2k) = 2^{-(2k-1)} = 2/n^2$,
and hence a union bound gives
\[
\pr(\cE_1) = 1 - n\cdot(2/n^2) = 1-2/n \ .
\]

Let $L := n^{3/4}$, and define $\cE_2$ to be the event that the
algorithm finishes at most $n-\nf{L}2$ jobs during the first
$\tau = \floor{3(n-L)}$ timesteps. Thus, under $\cE_2$ we have $\delta(\tau,1) \geq \nf{L}{2}$.

In the following, we condition on $\cE_1$ and assume that all jobs are short.
Consider another setting where size of the job is revealed
at time $Y_j$, and the algorithm still processes the jobs serially
one after the other. Observe that this does not change the process
when all jobs are short.
To analyze $\cE_2$ under this process, consider the complementary event, and suppose
the algorithm finishes $m = n-\nf{L}2$ jobs in the first $\tau$
timesteps. Say these completed jobs are numbered $1, 2, \ldots, m$,
this means that $\sum_{j = 1}^m P_j \leq \tau$. Using
$P_j = 1 + Y_j$, we get that
$\sum_j Y_j \leq \tau - m = \floor{3(n-L)} -
(n-\nf{L}2)$. Rephrasing, this means that the first $2n - \nf{5L}2$
unbiased coin flips contain at least $n-L$ heads, whereas the
expected number is only $n - \nf{5L}4$. A Chernoff bound gives the
probability of this being $\exp(-O(n^{1/2}))$, and hence $\pr(\cE_2 \mid \cE_1)
 = 1-o(1)$.

Since also $\Pr(\cE_1) = 1-o(1)$, we conclude $\Pr(\cE_1 \cap \cE_2) = 1-o(1)$.
Then
  \[
  \EX[\delta(\tau,1)] \geq \frac{L}{2} \cdot \pr(\cE_1 \cap \cE_2) \geq
    \Omega(L) = \Omega(n^{3/4}) \ ,
\]
which proves the lemma.
\end{proof}

From \Cref{lemma:randomized-lb-alg} and \Cref{lemma:randomized-lb-opt}, we can conclude that
\[
    \frac{\EX[\delta(\tau,1)]}{\EX[\delta^*(\tau)]} \geq \Omega(\log n) = \Omega(k) = \Omega\big( \factorWithoutCeil \big) \ .
\]
Thus, applying \Cref{prop:dos} to every realization of our distribution and finally applying Yao's principle
implies the $\Omega(\factorWithoutCeil)$ gap, as claimed in \Cref{theorem:randomized-lb}.

We finally show that similarly to the deterministic setting, even with randomization, one cannot approach an optimal solution if $\eps$ approaches $1$. In that sense, the power of randomization for $\eps$-clairvoyant flow time minimization seems to be restricted to constants.

\begin{theorem}
	For every $\eps \in (0,1)$, every $\eps$-clairvoyant randomized algorithm has a competitive ratio of at least $\frac{3}{2}$.
\end{theorem}

\begin{proof}

    Let $k \gg 1$ and let $\lambda = \frac{5-\eps}{1-\eps}$. Fix any deterministic algorithm. We construct a distribution of instances as follows. We have $k$ consecutive phases indexed by $k,\ldots,1$, where phase $i$ has length $\lambda^i$. Let $t$ denote the end of the last phase.

At the start $s_i$ of each phase $i$, the adversary releases two jobs $j_1^i$ and $j_2^i$, and selects one job $j^i_{s} \in \{j_1^i,j_2^i\}$ uniformly at random, which we call the \emph{short} job of phase $i$. We call the other job $j_\ell^i$ the \emph{long} job of phase $i$.
    The adversary sets $p_{j^i_s} = \lambda^i$ and $p_{j^i_\ell} = 2\lambda^i$.

	At any point in time during the first $\myAlfa\lambda_i$ time units of a phase $i$, the probability that the algorithm executes the short job is at most $\frac{1}{2}$. This is because none of $\{j_1^i,j_2^i\}$ can become known during the first $\myAlfa\lambda_i$ time units of phase $i$, so the algorithm cannot distinguish the jobs.
	Thus, the probability that $j_s^i$ is processed for at most $\frac{1-\eps}{2}\lambda_i$ time units within the first $\myAlfa\lambda_i$ time units of phase $i$ is a least $\frac{1}{2}$.
    Therefore, with probability of at least $\frac{1}{2}$,
    $r_{j_s^i}(s_{i-1}) \geq \lambda^i - \frac{1-\eps}{2}\lambda_i - \eps \lambda^i = \frac{1-\eps}{2}\lambda_i$. For the long job, we have $r_{j_\ell^i}(s_{i-1}) \geq 2\lambda^i - \lambda^i \geq \lambda^i$.
    Moreover, since
     \[
        t - s_{i-1} \leq \sum_{i'=0}^{\infty} \lambda^{i-1-i'} = \lambda^{i-1} \sum_{i'=0}^{\infty} \lambda^{-i'} = \lambda^{i-1} \cdot \frac{1}{1 - \lambda^{-1}} = \lambda^{i-1} \cdot \frac{5-\eps}{4} = \frac{1-\eps}{4}\lambda^i \ ,
    \]
    we have that $r_{j_\ell^i}(t) \geq \lambda^i - \frac{1-\eps}{4}\lambda^i \geq  \frac{1-\eps}{4}\lambda^i > 1$, and
    with probability at least $\frac{1}{2}$, we have that $r_{j_s^i}(t) \geq \frac{1-\eps}{2} \lambda_i - \frac{1-\eps}{4} \lambda_i = \frac{1-\eps}{4} \lambda_i > 1$.

	Thus, the expected number of jobs released in phase $i$ that have a remaining processing time of at least $1$ at time $t$ is at least $\frac{3}{2}$.
	For time $t$, this implies that $\EX[\delta(t, 1)] \geq \frac{3}{2}k$.

    An optimal solution can finish the short job in every phase, and thus, has at time $t$ only $k$ active jobs; that is, $\delta^*(t) = k$. Finally, applying \Cref{prop:dos} to every realization of our distribution and finally applying Yao's principle gives the claimed bound.
\end{proof}

\section{Reduction from Speed Augmentation to $\eps$-Clairvoyance}
 \label{sec:reduction}

In this section, we prove \Cref{lem:reduction}. We restate it here for convenience.

\lemReduction*

We have argued the proof of the first part in the introduction. We focus on proving the second part. To do so, let us define a new non-clairvoyant algorithm $\SETFI$,
which takes both a set of jobs $\calI$ arriving online, and a set of
forbidden times $I \sse [0,\infty)$. For each time in $I$, the
\SETFI algorithm is forced to idle, else it runs jobs with the shortest
elapsed time. Hence, setting $I = \emptyset$ gives us back \SETF. We show the following ``compression'' statement in the next section:
\begin{restatable} {lemma}{lemdominancecompactify} \label{lem:dominance compactify}
  Suppose $S$ and $S'$ are the schedules produced by \SETFI and \SETF
  on some set of jobs $\calJ$ and forbidden times $I$. Then for any set
  of jobs $\calJ$ and any set $I$ of forbidden times, and any time $t$,
  $e^{\SETFI}_j(t) \leq e^{\SETF}_j(t)$. Therefore, $|\SETF_{\calJ}(t)| \leq |\SETFI_{\calJ}(t)|$ at all time $t$.
\end{restatable}

We are now ready to prove \Cref{lem:reduction}.
\begin{proof}[Proof of \Cref{lem:reduction} (Part 2)]
    Let $\eps, \delta > 0$ be two positive constants where $\eps = \frac{\delta}{1+\delta}$.
    Fix a job instance $\cJ$ where $\SETF$ runs on $\cJ$ with
    $1+\delta$ speed and $\ALG$ runs on $\cJ$ in $\eps$-clairvoyant
    model.  Let $I \subseteq [0,\infty)$ be a set of time intervals
    for which $\ALG$ was working on a known job on $\cJ$. We use $I$
    as a set of forbidden times for $\SETFI$ running on $\cJ'$ where
    $\cJ'$ be $\cJ$ but each job $i \in \cJ'$ has processing time
    $p'_i = (1-\eps)p_i$ where $p_i$ is the processing time of job
    $i \in \cJ$.  Since $\SETFI$'s schedule is identical to $\SLF$ but
    excludes the known jobs in $\ALG$ during the forbidden times $I$,
    we have that for all times $t$,
    \begin{align}\label{eq:sfl-sfl'}
      |\SETFI_{\cJ'}(t)| \leq |\ALG_{\cJ}(t)|.
    \end{align}
    Finally, since we shrink the jobs in $\calJ'$ by $(1-\eps)$, we
    get that $\SETF$ running on $\cJ$ with $1+\delta = 1/(1-\eps)$ speed is
    equivalent to $\SETF$ with 1-speed on $\cJ'$
    Therefore, for all $t$,
    \[ |\SETF_{\cJ,1+\delta}(t)| =
      |\SETF_{\cJ'}(t)| \overset{(\textrm{\Cref{lem:dominance compactify}})}{\leq} |\SETFI_{\cJ'}(t)| \overset{(\ref{eq:sfl-sfl'})}{\leq}  |\ALG_{\cJ}(t)|.\]
    This concludes the reduction. \qedhere

\end{proof}

\subsection{Proof of \Cref{lem:dominance compactify}} \label{sec:dominance compactify}

To prove \Cref{lem:dominance compactify}, we use the following \emph{the water-filling process}:

\begin{defn} [Water-filling System]
Let $k > 0$ be an integer. Given three vectors $x,x',p \in \mathbb{R}_{\geq  0}^k$, we define the following $(x,x',p)$-\emph{water-filling} system $e(t),e'(t) \in \mathbb{R}_{\geq 0}^k$ that evolve over time $t$ as follows. At time $0$, we are given two sets of jars, where the jars in each set are labeled from 1 to $k$, so that the $i$-th jar in the first set and second set contains $e_i(0) := x_i$ and $e'_i(0) := x'_i$ units of water, respectively. In both sets of jars, jar $i$ has capacity $p_i$. The goal is to fill all the jars. Over time, both sets of jars are being filled with water at the same rate, as long as there is a non-full jar. In each set of jars, the non-full jars are filled with water starting from the least loaded jars first (i.e., in SETF manner).
In particular, if a set of jars contains multiple least loaded jars, then the rate of the set of jars is equally distributed among the least loaded jars.
Let $e_i(t)$ and $e'_i(t)$ be the amount of water in the $i$-th jar in the first and second sets, respectively, at time $t$.
\end{defn}

For two vectors $x,x' \in \mathbb{R}_{\geq  0}^k$, we write $x \preceq x'$ if $x_i \leq x'_i$ for all $i \in [k]$, and we write $x \prec x'$ if $x \preceq x'$ and there exists an $i \in [k]$ with $x_i < x'_i$.
We show that $e'(t) \succeq e(t)$ holds at all times $t$ if we start with vectors $x \preceq x' \prec p$.

\begin{restatable}{lemma}{lemwaterfilling} \label{lem:water filling}
    For all integer $k > 0$, let $x, x', p\in \mathbb{R}_{\geq 0}^k$ where $x \preceq x' \preceq p$, let $e$ and $e'$ be the water functions of jars in the $(x,x',p)$-water-filling system. Then, $e(t) \preceq e'(t)$ at all times $t$.
\end{restatable}
\begin{proof}

Suppose for the sake of contradiction that there exists a time $\bar{t}$ such that $e_i(\bar{t}) > e'_i(\bar{t})$ for at least one $i \in [k]$. Let $\bar{t}$ denote the earliest such point in time.

Let $I = [\bar{t}',\bar{t}]$ with $\bar{t}'<\bar{t}$ denote the interval of maximum length among the intervals that end in $\bar{t}$ such that the fill-rate, i.e., one divided by the number of least loaded jars, of both $e$ and $e'$ stays the same during $I$. Let $F \subseteq [k]$ and $F'\subseteq [k]$ denote the sets of elements that receive water during $I$ for $e$ and $e'$, respectively.  By definition of $I$, these sets stay the same during the complete interval $I$.

By choice of $\bar{t}$, we have $\min_{i \in F} e_i(\bar{t}') \le \min_{i \in F'} e'_i(\bar{t}')$ and $\min_{i \in F} e_i(\bar{t}) > \min_{i \in F'} e'_i(\bar{t})$, which implies $|F| < |F'|$ as the fill-up rate of $\frac{1}{|F|}$ for $e$ must be larger than the fill-up rate of $\frac{1}{|F'|}$ for $e'$.

In particular, this means that there is some $i \in F'\setminus F$ that does not receive water during $I$ for $e$, and thus, $e_i(\bar{t}') = e_i(\bar{t})$, but receives water during $I$ for $e'$. This implies
$$
e_i(\bar{t}) = e_i(\bar{t}') \le e'_i(\bar{t}') < e'_i(\bar{t}) = \min_{j \in F'} e'_j(\bar{t})  < \min_{j \in F} e_j(\bar{t}),
$$
where the first inequality follows from the choice of $\bar{t}$, the second inequality follows from $i \in F'$, the last equation holds by definition of the water-filling system, and the last inequality holds by assumption.

However, $e_i(\bar{t}') < \min_{j \in F} e_j(\bar{t})$ means that $i$ needs to receive water during $I$ for $e$ by definition of the water filling mechanism; a contradiction.
\end{proof}

Now, we are ready to prove \Cref{lem:dominance compactify}. We restate it here for convenience.
\lemdominancecompactify*
\begin{proof}
We define a sequence of water-filling processes so that the amount of water in a jar system corresponds to the elapsed times of the active jobs in $\ALGc$. The other system corresponds to the elapsed times of the active jobs in $\ALGi$, and we ``fast-forward" in time using \Cref{lem:water filling}.

The forbidden time $I$ can be represented as the set of maximal time intervals for which $\ALGi$ idles.

Let $P := \{t_{\max}\} \cup  B$ where $t_{\max}$ is the first time that $\ALGc$ completes all the jobs. Let $Q$ be the set of intervals formed by any two consecutive points in $P$.

At any time $t$, let $R(t)$ be the jobs released by time $t$. Let $\hat e^{\ALGi}(t)$ be $e^{\ALGi}(t)$ after restricting to the jobs in $R(t)$. We define $\hat e^{\ALGc}(t)$ similarly. Let $J$ be the set of jars so far (corresponding to the released jobs). Initially $J = \emptyset$.

We argue that the invariant $\hat e^{\ALGi}(t) \preceq \hat e^{\ALGc}(t)$ is maintained for all $t \in [a,b]$ and all $[a,b] \in Q$. To this end, consider the following loop and assume that the invariant holds for time $a$. For the first iteration, this assumption holds trivially, and for all subsequent iterations, the assumption is implied by the invariant for the previous iteration.

For each $[a,b]$ interval in $Q$ ordered by time:
\begin{enumerate} [nosep]
    \item If jobs arrive at time $a$, we add new jars corresponding to the new jobs with zero water to $J$.
    \item Let $I_{ab} = I \cap [a,b]$ be the set of intervals of the forbidden times inside $[a,b]$, and let $\bar I_{ab} = [a,b] \setminus I.$
    \item During $[a,b]$, $\SETFI$ alternates between working and idling modes depending on the intervals $I_{ab}$ and $\bar I_{ab}$.
    \item For each alternating interval $[a',b'] \subseteq [a,b]$ over time:
    \begin{enumerate} [nosep]
        \item If $[a',b']$ corresponds to the working mode for $\SETFI$, then  we have a $(\hat e^{\ALGi}(a),\hat e^{\ALGc}(a),p_J)$-water-filling system during $[a',b']$ where $p_J: R(t) \rightarrow \mathbb{R}_{\geq 0}$ is the processing time of the released jobs at time $a'$. Since $\hat e^{\ALGi}(a') \preceq \hat e^{\ALGc}(a')$,   \Cref{lem:water filling} implies that for all $t \in [a',b'], \hat e^{\ALGi}(t) \preceq \hat e^{\ALGc}(t)$. In particular, the invariant is maintained at time $b'$.
        \item If $[a',b']$ corresponds to the idling mode for $\SETFI$, then  for all $t \in [a',b'], \hat e^{\ALGi}(t) \preceq \hat e^{\ALGc}(t)$ because $\ALGi$ idles during $[a',b']$. In particular, the invariant is maintained at time $b'$.
    \end{enumerate}
\end{enumerate}
Since the invariant is maintained throughout the loop iteration, the lemma follows.
\end{proof}

\section{Simultaneous Release}\label{sec:simultaneous release}

In this section, we consider the special case where all jobs are available at time $0$, and no new jobs arrive over time. In this setting, the sum of flow times collapses to the sum of completion times.
We start with the upper bound of \Cref{thm:main2}. Its proof follows the lines of the original analysis of Round-Robin for the total completion time objective~\cite{MotwaniPT94}.

\begin{lemma}
    For every $\eps \in [0,1]$, the algorithm $\ALG$ is $(2-\eps)$-competitive for minimizing total flow time on a single machine if all jobs arrive at time $0$.
\end{lemma}

\begin{proof}
    Assume that $p_1 < \ldots < p_n$. Note that when all jobs arrive at time $0$, the objective of total flow time becomes the total completion time.
    For any two jobs $i$ and $j$, let $d(i,j)$ denote the total amount of processing on job $i$ in $\ALG$ before time $C_j$. Note that $d(i,j) \leq p_i$.
    We start with the following identity, which is well-known in literature~\cite{MotwaniPT94}.
    \begin{equation}
    \sum_{j=1}^n C_j = \sum_{j=1}^n p_j + \sum_{j=1}^n \sum_{i=1}^{j-1} d(i,j) + d(j,i) \ .
    \end{equation}
    Now consider two jobs $i$ and $j$ with $i < j$. By the description of $\ALG$, if $i$ becomes known at time $t$, $\ALG$ will process $i$ to completion between $t$ and $t + \eps p_i$ before working on any other job. This crucially uses the fact that during this time no new job arrives.
    Since before $t$, $\ALG$ has worked equally on $i$ and $j$, it process an amount equal to $\alpha p_i$ until $t$. Thus, we conclude $d(j,i) = (1-\eps) p_i$, and the above is at most
    \[
    \sum_{j=1}^n p_j + \sum_{j=1}^n \sum_{i=1}^{j-1} (2 - \eps) p_i = \sum_{j=1}^n p_j + (2 - \eps) \sum_{j=1}^n (n-j) p_j  \leq (2 - \eps) \sum_{j=1}^n (n-j+1) p_j = (2 - \eps) \opt \ ,
    \]
    where the final equality uses that an optimal solution schedules jobs in the order of their index.
\end{proof}

We now move to the proof of the randomized lower bound of \Cref{thm:main2}, which is based on the randomized non-clairvoyant lower bound for minimizing the total completion time~\cite{MotwaniPT94}.

\begin{lemma}
    For every $\eps \in [0,1]$, every randomized algorithm has a competitive ratio of at least $2-\eps$ for minimizing the total flow time, even if all jobs arrive at time $0$.
\end{lemma}

\begin{proof}
    Using Yao's principle, we fix a deterministic algorithm and consider a randomized instance with $n$ jobs $1,\ldots,n$ available at time $0$ with processing times drawn independently from the exponential distribution with scale $1$. That is, every job has a size of at least $x$ with probability $e^{-x}$, and $\EX[P_j] = 1$.

    Fix any two jobs $i$ and $j$.
    Let $d(i,j)$ denote the total amount of processing on job $i$ in the algorithm's schedule before the completion time of $j$.
    We compute their expected pairwise delay $\EX[d(i,j) + d(j,i)]$ in the algorithm's schedule. To this end, consider for any fixed processing amount $y$
    the event $\cE(y)$ that neither $i$ nor $j$ is known after the algorithm processed a total amount of $y$ on both $i$ and $j$ together. Let $y_i$ denote the part of $y$ used for processing job $i$. Note that
    \[
        \pr(\cE(y)) = \pr((1-\eps) P_i > y_i) \cdot \pr(\alpha P_j > y - y_i) = e^{-y/(1-\eps)} \ .
    \]
    Therefore, the expected pairwise delay of $i$ and $j$ while both are unknown is equal to $\int_{0}^{\infty} \pr(\cE(y)) \,dy = 1-\eps$.
    Furthermore, the expected pairwise delay of $i$ and $j$ while at least one is known is equal to
    \[
        \EX[\min\{\eps P_i,\eps P_j\}] = \eps \int_{0}^{\infty} \pr(P_i > x) \cdot \pr(P_j > x) \,dx = \frac{1}{2}\eps \ .
    \]
    Thus, $\EX[d(i,j) + d(j,i)] = 1 - \frac{\eps}{2}$. Since the total completion time of the algorithm's schedule $\alg$ can be written as $\sum_{j=1}^n P_j + \sum_{j=1}^n \sum_{i=1}^{j-1} d(i,j) + d(j,i)$, we conclude
    that $\EX[\alg] \geq n + \sum_{j=1}^n \sum_{i=1}^{j-1} (1 - \frac{\eps}{2})$.

    We now study the optimal objective value $\opt$ for this instance. In this scenario, an optimal schedule schedules job in non-decreasing order of their processing times.
    Thus, the expected pairwise delay of $i$ and $j$ in the optimal schedule is equal to $\EX[\min\{P_i,P_j\}] = \frac{1}{2}$. Hence, $\EX[\opt] = n + \sum_{j=1}^n \sum_{i=1}^{j-1} \frac{1}{2}$.

    We conclude that the competitive ratio is at least
    \[
        \frac{n + \sum_{j=1}^n \sum_{i=1}^{j-1} (1 - \frac{\eps}{2})}{n + \sum_{j=1}^n \sum_{i=1}^{j-1} \frac{1}{2}} \ ,
    \]
    which approaches $2-\eps$ as $n$ goes to $\infty$, and thus, implies the theorem.
\end{proof}

\section{Closing Remarks}

In this work, we studied the classical problem of minimizing the total
flow time on a single machine, in the $\eps$-clairvoyant model. In
this model, the algorithm does not know the job length up-front, but
instead ``learns'' the processing time of each job $j$ only when an
$\eps$ fraction of the job remains to be processed. (For instance, the
behavior of the job's execution for its first $(1-\eps)$ fraction may
allow the algorithm to get its processing time.)
Our main result
is an $\eps$-clairvoyant algorithm with competitive ratio $\factor$,
which we show is the best possible. We also relate this model to the
resource augmentation model, showing that our result implies a result
in the non-clairvoyant model with speed-augmentation.

Our work suggests several interesting directions in different settings: Can we get
$\eps$-clairvoyant algorithms for other scheduling problems, thereby
interpolating between the clairvoyant and non-clairvoyant settings? It
would be interesting to see if the models of $\eps$-clairvoyance and
semi-clairvoyance can be fruitfully combined: e.g., what if the
algorithm obtains only a noisy estimate of the true processing time at
some point in its execution?  Can we better understand the connections
between $(1+\eps)$-speed-augmentation and $\eps$-clairvoyance?

\section*{Acknowledgement}  This research was partially supported by the ERC Starting Grant (CODY 101039914), Dr. Max Rössler, the Walter Haefner Foundation, and the ETH Zürich Foundation, and by NSF awards CCF-1955785, CCF-2006953, and CCF-2224718. Part of this work was done at the Simons Institute for the Theory of Computing, UC Berkeley, during the program “Data Structures and Optimization for Fast Algorithms”, and at Google Research.  SY would like to thank Eric Torng and Bala Kalyanasundaram for their insightful discussion. AL was supported by the ``Humans on Mars Initiative'', funded by the Federal State of Bremen and the University of Bremen. The work of JS was supported by the research project \emph{Optimization for and with Machine Learning (OPTIMAL)}, funded by the Dutch Research Council (NWO), grant number OCENW.GROOT.2019.015.

\addcontentsline{toc}{section}{Acknowledgments}

{\small
\printbibliography
}
\appendix

\section{Omitted Proofs} \label{appendix:omitted proofs}

\begin{proof}[Proof of \Cref{obs:known-jobs-block-earlier-jobs}]
    Fix an arbitrary $j' \not= j$ with $q_{j'} < t'$ that is unknown at $t'$. Since the known $j$ is touched at $t'$, the algorithm's definition implies $r_j(t') \le \threshold e_{j'}(t')$. Thus, because $j$ is touched at $t',$ $r_j(t'^+) < \threshold e_{j'}(t^+)$ for the point in time $t'^+$ immediately after $t'$. As the remaining time of $j$ can only decrease over time, this implies that \ALG will always prefer $j$ over $j'$ until $j$ completes. Thus, $j'$ is not touched during $(t',c_j]$.
\end{proof}

\begin{proof}[Proof of \Cref{lem:new job joins RR}]
    Let $\ell$ be the first time that $i$ is touched after $t$. Observe that $t < \ell \leq t'$. If $j$ was completed by time $\ell$, then we are done. So, assume that $j \in \ALG(\ell)$.  We first claim that $j$ cannot be a known job at time $\ell$. Suppose otherwise that $j$ becomes known at time $\ell$. Let $t_k < \ell$ be the time at which $j$ became known.
 Therefore, \[ \eta_j(t_k) = r_j(t_k) = \threshold\cdot e_j(t_k).\] Also,
\[
\begin{array}{rcll}
    \eta_j(t_k)& \leq & \eta_i(t_k) & \text{(\ALG works on $j$ at time $t_k$)} \\[5pt]
    & \iff&  \\
    e_j(t_k) & \leq & e_i(t_k) & \text{} \\[8pt]
    & = & e_i(t) & \text{ (\ALG does not work on $i$ before time $\ell$ and $t_k < \ell$)}
\end{array}
\]

On the other hand,
\[
\begin{array}{rcll}
    r_j(t_k)& \geq & r_j(\ell) & \text{(Remaining time cannot increase)} \\[5pt]
     & > & \eta_i(\ell)  & \text{(\ALG works on $i$ at time $\ell$)} \\[8pt]
    & = & \threshold\cdot e_i(\ell)  & \\[8pt]
     & = & \threshold\cdot e_i(t)   & \text{(\ALG does not work on $i$ before time $\ell$)}
\end{array}
\]
Since $r_j(t_k) = \threshold\cdot e_j(t_k)$, we obtain $e_j(t_k) > e_i(t)$, contradicting the inequality $e_j(t_k) \leq e_i(t)$.

Thus, $j$ must be unknown at time $\ell$. Since $j$ arrives at time $t$ during which $i$ was unknown, and $i$ was touched at time $\ell$ while $j$ is unknown, we conclude that $i$ and $j$ must have the same elapsed time at time $\ell$.  From $\ell$ to $t'$, if $j$ becomes known at some point between $\ell$ and $t'$ then $j$ must be completed before $i$ is processed again at time $t'$. Otherwise, $j$ must have the same elapsed time with $i$ throughout until $t'$.
\end{proof}

\subsection{Proof of~\Cref{lem:prop H'}}\label{sec:omitted}

We proof~\Cref{lem:prop H'}. For the sake of readability, we restate the lemma here.

\lemPropHPrime*

We split the proof into the two cases (1) $\nu \leq \nu^*$ and (2) $\nu^* < \nu$.

\paragraph{Case (1):} Assume $\nu \leq \nu^*$. Thus, \Cref{alg:update assignment} calls \Cref{fig:update1} in Step~\ref{line:3}.

\begin{proof} [Proof of \Cref{lem:prop H'} $(\nu \leq \nu^*)$]

We prove each item of the lemma in the stated order.
\begin{enumerate}
       \item Fix $j \in J_{new}$. An edge $(j,j^*)$ of the matching $M^{(1)}$ is created in Step~\ref{line:1b} of \Cref{alg:update assignment} with weight $\vol_{M^{(1)}}(j)=p_j$, then its weight is reduced by $\Delta(j)$ at Step~\ref{line:2a} of \Cref{alg:update assignment}. We argue in the following about the change in its weight in \Cref{fig:update1}.
       \begin{enumerate}
       \item If $j \in O_{\ell}^+$, then $j$ is part of $M^{(2)}_p$. Its weight is further reduced by $\tau(j)$ at Step~\ref{line:3bi} of \Cref{fig:update1}. Thus, in $M^{(3)}$, the mass of $j$ has been reduced by $\Delta(j)+\tau(j)$ and we can conclude with $\vol_{M^{(1)}}(j)-\vol_{H'}(j) = \Delta(j)+\tau(j)$.
       \item If $j \in A_{\ell}^+$, then $\tau(j) = 0$ by \Cref{def:O ell plus}. In this case, $j$ is part of $M^{(2)}_s$ and merged into $H^{(3)}$. By \Cref{lem:merge},  $\vol_{H^{(3)}}(j) = \vol_M(j)$ since $M$ and $H$ are disjoint.
       Thus, the weight of $j$ in $H^{(3)}$ is $\vol_{H^{(3)}}(j) = \vol_M(j) = p_j - \Delta(j)$.  We argue that the weight of $j$ does not change after this step. Ultimately, $j$ will end up in $G \cup H^{(2)}_p$. The split in Step~\ref{line:3bii} in \Cref{fig:update1}, reduces the weight of $j$ in $H_p^{(3)}$ to $\vol_{H_p^{(3)}}(j) = \vol_{H^{(3)}}(j) - \vol_{H_s^{(3)}}(j)$. Afterwards, the $\greedy$ in Step~\ref{line:3biii} increases the weight of $j$ in $G$ to $\vol_{G}(j) = \vol_{H_s^{(3)}}(j)$. Hence, $\vol_{H'}(j) = \vol_{H_p^{(3)} \cup G} = \vol_{H^{(3)}}(j) = p_j - \Delta(j)$ and $\vol_{M^{(1)}}(j)-\vol_{H'}(j) = \Delta(j) = \Delta(j)+\tau(j)$ using $\tau(j) = 0$.

       \item Else, we have $j \not\in O_\ell^+ \cup A_\ell^+$, which implies $j \in D_\ell$. By Step~\ref{line:2a} of \Cref{alg:update assignment}, we have $\vol_{M^{(2)}_d}(j) = \vol_{M^{(2)}}(j) = p_j - \Delta(j)$.  Since $j \in D_\ell$ implies $j \not\in V(G) \cup V(M^{(3)}) \cup V(H_p^{(3)})$, this means $\vol_{H'}(j) = \vol_{M^{(2)}_d}(j) = p_j - \Delta(j)$. Using that $\tau(j) = 0$, we can conclude with $\vol_{M^{(1)}}(j)-\vol_{H'}(j) = \Delta(j) = \Delta(j) + \tau(j)$.
       \end{enumerate}
       \item After Step~\ref{line:2} of \Cref{alg:update assignment}, the suffix of $V(H^{(1)})$ in $H^{(1)}$ is peeled off by  $\min\{\nu,\nu^*\} = \nu$. By \Cref{fact:volumes nu}, this corresponds to the total volume in $K(s) \setminus K(\ell)$ in $\ALG$ at time $s$. Since the suffix of $H^{(1)}$ has the same order as the jobs in $K(s) \cup U(s)$ processed by $\ALG$, and $\ALG$ does not process $U(s)$, we conclude that every job $j \in K(s) \setminus K(\ell)$ in $H^{(1)}$ is split into $H_{s}^{(1)}$ right after Step~\ref{line:2} and we are done. Note that \Cref{fig:update1} does not affect these volumes.
       \item Fix $j \in \ALG_{J}(s) \setminus (K(s) \setminus K(\ell))$. By the argument in the previous case, $j$ was not affected by Step~\ref{line:2} of \Cref{alg:update assignment}. We argue that its volume remains unchanged after Step~\ref{line:3}. Indeed, if $j \not \in X$, then there is no change to $j$ in \Cref{fig:update1}. If $j \in X$, then it is enough to show that volume of $j$ does not change in \Cref{fig:update1}.
       This follows because, in $G \cup H^{(3)}_p$, $\textsc{Split}$ in Step~\ref{line:3bii} of \Cref{fig:update1} reduces the weight of $j$ in $ H^{(3)}_p$ to $\vol_{H_p^{(3)}}(j) = \vol_{H^{(3)}}(j) - \vol_{H_s^{(3)}}(j)$, and  $\greedy$ in Step~\ref{line:3biii} of \Cref{fig:update1} increases the weight of $j$ in $G$ to $\vol_{G}(j) = \vol_{H_s^{(3)}}(j)$.
       Hence, $\vol_{H'}(j) = \vol_{H_p^{(3)} \cup G} = \vol_{H^{(2)}}(j)$ and the weight of $j$ is not changed in \Cref{fig:update1}.

       \item   Fix $j \in J_{new}$.  A edge $(j,j^*)$ of the matching $M^{(1)}$ is created at Step~\ref{line:1b} of \Cref{alg:update assignment} with weight $\vol^*_{M^{(1)}}(j) = p_j$, then the weight is reduced by $\Delta(j)$ at Step~\ref{line:2a} of \Cref{alg:update assignment}. We argue the change in weight of $j$ in \Cref{fig:update1}.
       \begin{enumerate}
       \item If $j \in O_{\ell}^+$, then $j$ is part of $M^{(3)}$ for which we reduce its weight by another $\tau(j)$ at Step~\ref{line:3bi} of \Cref{fig:update1}. At Step~\ref{line:3biii}, $\greedy$ adds the weight $\tau(j)$, and so the total weight reduction of $j$ is $\Delta(j)+\tau(j) - \tau(j) = \Delta(j) = \Delta + \tau^*(j)$ using that $j \in O_{\ell}^+$ implies $\tau^*_j = 0$
       \item If $j \in A_{\ell}^+$, then
        in this case, $j$ is part of $M^{(2)}_s$ and merged with $H^{(2)}$ into $H^{(3)}$ in \Cref{fig:update1}. By \Cref{lem:merge}, we have $\vol^*_{H^{(3)}}(j) = \vol^*_{M^{(2)}}(j) +  \vol^*_{H^{(2)}}(j) =  \vol^*_{M^{(2)}}(j)$, where the last inequality uses that $M^{(2)}$ and $H^{(2)}$ are disjoint. This means $\vol^*_{H^{(3)}}(j) =  \vol^*_{M^{(2)}}(j) = p_j - \Delta(j)$.

        We argue that all such jobs are also part of $V^*(H^{(3)}_s)$. To see this, notice that
 $$\vol_{H_s^{(3)}}^*(V^*(H_s^{(3)})) + \vol_{H_s^{(1)}}^*(V^*(H_s^{(1)})) = \nu + T = \ell - s - \Delta(J_{new})$$
            by~\Cref{fact:total work ALG OPT}. Thus, $\vol_{H_s^{(3)}}^*(V^*(H_s^{(3)})) =  \ell - s - \Delta(J_{new}) -  \vol_{H_s^{(1)}}^*(V^*(H_s^{(1)}))$, which means that $\vol_{H_s^{(3)}}^*(V^*(H_s^{(3)}))$ contains the total processing volume of $\OPT$ during $[s,\ell]$ apart from the already shaved of $\vol_{H_s^{(1)}}^*(V^*(H_s^{(1)}))$ and the already removed common volume $\Delta(J_{new})$. Since all $i \in A_\ell^+$ have $\tau^*(i) > 0$, they all receive processing by $\OPT$ during $[s,\ell]$. By the fact that $\OPT$ runs SRPT and by the assumption that no new jobs are released during $(s,\ell]$, this means that they must be part of the suffix $V^*(H_s^{(3)})$ as otherwise they would not be processed by $\OPT$.

           The fact that $j \in V^*(H^{(3)}_s)$ means that the volume of $j$ is further reduced by $\vol^*_{H^{(3)}_s}(j)$. For all members $j$ of  $V^*(H^{(3)}_s)$, except for possibly the very first one, we have $\vol^*_{H^{(3)}_s}(j) = \vol^*_{H^{(3)}}(j) = p_j - \Delta(j)$ and the total reduction is $p_j$, which is equal to $\Delta(j) + \tau^*(j)$ using the fact that $\OPT$ finishes all jobs it touches during $[s,\ell]$ except for possibly one. If $j$ is the first job of  $V^*(H^{(3)}_s)$, then we might have $\vol^*_{H^{(3)}_s}(j) <  p_j - \Delta(j)$ if $\OPT$ does not finish $j$ by time $\ell$. However, then the volume bound  $\vol^*(V^*(H_s^{(3)})) =  \ell - s - \Delta(J_{new}) -  \vol^*(V^*(H_s^{(1)}))$ implies $\vol^*_{H^{(3)}_s}(j) = \tau^*(j)$, and the total reduction in the weight of $j$ is also $\Delta(j)+ \tau^*(j)$.

       \item Else, we have $j \not\in O_\ell^+ \cup A_\ell^+$, which implies $j \in D_\ell$. By Step~\ref{line:2a} of \Cref{alg:update assignment}, we have $\vol^*_{M^{(2)}_d}(j) = \vol^*_{M^{(2)}}(j) = p_j - \Delta(j)$.  Since $j \in D_\ell$ implies $j \not\in V^*(G) \cup V^*(M^{(3)}) \cup V^*(H_p^{(3)})$, this means $\vol^*_{H'}(j) = \vol^*_{M^{(2)}_d}(j) = p_j - \Delta(j)$. Using that $\tau^*(j) = 0$, we can conclude with $\vol_{M^{(1)}}(j)-\vol_{H'}(j) = \Delta(j) = \Delta(j) + \tau^*(j)$.
       \end{enumerate}
       \item It is enough to prove that, at the end, the algorithm
         peels the suffix of $V^*(H^{(1)})$ in $H^{(1)}$ for $\nu^*$
         unit of volume. Indeed, Step~\ref{line:2} of \Cref{alg:update assignment} peels off for
         $\min\{\nu,\nu^*\} = \nu$, and \Cref{fig:update1} peels off
         the remaining $\tau(J_{new}) - \tau^*(J_{new})  = \nu^* - \nu$
         units.
        The last equality follows from \Cref{fact:total
           work ALG OPT}. \qedhere
 \end{enumerate}
 \end{proof}

\paragraph{Case (2):} Assume $\nu^* < \nu $. Thus, \Cref{alg:update assignment} calls \Cref{fig:update2} in Step~\ref{line:3}.

\begin{proof} [Proof of \Cref{lem:prop H'} $(\nu^* < \nu)$]  We prove each item of the lemma in the stated order.

\begin{enumerate}
   \item Fix $j \in J_{new}$. We want to show that $\vol_{M^{(1)}}(j) - \vol_{H'}(j) = \Delta(j) + \tau(j)$.
       An edge $(j,j^*)$ of the matching $M^{(1)}$ is created in Step~\ref{line:1b} of \Cref{alg:update assignment} with weight $p_j$. Hence, $\vol_{M^{(1)}}(j) = p_j$.
       Then the weight of $j$ in $M^{(2)}$ is reduced by $\Delta(j)$ in Step~\ref{line:2a} to $\vol_{M^{(2)}}(j) = p_j - \Delta(j)$. We argue about the change of its weight in \Cref{fig:update2}.

       \begin{enumerate}
       \item If $j \in O_{\ell}^+$, then Step~\ref{line:4bi} of \Cref{fig:update2} reduces  $j$'s weight in $M^{(3)}$ by another $\tau(j)$ to $\vol_{M^{(3)}}(j) = p_j - \Delta(j) - \tau(j)$.
        Furthermore, we have $\vol_{H_p^{(2)}}(j) = 0$ since $j$ does not appear in $H_p^{(2)}$ and $\vol_{G}(j) = 0$ since $j \not\in V(G)$.
        Hence, $\vol_{H'}(j) =\vol_{M^{(3)}}(j) + \vol_{H_p^{(2)}}(j) + \vol_{G}(j) = p_j - \Delta(j) - \tau(j)$ and we can conclude with $\vol_{M^{(1)}}(j) - \vol_{H'}(j) = \Delta(j) + \tau(j)$.

       \item If $j \in A_{\ell}^+$, then $\tau(j) = 0$ by \Cref{def:O ell plus}. It is enough to prove that there is no change in the weight of $j$ in \Cref{fig:update2}.
        In fact, we have $\vol_{M^{(3)}}(j) = p_j - \delta(j) - \tau^*(j)$ by Step~\ref{line:4bii}, and $\vol_G(j) = \tau^*(j)$ by definition of $\greedy$ by Step~\ref{line:4biii} of \Cref{fig:update2}.
        Furthermore, we have $\vol_{H_p^{(2)}} = 0$ since $j \not\in V(H_p^{(2)})$.
        We can conclude that $\vol_{H'}(j) = \vol_{M^{(3)}}(j) + \vol_{H_p^{(2)}}(j) + \vol_{G}(j) =  \vol_{M^{(3)}}(j) + \vol_{G}(j) = p_j - \Delta(j)$ and thus  $\vol_{M^{(1)}}(j) - \vol_{H'}(j) = \Delta(j) = \Delta(j) + \tau(j)$.

              \item Else, we have $j \not\in O_\ell^+ \cup A_\ell^+$, which implies $j \in D_\ell$. By Step~\ref{line:2a} of \Cref{alg:update assignment} and Step~\ref{line:4b} of \Cref{fig:update2}, we have $\vol_{M^{(3)}} = \vol_{M^{(2)}}(j) = p_j - \Delta(j)$.  Since $j \in D_\ell$ implies $j \not\in V(G) \cup V(H_p^{(2)})$, this means $\vol_{H'}(j) = \vol_{M^{(3)}}(j) = p_j - \Delta(j)$. Using that $\tau(j) = 0$, we can conclude with $\vol_{M^{(1)}}(j)-\vol_{H'}(j) = \Delta(j) = \Delta(j) + \tau(j)$.
       \end{enumerate}

              \item After Step~\ref{line:2} of \Cref{alg:update assignment}, the suffix of $V(H^{(1)})$ in $H^{(1)}$  is peeled off by  $\min\{\nu,\nu^*\} = \nu^*$. After Step~\ref{line:4biii} of \Cref{fig:update2}, the suffix of $V(H^{(1)})$ in $H^{(1)}$ is  peeled off further by $d = \nu - \nu^*$. Therefore, in total, the suffix of $V(H^{(1)})$ is peeled off by $\nu$.  By \Cref{fact:volumes nu}, this corresponds to the total volume in $K(s) \setminus K(\ell)$ in $\ALG$ at time $s$. Since the suffix of $H^{(1)}$ has the same order as the jobs in $K(s) \cup U(s)$ processed by $\ALG$ and $\ALG$ does not process $U(s)$, we conclude that the volume of each job $j \in K(s) \setminus K(\ell)$ in $H^{(1)}$ at the end is reduced to zero.

    \item Fix $j \in \ALG_{J}(s) \setminus (K(s) \setminus K(\ell))$. By the argument in the previous line, the weight of $j$ was not affected.

       \item Fix $j \in J_{new}$.  An edge $(j,j^*)$ of the matching $M^{(1)}$ is created at Step~\ref{line:1b} of \Cref{alg:update assignment} with weight  $\vol^*_{M^{(1)}}(j) = p_j$, then its weight in $M^{(2)}$ is reduced by $\Delta(j)$ to $\vol^*_{M^{(2)}}(j) = p_j - \Delta(j)$ in Step~\ref{line:2a} of \Cref{alg:update assignment}. We argue about the change in its weight in \Cref{fig:update2}:
       \begin{enumerate}

       \item If $j \in O_{\ell}^+$, then $\tau^*(j) = 0$. Step~\ref{line:4bi} of \Cref{fig:update2} reduces the weight of $j$ in $M^{(3)}$ by $\tau(j)$ to $\vol^*_{M^{(3)}}(j) = p_j - \Delta(j) - \tau(j)$. However, the $\greedy$ in Step~\ref{line:4biv} increases the weight of $j$ in $G$ to $\vol_G(j) = \tau(j) + \vol^*_{H_s^{(2)}}(j) = \tau(j)$, where $\vol^*_{H_s^{(2)}}(j) = 0$ follows from $j \not\in V^*(H_s^{(2)})$.
       Hence, $\vol^*_{H'}(j) = \vol^*_{M^{(3)}}(j) + \vol^*_{G}(j) + \vol^*_{H_p^{(2)}}(j) = p_j - \Delta(j)$, where we use that $\vol^*_{H_p^{(2)}}(j) = 0$ follows from $j \not\in V^*(H_p^{(2)})$.
       We can conclude with $\vol^*_{M^{(1)}}(j)-\vol^*_{H'}(j) = \Delta(j) = \Delta(j) + \tau^*(j)$.
       \item If $j \in A_{\ell}^+$, then Step~\ref{line:4bii} reduces the weight of $j$ in $M^{(3)}$ to $\vol_{M^{(3)}}^*(j) = p_j - \Delta(j) - \tau^*(j)$. Since $j \not\in V^*(G)$ and $j \not\in V^*(H_p^{2})$, we get $\vol^*_{H'}(j) = \vol^*_{M^{(3)}} = p_j - \Delta(j) - \tau^*(j)$. Hence, $\vol^*_{M^{(1)}}(j)-\vol^*_{H'}(j) = \Delta(j) + \tau^*(j)$
       \item Else, we have $j \not\in O_\ell^+ \cup A_\ell^+$, which implies $j \in D_\ell$. By  Step~\ref{line:2a} of \Cref{alg:update assignment} and Step~\ref{line:4b} of \Cref{fig:update2}, we have $\vol^*_{M^{(3)}} = \vol^*_{M^{(2)}}(j) = p_j - \Delta(j)$.  Since $j \in D_\ell$ implies $j \not\in V^*(G) \cup V^*(H_p^{(2)})$, this means $\vol^*_{H'}(j) = \vol^*_{M^{(3)}}(j) = p_j - \Delta(j)$. Using that $\tau^*(j) = 0$, we can conclude with $\vol^*_{M^{(1)}}(j)-\vol^*_{H'}(j) = \Delta(j) = \Delta(j) + \tau^*(j)$.
       \end{enumerate}
       \item It is enough to prove that at the end the algorithm peels
         the suffix of $V^*(H^{(1)})$ in $H^{(1)}$ for
         $\nu^*$. Indeed, Step~\ref{line:2} of \Cref{alg:update assignment} peels off for
         $\min\{\nu,\nu^*\} = \nu^*$.  \qedhere
       \end{enumerate}
\end{proof}

\end{document}